%% file: ouazad_kahn_2023.tex
\documentclass[11pt,english]{article}
\usepackage[T1]{fontenc}
\usepackage[latin9]{inputenc}
\usepackage{geometry}
\usepackage{babel}
\usepackage{float}
\usepackage{amsmath}
\usepackage{amsthm}
\usepackage{graphicx}
\usepackage{setspace}
\usepackage[unicode=true]
 {hyperref}

\makeatletter

\providecommand{\tabularnewline}{\\}

\usepackage{color}
\usepackage{rotfloat}
 
 \usepackage{array}
 \newcolumntype{L}[1]{>{\raggedright\let\newline\\\arraybackslash\hspace{0pt}}m{#1}}
 \newcolumntype{C}[1]{>{\centering\let\newline\\\arraybackslash\hspace{0pt}}m{#1}}
 \newcolumntype{R}[1]{>{\raggedleft\let\newline\\\arraybackslash\hspace{0pt}}m{#1}}

\usepackage{caption}
\usepackage{harvard}
\usepackage{booktabs}
\usepackage{mathtools}

\usepackage{titling}

\usepackage{todonotes}

\usepackage{longtable}

\usepackage{mathtools, nccmath}

\DeclarePairedDelimiter{\nint}\lfloor\rceil

\@ifundefined{showcaptionsetup}{}{%
 \PassOptionsToPackage{caption=false}{subfig}}
\usepackage{subfig}
\makeatother

\begin{document}
\title{Mortgage Securitization Dynamics in the Aftermath of Natural Disasters:
A Reply\thanks{We thank Robert Huang for research assistance. We thank Benjamin Keys
for stimulating discussions about the emerging empirical literature.
Code for this paper is available at \emph{\protect\href{https://github.com/aouazad/Mortgage-Securitization-Natural-Disasters-Reply.git}{https://github.com/aouazad/Mortgage-Securitization-Natural-Disasters-Reply.git}.}}}
\author{Amine Ouazad\thanks{Associate professor, Rutgers Business School, New Brunswick and Newark,
100 Rockafeller Rd, Piscataway, NJ 08854 \emph{aouazad@business.rutgers.edu},
and Associate professor, HEC Montreal, 3000 Chemin de la C�te Sainte
Catherine, Montreal, H3T 2A7, Canada.}~~~~~Matthew E. Kahn\thanks{Provost Professor, Dana and David Dornsife College of Letters, Arts
and Sciences, Department of Economics, University of Southern California.
Kaprielian Hall, 3620 S Vermont Ave, Los Angeles, CA 90089 \emph{kahnme@usc.edu.}}}
\date{May 2023}
\maketitle
\begin{abstract}
\onehalfspacing Climate change poses new risks for real estate assets.
Given that the majority of home buyers use a loan to pay for their
homes and the majority of these loans are purchased by the Government
Sponsored Enterprises (GSEs), it is important to understand how rising
natural disaster risk affects the mortgage finance market. The climate
securitization hypothesis (CSH) posits that, in the aftermath of natural
disasters, lenders strategically react to the GSEs conforming loan
securitization rules that create incentives that foster both moral
hazard and adverse selection effects. The climate risks bundled into
GSE mortgage-backed securities emerge because of the complex securitization
chain that creates weak monitoring and screening incentives. We survey
the recent theoretical literature and empirical literature exploring
screening incentive effects. Using regression discontinuity methods,
we test key hypotheses presented in the securitization literature
with a focus on securitization dynamics immediately after major hurricanes.
Our evidence supports the CSH. We address the data construction issues
posed by LaCour-Little et. al. and show that their concerns do not
affect our main results. Under the current \textquotedblleft rules
of the game,\textquotedblright{} climate risks exacerbates the established
lemons problem commonly found in loan securitization markets. 
\end{abstract}
\vfill{}

\pagebreak{}

\clearpage{}

\section{Introduction: The Climate Securitization Hypothesis (CSH)}

Climate change poses new risks for real estate assets. When a major
natural disaster, such as Hurricane Ian in 2022, strikes an area,\footnote{NOAA estimates suggest a cost of 112.9 billion dollars in the United
States. In Fort Myers Beach, 3,100 structures were either destroyed
or damaged. Census data suggests there were 9,692 housing units in
this Census place, 89\% owner-occupied. } there is considerable uncertainty about how the dynamic of the local
real estate market is and will be affected. The demand to live in
the area could decline as home buyer expectations about short run
and long run local quality of life changes~\cite{keys2020neglected}.
The disaster causes physical damage to homes and infrastructure. Local
businesses may close. The pace of Federal disaster transfers is unknown.
Unprepared home owners may be caught in a cash crunch that limits
their ability to repay their debts. 

A new and expanding literature provides convincing evidence of the
impact of natural disasters on delinquencies, defaults, foreclosures,
and prepayments \cite{ratcliffe2020bad,issler2021housing,holtermans2022climate,biswas2023california,ho2023we}.
The seminal foreclosure externality literature~\cite{gerardi2015foreclosure}
has documented evidence of the association between neighbor foreclosures
and reductions in home prices. Given that a natural disaster creates
spatially clustered damage to homes, the foreclosure effect has the
potential to create a type of \textquotedblleft domino effect\textquotedblright{}
such that an increased count of affected home owners may be at risk
of defaulting on mortgage payments. Lenders have strong incentives
to be aware of these potential short run risks they face in terms
of being repaid for their residential real estate loans. 

The Climate Securitization Hypothesis states that, for lenders, the
\emph{value of the securitization option} increases in the face of
rising climate risk. The CSH empirically translates into the causal
impact of climate risk on lenders' strategic decision to sell a larger
share of their mortgages to securitizers.\footnote{The choice of holding vs securitizing based on the profit of each
option is similar to the Roy model of \citeasnoun{heckman1990empirical},
modelled in equations (13)\textendash (16) of Section~4.1.3 of \citeasnoun{ouazad2022mortgage}.} A seminal literature~\cite{keys2010did} has established that securitization
may foster both adverse selection and moral hazard effects. This hypothesis
suggests that mortgages flow from lenders to securitizers, and that
lenders face a portfolio problem of keeping loans within the portfolio
or selling them without significant shifts in prices; this is consistent
with the new finance literature focusing on portfolio allocation problems
rather than shifts in prices~\cite{koijen2019demand,alekseev2022quantity}.

The aftermath of major natural disasters provides new information
about the location, severity, and trends of future climate risk. When
faced with such new news, lenders have many margins of adjustment.
First, they can originate and hold the mortgages on their balance
sheets, potentially adjusting the amortization structure of the mortgages
(ARM vs. FRM, IO, negative amortiziation, fully amortizing, balloon
payment). Second, they can originate and sell the mortgages to private-label
securitizers. Third, lenders can originate loans guaranteed by the
Federal Housing Administration, or other federal agencies, in turn
guaranteed by Ginnie Mae in pools. Fourth, in the face of climate
risk, lenders can originate loans and sell the loans to Fannie Mae
and Freddie Mac when they satisfy the criteria for mortgage securitization.
At 6.5 trillion dollars, the total volume of outstanding loans in
Fannie Mae and Freddie Mac pools represents the largest volume compared
to Ginnie Mae loans (2.3 trillion dollars), private-label mortgage
backed securities (400 billion dollars), unsecuritized first-liens
(1.1 trillion dollars). 

\citeasnoun{ouazad2022mortgage} focuses on the identification of
the value of the GSE securitization option for lenders. It does so
in the aftermath of the 15 billion-dollar disasters with the largest
estimated damages, for conventional mortgages backed by owner-occupied
single-family housing units. 

Estimating the causal impact of climate risk on the value of the securitization
option is challenging. The fundamental identification problem is the
lack of observability of counterfactuals: for a given mortgage exposed
to climate risk, the `counterfactual mortgage' that differs only in
its exposure to risk, is not observed. The baseline correlation between
securitization and climate risk is also not helpful in the identification
of the causal impact of climate risk, as securitized loans differ
from mortgages held on the balance sheet in dimensions such as the
creditworthiness of borrowers and the characteristics of the collateral. 

There is a long tradition in economics of using policy rules' discontinuities
to identify causal impacts.\footnote{This includes \citeasnoun{card2004using}, \citeasnoun{dinardo2004economic},
\citeasnoun{cellini2010value}.} The rules of the Government Sponsored Enterprises provide such a
policy instrument: there is a sharp discontinuity in the ability to
securitize when loan amounts cross the conforming loan limit. In the
aftermath of natural disasters, focusing on the dynamic of the discontinuity
in approvals, originations, and securitizations at the conforming
loan limit provides an identification strategy in the spirit of the
credibility revolution of economics~\cite{angrist2010credibility}.
Policy rules such as the conforming loan limit provide us with an
opportunity to observe the strategic decision of lenders to originate
conforming loans (with loan amounts below such conforming loan limit)
vs. originate jumbo loans (with loan amounts above such limit); thus
heading towards the identification of the discrete choice problem
that lenders face. The Climate Securitization Hypothesis should thus
be particularly testable at this sharp discontinuity in the ability
to securitize conforming loans. \citeasnoun{ouazad2022mortgage} combines
this regression discontinuity with a difference-in-differences approach
controlling for fixed effects for 5-digit ZIP codes, years, years$\times$conforming
segment, disaster \textendash{} thus combining two popular identification
strategies with well-defined methodologies.\footnote{For Regression Discontinuity Designs (RDDs), references include \citeasnoun{imbens2008regression},
\citeasnoun{lee2010regression}, \citeasnoun{cattaneo2022regression},
which provide detailed guidance on bandwidths and kernels. For difference-in-differences
(DiD), references include \citeasnoun{de2020two}, \citeasnoun{callaway2021difference},
\citeasnoun{roth2023s}.} 

The CSH is tested using publicly-available mortgage-level data from
the Home Mortgage Disclosure Act for the Atlantic states combined
with NOAA's measures of hurricanes wind speeds, USGS's Digital Elevation
Model and National Land Cover data. This test relies on freely available
data, fostering a public debate and a possible investigation of heterogeneous
treatment effects.

This paper and \citeasnoun{ouazad2022mortgage} report three main
results establishing the empirical importance of the CSH. First, there
is a substantial increase in the discontinuity in the rates of approval,
origination, and GSE securitization conditional on origination at
the conforming loan limit, in the 4 years following a top 15 billion-dollar
disaster. Second, the discontinuity in approval and origination rates
increases significantly in years 1 to 3, before tapering off as the
volume of loans increases. Third, the discontinuity in GSE securitization
probabilities picks up in years 3 and 4, as the lenders return to
their approval standards pre-hurricane while increasing the share
of their mortgages that they sell to Fannie Mae and Freddie Mac.

Recently \citeasnoun{lacour2022adverse} have issued a critique focused
on data construction issues. In this reply, we provide an in-depth
analysis of their study and find that very serious fundamental mistakes
in the analysis render their results uninterpretable. We also study
the scientific basis for each of their claims. We find that our results
are robust to addressing their concerns. 

We also use this opportunity to revisit our core question using a
regression discontinuity design (RDD) estimator that has three key
advantages. First, this RDD estimator provides a detailed and visual
description of the treatment effect of billion-dollar disasters on
discontinuities. It explains where the effects of this paper and \citeasnoun{ouazad2022mortgage}
are coming from: they are driven by economically and statistically
significant discontinuities exactly at the conforming loan limit.
Second, using the latest regression discontinuity approach allows
us to estimate the effects for a large range of bandwidths. We find
that effects are robust to bandwidths from 2\% to 20\% around the
conforming loan limits, with significance levels from 95\% to 99\%.
Third, the effects are located in a narrow 2-3\% bandwidth in year
$t+1$ (approval and origination rates) and in year $t+1$ and year
$t+2$ (securitization rates), but are broader and affect the entire
window in years $t+2$ to $t+3$ (approval and origination rates)
and in years $t+3$ to $t+4$ (securitization rates).

Such evidence for the climate securitization hypothesis invites us
to simulate mortgage and housing markets without the GSEs given the
identification of the Roy model of originating and holding vs. originating
and securitizing. We are not the first to simulate such counterfactual
scenario~\cite{elenev2016phasing} and/or to simulate changes in
the `rules of the game' of the agency securitization market \cite{richardson2017gses}.

This paper proceeds as follows. Section~\ref{sec:Evidence-for-the}
presents the current evidence for the Climate Securitization Hypothesis
in the United States, with four different identification strategies
in four different contexts. Section~\ref{sec:The-LaCour-Little-et}
analyzes the \citeasnoun{lacour2022adverse} critique, an independent
construction of the data and econometric specifications. It finds
very serious mistakes in the construction of the data, rendering the
estimates uninterpretable. Section~\ref{subsec:Rounding-of-Loan}
discusses the point that loan amounts are rounded and the fix suggested
of rounding conforming limits up. We find that \possessivecite{lacour2022adverse}
approach yields a systematically upward biased and imprecise count
of conforming loans. Section~\ref{subsec:Point-=0000232:-The} shows
that the small number of high-cost counties, and their location in
the mid-west and the West coast makes it unlikely that this affects
results. Section~\ref{ssec:specification_main} tests the Climate
Securitization Hypothesis, replicates \citeasnoun{ouazad2022mortgage},
and provides a granular, non-parametric, point by point description
of \emph{where the effects come from}. The treatment effects display
sharp discontinuities in approval, origination, securitization at
the limit. Section~\ref{sec:Policy-Trade-Offs:-Market} presents
the key policy trade-off that the Federal Housing Finance Agency faces:
flexible guarantee fees pricing climate risk without causing bluelining;
such bluelining, reminiscent of the redlining, may cause declines
in homeownership rates, affordability, and overall welfare of incumbent
residents. Section~\ref{sec:Conclusion:-Adaptation-and} concludes
and provides a forward-looking research agenda for research in the
asset pricing and portfolio allocation in the face of physical climate
risk.

\section{Evidence for the \emph{Climate Securitization Hypothesis\label{sec:Evidence-for-the}}}

Suppose that lenders could not sell any loans from their portfolios.
In this case, lenders would have strong incentives to devote effort
screening loans in terms of the risks posed by the borrower and the
location. If borrowers prone to engage in strategic default tend to
purchase homes in risky locations, then lenders would adjust their
loan terms to reflect this distribution of differentiated risks. If
lenders can sell loans and if buyers and sellers in the loan market
have symmetric information then the logic of hedonic differentiated
product markets predicts that an assignment problem would arise as
heterogeneous loan sellers and buyers would pair off. Lenders would
specialize in issuing loans with specific risk exposure profiles and
loan buyers with a taste for higher risk/higher return assets would
purchase these. The climate securitization hypothesis, in the context
of Fannie Mae and Freddie Mac, posits that the GSEs' loan purchasing
rules introduce both moral hazard effects and adverse selection effects.
This bears a resemblance to the literature on rules in the health
insurance market. \footnote{\cite{einav2013selection} documents that rules for menus of health
insurance lead to adverse selection and moral hazard as patients have
private information about their health. When offered a menu of health
insurance plans, those with the greatest demand for health insurance
choose plans that have a lower marginal price for service and then
heavily utilize these services.} 

\citeasnoun{ouazad2022mortgage} estimates the impact of the exposure
of billion-dollar natural disasters on the approval, origination,
and securitization rates for mortgage in the conforming segment \textendash{}
where loans can be sold to the Government Sponsored Enterprises Fannie
Mae and Freddie Mac \textendash{} relative to mortgages in the jumbo
segment, where loans are either held on the balance sheet or privately
securitized in Private Label Mortgage-Backed Securities. The paper
finds that, in the aftermath of such natural disaster, there is a
significant increase in the probability of approval, origination and
securitization in the conforming segment while there is a decline
of such probabilities in the jumbo segment.

\citeasnoun{sastry2021bears} demonstrates how mortgage lenders transfer
flood risk to the government and under-insured households by taking
advantage of strict flood insurance coverage limits and staggered
flood map updates. The paper shows that lenders' risk management proceeds
by equalizing delinquency rates inside and outside of flood zones.
This is achieved through a combination of insurance requirements and
credit rationing, which results in a shift in the types of mortgages
offered in flood zones towards borrowers who are wealthier and have
higher credit quality. 

Figure 3 of the December 2022 version of the paper presents empirical
support for the \emph{climate securitization hypothesis, as} the adjustment
of leverage (LTV at origination) only occurs at a statistically significant
level when a mortgage is either privately securitized (other than
through the Government Sponsored Enterprises) or held on the balance
sheet.

\citeasnoun{nguyen2022climate} finds that, on average, lenders charge
higher interest rates for mortgages in areas projected at risk of
sea level rise. The effects are driven by long-term loans. The paper
explores whether such a sea level rise (SLR) premium depends on a
loan's eligibility to be securitized by Fannie Mae or Freddie Mac.
As the pricing of securitization in guarantee fees (g-fees) by Fannie
Mae and Freddie Mac depends on a Loan Level Performance Adjustment
(LLPA) matrix that is independent of SLR or other forms of climate
risk, the SLR premium is likely to be smaller for loans eligible for
securitization to the agencies. The paper finds evidence consistent
with the \emph{Climate Securitization Hypothesis }as the SLR premium
is significantly (economically and statistically) higher for jumbo
mortgages that are not eligible for securitization by the agencies
(Table VIII, page 1538). 

\citeasnoun{bakkensen2023leveraging} presents both a theoretical
model and an empirical analysis of the choice of debt when agents
have beliefs over the future evolution of risk in a specific area.
The model suggests that pessimistic agents take on more leveraged
loans when the collateral is risky. This is consistent with the evidence
in \citeasnoun{hertzberg2016adverse} suggesting a positive correlation
between the leverage and the pessimistic beliefs of the agents. In
\citeasnoun{bakkensen2023leveraging}, this leverage also manifests
in longer maturity loans. The paper provides empirical evidence for
the climate securitization hypothesis. First, the authors find robust
evidence of higher leverage in places affected by Sea Level Rise risk.
Second, this effect (SLR exposure interacted with climate belief)
is stronger for the conforming loan segment \textendash{} which banks
can securitize loans and sell to the GSEs \textendash{} than for the
nonconforming loan segment. This latter point is consistent with the
Climate Securitization Hypothesis.

\section{The LaCour-Little et al. (2022) Critique\label{sec:The-LaCour-Little-et}}

\citeasnoun{lacour2022adverse} suggests that lenders do not systematically
increase the approval, origination and securitization rates of conforming
loans relative to jumbo loans in the aftermath of natural disasters.
As such they argue that there is insufficient evidence rejecting the
null hypothesis that lenders have the same lending standards before
and after a natural disaster. The authors' argument is based on a
reexamination of \possessivecite{ouazad2022mortgage}, which found
that approval, origination and securitization rates increase in the
conforming segment relative to the jumbo segment. \citeasnoun{lacour2022adverse}
make two claims. First, that as loan amounts in thousands in the publicly-available
mortgage data are rounded to the nearest integer, the conforming loan
limits need to be rounded \emph{up}. Second, that, in high cost counties,
the geographic variation in the conforming loan limit affects the
statistical test of the hypothesis. 

This paper examines this alternative evidence thanks to the replication
files provided by the authors of \citeasnoun{lacour2022adverse}.
Close inspection of these data suggests significant errors in basic
data construction of the \citeasnoun{lacour2022adverse}, such as
incorrectly coded treatment years, missing hurricanes, and an incorrect
event-study design. Such errors could have been averted with simple
descriptive statistics, which are absent from the paper. This paper
also replicates \possessivecite{ouazad2022mortgage} findings, and
conducts additional robustness checks to show the sensitivity of statistical
tests of the climate securization hypothesis to choices such as the
regression discontinuity bandwidth. 

\subsection{Significant Errors in Data Construction in LaCour-Little et al. (2022):
The Miscoding of Hurricane Treatment Years\label{subsec:Significant-Errors-in}}

Data for \citeasnoun{lacour2022adverse} was accessed in March 2023
and stored at the link in this footnote.\footnote{\href{http://www.ouazad.com/papers/lacour_little_data_archive.zip}{http://www.ouazad.com/papers/lacour\_little\_data\_archive.zip}}
We find that the event study has incorrectly coded hurricane years
for a significant share of the observations. This is presented on
Table \ref{tab:miscoding_years}, which suggests that the problem
is broadly affecting all hurricanes of the sample. For hurricane Frances,
which occured between Aug 24, 2004 \textendash{} Sep 10, 2004, in
time t+1, 32\% of the observations are coded as treated in 2005 or
in 2016. \citeasnoun{lacour2022adverse} has no observation for hurricane
Charley (2004) and hurricane Dennis (2005), while \citeasnoun{ouazad2022mortgage}
has 7,108 observations for these hurricanes. For hurricane Jeanne
(2004), 33\% of the observations are coded as treated in 2005 and
2016. For hurricane Katrina (2005), treatment years are coded as 2004,
2005, 2008, 2012. \citeasnoun{lacour2022adverse} has a very small
number of observations for Hurricanes Rita, Dolly, and Ike. 

This miscoding of treatment years in the ``as\_of\_year'' variable
has direct impacts on the estimation as this variable ``as\_of\_year''
is used in two positions of the core regression of \emph{``run\_regressions.R'':
}as a fixed effect, that controls for, for instance, the impact of
the great financial crisis on origination, approval, securitization
rates; and as an interaction term to capture how the level of discontinuity
at the conforming loan limit varies across years.

This miscoding is also not driven by a desire to code years differently
for hurricanes happening later in the year: the publicly-available
Home Mortgage Disclosure Act data over this time does not include
variables for the month of origination. Many miscoded years are many
years before or after the actual year of the hurricane. 

Perhaps even more concerning is the correlation between this miscoding
of treatment years and the conforming / jumbo loan status of a mortgage
application. Figure~\ref{tab:miscoding_years} suggests that the
miscoding peaks at the conforming loan limit, especially pronounced
for Hurricane Katrina (2005). There the share with the wrong treatment
year peaks at 35\% of the observation at the limit, and then drops
to almost 0 right above the limit. 

We provide a formal test that this miscoding of treatment errors peaks
at the limit, using a regression discontinuity design. Table \ref{tab:formal_test}
presents the results of such RD design where the dependent variable
is 1 when the treatment year is incorrect. For instance, hurricane
Katrina's observations coded for any year other than 2005. Column
1 is the simple regression with the ``Below the Conforming Loan Limit''
dummy variable as the sole explanatory variable. Columns 2,3,4 use
a polynomial of the log distance of the loan amount to the conforming
loan limit. Standard errors are double clustered at the ZIP and year
levels, as in \citeasnoun{cameron2008bootstrap}. In all 4 columns,
there is a bunching of the coding errors at the limit, significant
at 5\%. It is intriguing that apparently random coding errors could
be so related to the main focus of the analysis, the conforming loan
limit. 

We note that there is no evidence of willful manipulation of the data
to switch observations from the conforming to the jumbo segment. Yet
we also remark that \citeasnoun{lacour2022adverse} updated their
archive in April 2023 to remove the main CSV data file \textendash{}
thus forcing researchers to launch their code before being able to
inspect the evidence. We accessed their archive before such removal,
and also remark that the code is identical in the previous archive
and the current one. The only difference is the removal of the CSV
file. In contrast, the main data frame for \citeasnoun{ouazad2022mortgage}
has been available throughout on the RFS Dataverse since 2021 and
will remain so. In addition, the code for this current paper is available
at \emph{https://github.com/aouazad/Mortgage-Securitization-Natural-Disasters-Reply.git}. 

\begin{table}
\caption{Errors in LaCour-Little et al. (2022) \textendash{} Miscoding of Hurricane
Treatment Years }
\label{tab:miscoding_years}

\emph{This table presents the analysis of the file ``originated\_05\_CT\_treatment.csv''
provided by the LaCour-Little coauthorship team. This file is used
in the main regression of their paper, titled ``run\_regressions.R.''
The replication package of these authors was accessed in April 2023
and is stored for your convenience at }\href{http://www.ouazad.com/papers/lacour_little_data_archive.zip}{http://www.ouazad.com/papers/lacour\_little\_data\_archive.zip}.

\begin{center}

{\footnotesize \begin{tabular}{lccccc}   
\toprule
& & \multicolumn{4}{c}{Time After Treatment} \\
\cmidrule(lr){3-6} 
Hurricane & Year of Treatment in LaCour-Little et al. & t+1 & t+2 & t+3 & t+4 \\   
\midrule 
\textbf{Frances (2004)} & 2004 & 1288 & 837 & 698 & 463 \\
 & 2005 & 620 & 537 & 338 & 212 \\
 & 2016 & 13 & 14 & 13 & 19 \\   
\midrule 
\textbf{Charley (2004)} & \multicolumn{5}{c}{No Observation in \citeasnoun{lacour2022adverse}} \\   
\midrule 
\textbf{Ivan (2004)} & 2004 & 239 & 145 & 165 & 148 \\
 & 2005 & 1 &  & 1 & 2 \\   
\midrule 
\textbf{Jeanne (2004)} & 2004 & 653 & 433 & 398 & 220 \\
 & 2005 & 314 & 290 & 143 & 104 \\
 & 2016 & 13 & 14 & 13 & 19 \\   
\midrule 
\textbf{Dennis (2005)} & \multicolumn{5}{c}{No Observation in \citeasnoun{lacour2022adverse}} \\   
\midrule 
\textbf{Wilma (2005)} & 2004 & 991 & 620 & 537 & 338 \\
 & 2005 & 2468 & 2199 & 1322 & 643 \\   
\midrule 
\textbf{Katrina (2005)} & 2004 & 2 & 1 &  & 1 \\
 & 2005 & 862 & 841 & 554 & 286 \\ 
 & 2008 & 77 & 99 & 92 & 100 \\
 & 2012 & 281 & 238 & 303 & 341 \\   
\midrule 
\textbf{Rita (2005)} & 2005 & 5 & 3 & 2 & 5 \\
 & 2008 & 5 & 1 & 2 &  \\   
\midrule 
\textbf{Ophelia (2005)} & 2005 & 121 & 120 & 179 & 99 \\   
\midrule 
\textbf{Gustav (2008)} & 2005 & 86 & 73 & 85 & 77 \\
 & 2008 & 86 & 108 & 98 & 109 \\
 & 2012 & 138 & 89 & 131 & 151 \\   
\midrule 
\textbf{Ike (2008)} & 2005 & 5 & 3 & 2 & 5 \\
 & 2008 & 38 & 32 & 28 & 46 \\   
\midrule 
\textbf{Dolly (2008)} & 2008 & 2 & 2 &  &  \\   
\midrule 
\textbf{Irene (2011)} & 2005 & 12 & 5 & 15 & 15 \\
 & 2011 & 20 & 66 & 59 & 83 \\
 & 2012 & 32 & 31 & 36 & 42 \\   
\midrule 
\textbf{Sandy (2012)} & 2011 & 14 & 32 & 31 & 36 \\
 & 2012 & 1254 & 980 & 1180 & 1476 \\   
\midrule 
\textbf{Isaac (2012)} & 2005 & 144 & 136 & 157 & 147 \\
 & 2008 & 76 & 96 & 91 & 99 \\
 & 2012 & 281 & 238 & 303 & 343 \\   
\midrule 
\textbf{Matthew (2016)} & 2004 & 13 & 10 & 6 & 13 \\
    & 2016 & 740 & 900 & 819 & 1428 \\     
\bottomrule \end{tabular}}

\end{center}
\end{table}

\begin{figure}
\caption{Errors in LaCour-Little et al. (2022) \textendash{} Miscoding of Treatment
Year by Distance to the Conforming Loan Limit}
\label{fig:miscoding_by_distance}

\emph{These figures plot, for each hurricane, the share of mortgages
with the wrong treatment year in LaCour-Little et al. (2022). For
instance, Table \ref{tab:miscoding_years} shows that 29.5\% of the
observations for hurricane Katrina have treatment years in 2004, 2008,
and 2012. These graphs below plot such share at each distance of the
conforming loan limit. They suggest that misclassification tends to
peak right before the conforming loan limit, and then drop. }

\begin{center}

\subfloat[Hurricane Frances (2004)]{

\includegraphics[scale=0.5]{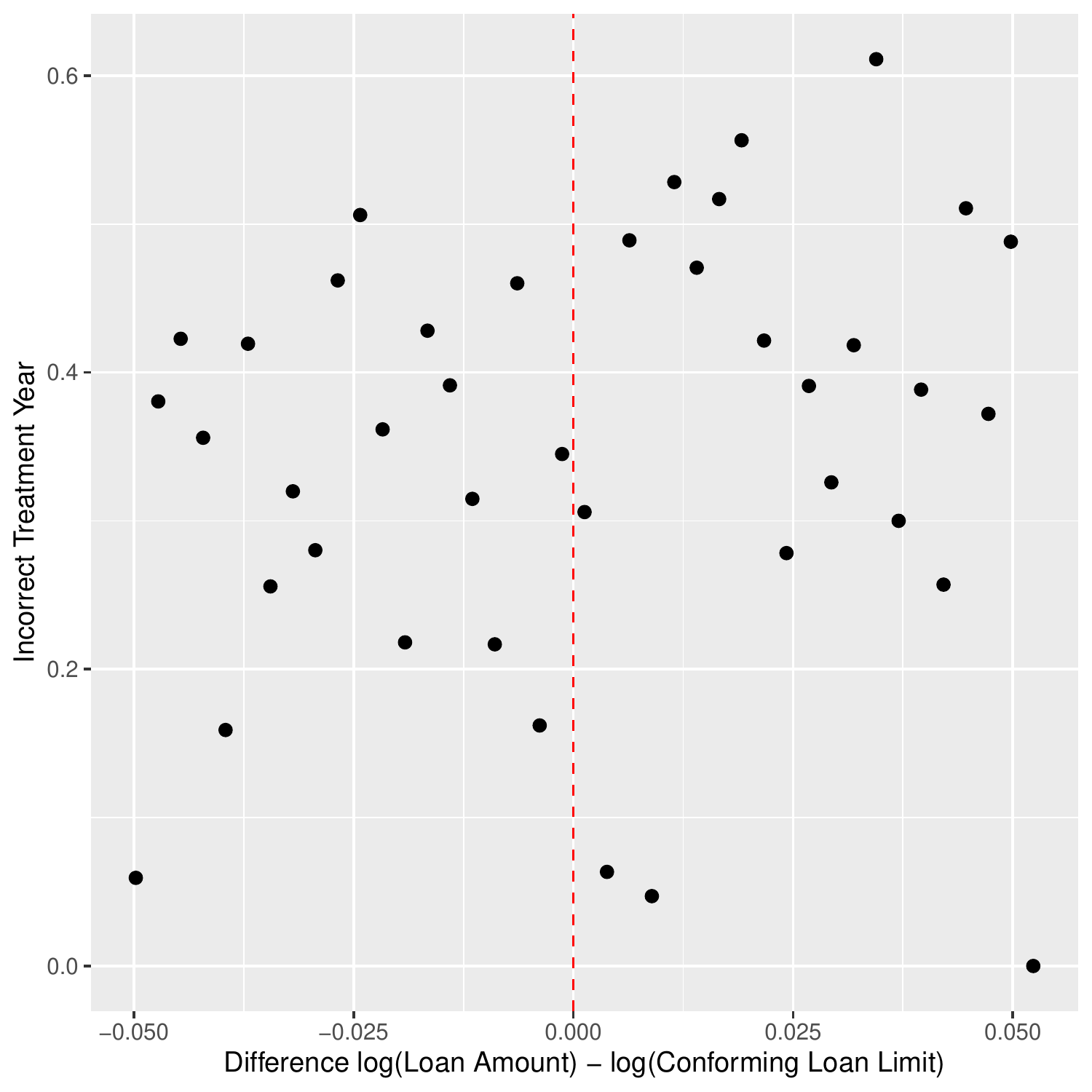}}\subfloat[Hurricane Katrina (2005)]{

\includegraphics[scale=0.5]{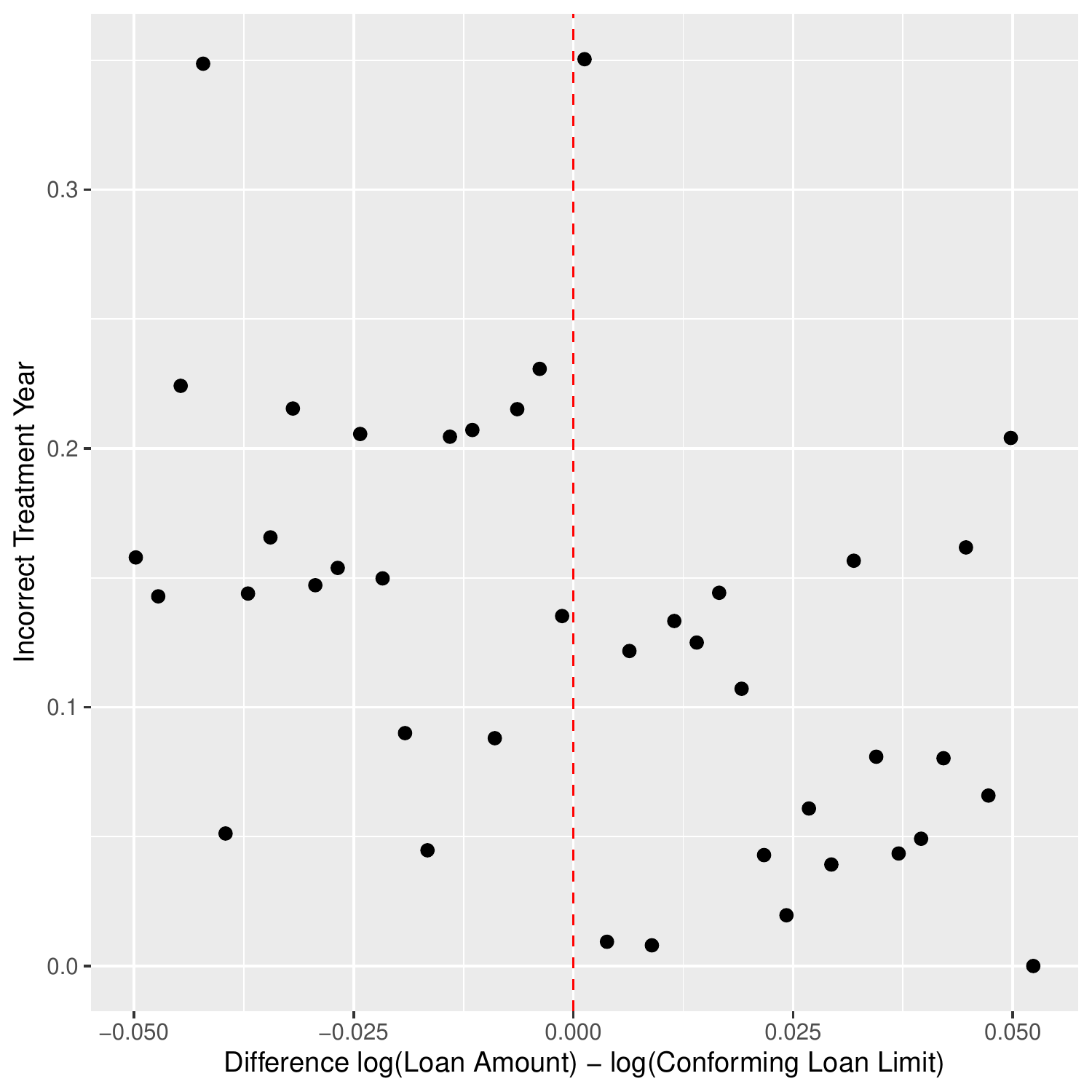}}

\subfloat[Hurricane Isaac (2012)]{

\includegraphics[scale=0.5]{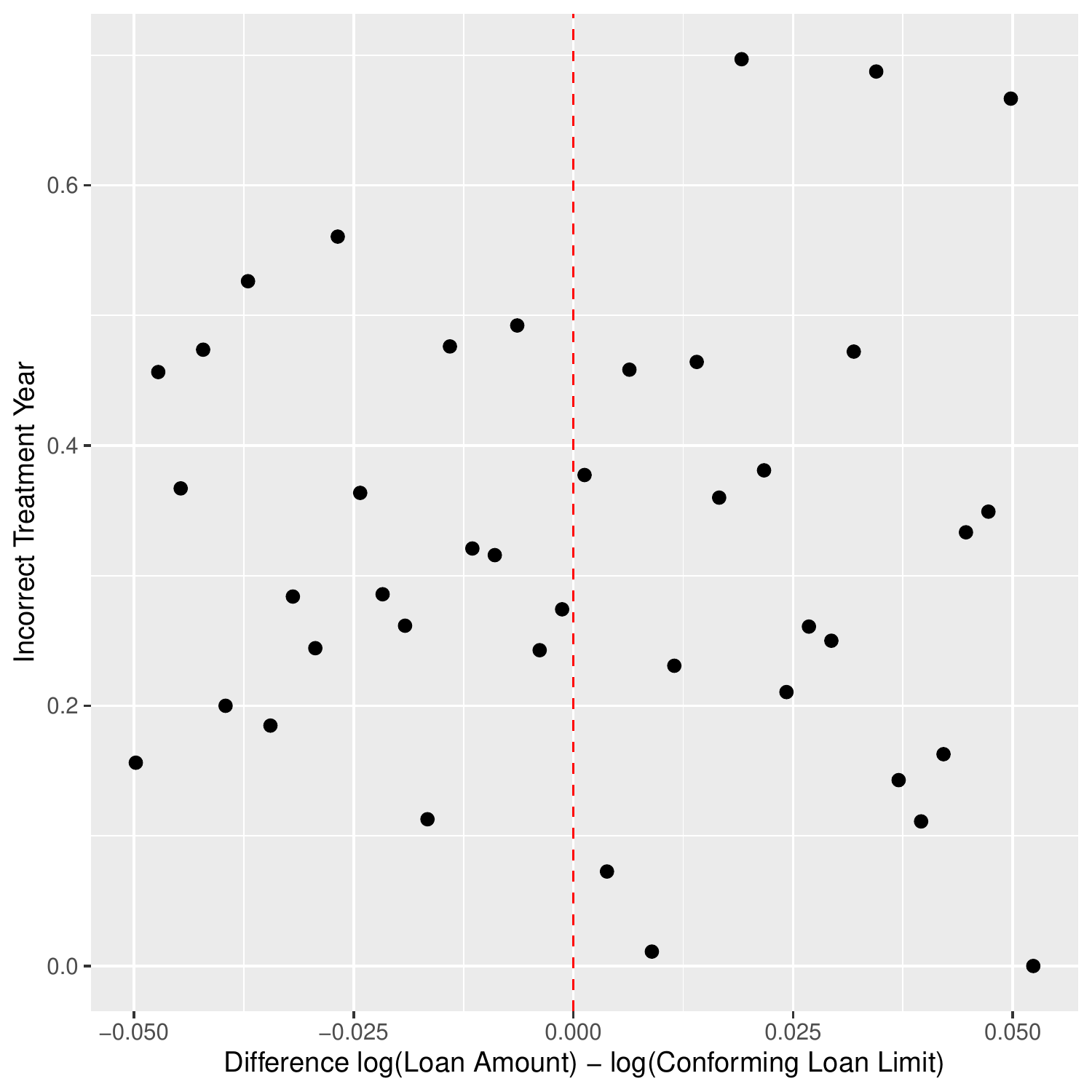}}\subfloat[Hurricane Matthew (2016)]{

\includegraphics[scale=0.5]{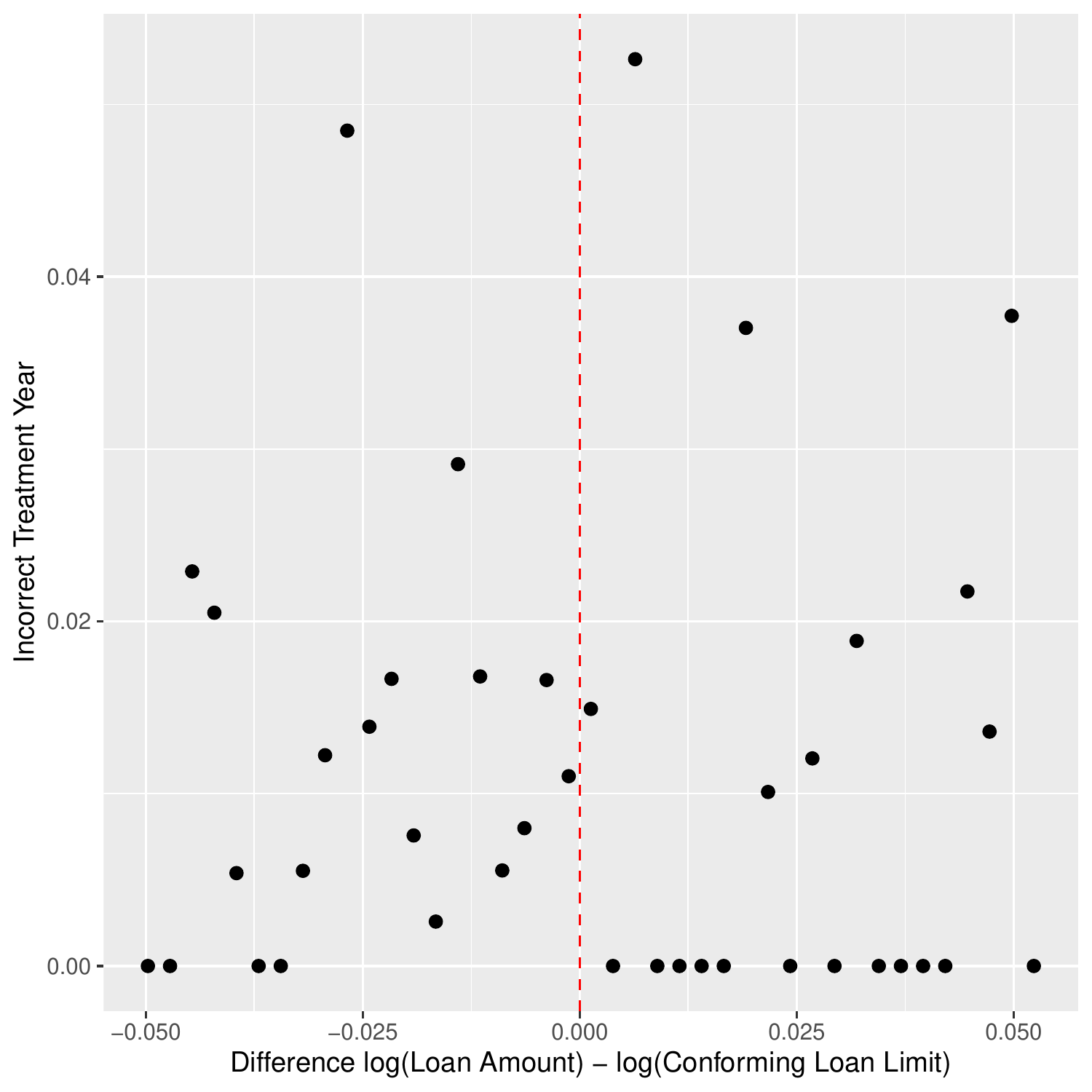}}

\end{center}
\end{figure}

\begin{table}

\caption{Errors in LaCour-Little et al. (2022) \textendash{} A Formal Test
of the Peak of Hurricane Treatment Year Errors at the Conforming Loan
Limit}
\label{tab:formal_test}

\emph{This table presents a formal test that the miscoding of treatment
years peaks at the conforming loan limit in the data of }\citeasnoun{lacour2022adverse}\emph{.
We perform a regression discontinuity design estimation at the conforming
loan limit. The dependent variable is 1 if the year of treatment was
miscoded. For instance, as Table~\ref{tab:miscoding_years} shows,
observations for many hurricanes in }\citeasnoun{lacour2022adverse}
\emph{are mistakenly coded in a different year. Many Katrina observations
are treated in 2004, 2008, and 2012. In this case, the dependent variable
is 1. The right-hand side of each regression has the discontinuity
and a polynomial of order 1 (column 2), 2 (column 3), 3 (column 4)
in the log difference of the loan amount with the conforming loan
limit. Standard errors are double-clustered at the ZIP and year level.
The number of observations is the number of treated observations at
any point between $t-4$ and $t+4$ in }\citeasnoun{lacour2022adverse}.

\begin{center}

{\footnotesize \input{tables/RD_miscoding_1.tex}}

\end{center}

\emph{The replication package of these authors was accessed in March
2023 and is stored for your convenience at }\href{http://www.ouazad.com/papers/lacour_little_data_archive.zip}{http://www.ouazad.com/papers/lacour\_little\_data\_archive.zip}.
\end{table}

There are other important errors in the paper. A major issue is the
incorrect event study design. As the data set has a flat longitudinal
structure, this subjects the analysis to the classic issue of the
staggered difference-in-differences problem. We display in the table
below the sum of the indicator variables for $Time=-4$ to $Time=+4$.
In \possessivecite{ouazad2022mortgage}, this sum is always equal
to 1 in the treatment group. Surprisingly, in the \citeasnoun{lacour2022adverse},
this sum can be 0, 2, or 3. This is not due to the reference dummy
variable since $\mathbf{1}(Time=0)$ and $\mathbf{1}(Time=-1)$ are
included in the sum. Therefore we are observing two major anomalies
in the analysis: first, that there are observations with multiple
time dummy variables equal to 1 at the same time. 653 mortgages have
$\mathbf{1}(Time=+1)$ and $\mathbf{1}(Time=+2)$ equal to 1 at the
same time. 568 mortgages have $\mathbf{1}(Time=+2)$ and $\mathbf{1}(Time=+3)$
equal to 1 at the same time. Perhaps even more surprising is that
a large chunk of treated observations ($Treatment=1$) may have no
dummy at all equal to 1 at any point from $Time=-4$ to $Time=+4$
, i.e. $\ensuremath{\sum_{k}\mathbf{1}(Time_{it}=k)}=0$. This suggests
that the regression's reference point is incorrectly specified, since
the control group will include observations in the pre- and post-treatment
periods.

\begin{table}[h]
\caption{Errors in LaCour-Little et al. (2022) -- Missing Time Indicator Variables}
\emph{This table presents, for the treatment group only, the sum of the indicator variables for the number of years relative to the hurricane. In a well-designed event study, each treated observation should have only one time dummy variable equal to 1. Except for the reference time period (e.g. $-1$). In contrast, in LaCour-Little (2022), more than 93,000 observations are in the treatment but have no corresponding time dummy. And more than 6,200 observations have multiple time dummies equal to 1 at the same time. There is no dummy variable for times before $-4$ and no dummy variable for times after $+4$.}
\begin{center}
\begin{tabular}{lcccc}
\toprule
& \multicolumn{4}{c}{$\sum_{k=-4}^{k=+4} \mathbf{1}(Time_{it} = k)$} \\
\cmidrule(lr){2-5}
    &  0    &   1  &     2   &    3  \\
\midrule
Number of Treated Observations & 93,231   & 35,885  &  6,057  &   161   \\
\bottomrule
\end{tabular}
\end{center}
\end{table}

Finally, although the sample is extended to 2020, it does not include
any billion-dollar hurricane for that time period such as hurricane
Harvey (2017). This means that treated observations may be part of
the control group.

The evidence presented in this section suggests that \citeasnoun{lacour2022adverse}
exhibits major flaws that render the results uninterpretable.

\subsection{Rounding of Loan Amounts in Home Mortgage Disclosure Act Data\label{subsec:Rounding-of-Loan}}

The previous section has displayed major errors in the empirical work
of \citeasnoun{lacour2022adverse}. There are interesting points worthy
of further investigation in \citeasnoun{lacour2022adverse}. The first
one is that the rounding of loan amounts in HMDA affects the estimation
and the fix suggested by \citeasnoun{lacour2022adverse}. The second
point is that high cost counties' conforming loan limits play a significant
role in the estimation. We investigate both of these points in turn
to see if they affect evidence on the \emph{Climate Securitization
Hypothesis.}

\subsubsection*{Intuition and Descriptive Statistics}

In Home Mortgage Disclosure Act data, loan amounts are reported by
rounding loan amounts \emph{to the nearest thousand}. For instance,
on page 12 of the 2013 ``Guide to HMDA Reporting, Getting It Right!'':
\begin{quote}
\textbf{Loan amount.} Report the dollar amount granted or requested
in thousands. For example, if the dollar amount was \$95,000, enter
95; if it was \$1,500,000, enter 1500. Round to the nearest thousand;
round \$500 up to the next thousand. For example, if the loan was
for \$152,500, enter 153. But if the loan was for \$152,499, enter
152.
\end{quote}
Figure~\ref{fig:Rounding-and-Loan}, case \#1, shows what this implies
for the difference between actual loan amounts and observed loan amounts
when zooming in on, for instance the range of loan amounts between
\$424,000 and \$425,000. This range is relevant for Clay County, Florida,
which had a conforming loan limit of \$424,100. Loan amounts between
424 and 424.5 (not included) are reported as 424, and loan amounts
between 424.5 (included) and 425 are reported as 425. The conforming
loan limit is the vertical green dotted line. In this case, only loans
between 424 and 424.1 can be conforming loans, but when rounding,
all loans between 424 and 424.5 (not included) are reported as conforming.
This \emph{overestimates }the number of conforming loans.

Case \#2 suggests that the rounding can also lead to an \emph{underestimation}
of the number of conforming loans. In the case of Collier Count, Florida,
the limit is 450.8. Loans between 450 and 450.5 (not included) are
counted as conforming, while the true count is larger, as it should
include 450 to 450.8. On balance thus, the approach employed in Ouazad
and Kahn (2022) leads to symmetric noise. In practice also, the $\pm10\%$
or $\pm5\%$ around loan amounts is substantially wider than \$1000,
ranging from 381 to 467, leading to fewer rounding issues than Figure~\ref{fig:Rounding-and-Loan}
would suggest.

\citeasnoun{lacour2022adverse} argues that:
\begin{quote}
``The correct comparison would \textbf{round the conventional (sic)
}\footnote{This excerpt of \citeasnoun{lacour2022adverse} should, of course,
say the \emph{conforming }loan limit.}\textbf{ loan limit in the second term }\textbf{\emph{above to the
nearest \$1000}}, so that the variable \textquotedbl diff\_log\_loan\_amount\textquotedbl{}
becomes zero and the loan is correctly classified as being at or below
the FHA conventional limit.'' (page 34 of the October 31, 2022 version,
accessed in January 2023, emphasis is ours).
\end{quote}
We show below that this approach is incorrect as the share of conforming
loans systematically upward biased. In both cases \#1 and \#2 of Figure~\ref{fig:Rounding-and-Loan},
the \citeasnoun{lacour2022adverse} approach would count all loans
between 424 and 425 (case \#1) and between 450 and 451 (case \#2).
This uses the red dotted line in both graphs. While it generates bunching
graphs with no overlapping point (and thus ``seem cleaner'') it
leads to high and systematic levels of misclassification. 

The econometric exercise performed below estimates the properties
of the discontinuity estimator implied by Ouazad and Kahn's (2022)
approach with those estimators implied by the \citeasnoun{lacour2022adverse}
approach. The exercises suggests that LLPW estimators exhibit larger
variances, less precision, higher absolute biases, and slower speeds
of convergence to their asymptotic value.

A third approach, not suggested in \citeasnoun{lacour2022adverse},
is to round the conforming loan limit in the same way as for HMDA
loan amounts, to the nearest integer. It is easy to see that it also
does not address this issue as it leads to greater over- or underestimation
of the share of conforming loans. Overall, the approach of \citeasnoun{ouazad2022mortgage}
leads to the smallest bias and variance across the three approaches.

\subsection{The Rounding of Loan Amounts: Simple Notations and the Economics
of Rounding}

We can deepen our understanding of the impact of the rounding of loan
amounts by (a) understanding under what restrictive set of assumptions
the \citeasnoun{lacour2022adverse} approach is correct (always bunching)
and (b) by performing an econometric analysis of the estimators using
both the \citeasnoun{lacour2022adverse} approach and the \citeasnoun{ouazad2022mortgage}
approach. Such econometric analysis will measure the bias and the
precision of the estimators.

The true amount of loan $i$ denoted $L_{i}^{*}$ is a latent (unobserved)
variable in HMDA, unless one relies on private data sets such as those
sold by Corelogic. The observed loan amount $L_{i}$ is the true loan
amount $L_{i}^{*}$ rounded to the nearest integer. This can be denoted
as $L_{i}=\nint{L_{i}^{*}}$. 

Let's then denote by $\overline{L}^{*}$ the conforming loan limit
in the county of loan $i$. A necessary (but not sufficient) condition
for a loan to be conforming is that the true loan amount $L_{i}^{*}\leq\overline{L}^{*}$.
We denote this binary variable by $C_{i}^{*}=\mathbf{1}(L_{i}^{*}\leq\overline{L}^{*})$
the classification of loan $i$. One can see that comparing the reported
HMDA loan amount $L_{i}$ to the exact conforming loan limit $\overline{L}^{*}$
may lead to an imperfect classification. We denote a HMDA-based classification
as$C_{i}^{H}=\mathbf{1}(L_{i}\leq\overline{L}^{*})$. Thus at this
stage we have two binary indicator variables: $C_{i}^{*}\in\{0,1\}$
for the true classification (conforming is 1), and $C_{i}^{H}\in\{0,1\}$
for the HMDA-based classification.

Using data where numbers are not rounded is an obvious albeit non-free
solution. Using expensive data hinders the public debate over the
climate securitization hypothesis.

\citeasnoun{lacour2022adverse} suggests rounding the limit up. This
leads to a third classification, denoted $C_{i}^{LL}=\mathbf{1}(L_{i}\leq\nint{\overline{L}^{*}})$.
It is easy to see that across all classifications $C_{i}^{*}$ (the
true one), $C_{i}^{H}$ (the one inferred from HMDA), $C_{i}^{LL}$
(the LaCour-Little), the LaCour-Little approach generates the highest
share of conforming loans:
\begin{equation}
C_{i}^{LL}\geq\max\{C_{i}^{*},C_{i}^{H}\}\label{eq:lacour-little-highest-share}
\end{equation}
This is visible on Figure~\ref{fig:rounding}. Meanwhile, the HMDA-based
classification $C_{i}^{H}$ can either be higher or lower than the
true classification $C_{i}^{*}$:
\begin{equation}
\text{There are mortgages \ensuremath{i} and \ensuremath{j} such that }C_{i}^{H}\geq C_{i}^{*},\quad\text{and}\quad C_{j}^{*}\geq C_{j}^{H}\label{eq:HMDA-based-classification}
\end{equation}
 This is a first hint that the measure $C_{i}$ will be less biased
than the LaCour-Little approach.

The gap between the HMDA-based classification $C_{i}^{H}$ and the
LaCour-Little specification $C_{i}^{LL}$ depends on the share of
mortgages in the jumbo segment. Perhaps surprisingly, the LaCour-Little
approach $C_{i}^{LL}$ is equal to the true classification $C_{i}^{*}$
only when \emph{all mortgages are always bunched in the conforming
segment, }i.e. when there is no jumbo loan in the window. This is
unlikely to be true.

The main reason that the is that there are sizeable jumbo applications
and originations around the conforming loan limit even for narrow
windows. The volume of jumbo mortgage applications and originations
does not decline as the window narrows. This is likely due to a few
factors. First, the agencies retreated from the mortgage market in
2004-2008, in the wake of the accounting irregularities; this matches
the expansion of the private-label mortgage market.\footnote{\textquotedbl Accounting Irregularities at Fannie Mae\textquotedbl{}
by Chairman Christopher Cox U.S. Securities \& Exchange Commission.
Accessible at https://www.sec.gov/news/testimony/2006/ts061506cc.htm} Second, a significant share of borrowers choose to borrow using jumbo
loans due to either the characteristics of their house or their FICO
or LTV.

\begin{table}
\caption{The Share of Jumbo Loans Remains Substantial Even Close to the Conforming Loan Limit}
\emph{This table presents the share (column 2) of loans with loan amounts above the conforming loan limit, for each size of the window around the conforming loan limit. This suggests that the method suggested by \citeasnoun{lacour2022adverse}, of rounding up the conforming loan limit, would misclassify approximately 27 to 30\% of mortgages as conforming when they are jumbo. The window size is the $\max \log(\textrm{Loan Amount})-\log(\textrm{Conforming Limit})$.}
\bigskip
\begin{center}
\begin{tabular}{cc}   
\toprule
(1) & (2) \\
Window Size & Share of Jumbo Loans $\in [0,1]$ \\
\cmidrule(lr){1-1} \cmidrule(lr){2-2}
   $\pm 1\%$ & 0.297 \\
   $\pm 2\%$ & 0.289 \\
   $\pm 3\%$ & 0.283 \\
   $\pm 4\%$ & 0.290 \\
   $\pm 5\%$ & 0.277 \\
   $\pm 10\%$ & 0.283 \\
   $\pm 15\%$ & 0.278 \\
   $\pm 20\%$ & 0.278 \\     
\bottomrule
\end{tabular}
\end{center}
\end{table}

\subsection{Mismeasurement of Bunching in the Cross-Section using LaCour-Little
et al.'s (2022) Rounding}

\citeasnoun{lacour2022adverse} suggests that conforming loan limits
should be rounded up, ``The correct comparison would round the conventional
(sic) loan limit in the second term above to the nearest \$1000.''
We examine the econometric properties of such an approach below and
suggest that this alternative approached yields biased and imprecise
estimators.

\subsubsection*{Econometric Framework and Monte Carlo Tests in the Cross-Sectional
Bunching Case}

We conduct Monte-Carlo tests to assess the econometric properties
of discontinuity estimators when using either the true measure, the
Ouazad and Kahn (2022) approach, and the \citeasnoun{lacour2022adverse}
approach. We do so by following standard econometric practice. We
postulate a true model and estimate the parameters using either of
the imperfect classification methods $C_{i}^{H}$ (Ouazad and Kahn
(2022)) or $C_{i}^{LL}$ (\citeasnoun{lacour2022adverse}). 

The true model here is one where the probability of approval $P(\text{Approval}_{i}=1)$
is discontinuous at the true conforming loan limit. This is modelled
as a logit:
\begin{equation}
P(\text{Approval}_{i}=1)=P(\text{Approval}_{i}^{*}\geq0)=P(\alpha^{*}+\beta^{*}C_{i}^{*}-\varepsilon\geq0)\label{eq:true_model}
\end{equation}
The true model depends on the true classification $C_{i}^{*}$. Of
course, in econometric practice, the probability of approval depends
on a host of covariates such as the creditworthiness of the borrower,
the location and characteristics of the house, the characteristics
of the lender. The insights of our approach extend to this more general
class of econometric models.

Bunching at the conforming loan limit can be estimated in multiple
ways. A popular approach is to consider an OLS regression. Another
approach, which we leave to the reader for further analysis, is a
logit regression. As logit regressions entail other issues such as
incidental parameter problems \cite{chamberlain1980analysis,lancaster2000incidental}
when including fixed effects in a more comprehensive regression. We
thus focus on the least squares approach.

The data is generated according to \ref{eq:true_model}. The regression
is:
\[
\text{Approval}_{i}=\hat{\alpha}^{s}+\hat{\beta}^{s}C_{i}^{s}+\epsilon_{i}^{s},
\]
where $s\in\{*,H,LL\}$ is the choice of the classification of the
loan: using the true discontinuity, using the HMDA discontinuity,
and using the conforming loan limit. $\hat{\beta}^{*}$ is the OLS
estimate based on the true discontinuity. We compare $\hat{\beta}^{H}$
and $\hat{\beta}^{LL}$ to $\hat{\beta}^{*}$.

\subsubsection*{Results for the Static Bunching Case}

We consider a case with $N=1,000$ observations, true values $\alpha=0.2$
and $\beta=0.1$, and $S=10,000$ simulations. Each approval decision
is 1 with probability $F(\alpha^{*}+\beta^{*}C_{i}^{*})$, where $F$
is the cdf of the logit. 

The first visible statistic is that across the two methods, the LaCour-Little
approach is that which misclassifies loans more extensively. In the
Monte Carlo samples, misclassifications are as follows:
\begin{center}
\begin{tabular}{lcc}
\textbf{County} & \textbf{Share Misclassified HMDA} & \textbf{Share Misclassified LLPW}\tabularnewline
\hline 
\textbf{Clay County, Florida} & 0.006 & 0.014\tabularnewline
\textbf{Collier County, Florida} & 0.010 & 0.010\tabularnewline
\end{tabular}
\par\end{center}

This table suggests that in the case of Clay County the share of misclassified
loans (1.4\%) with the \possessivecite{lacour2022adverse} approach
is more than double (0.6\%) that of the HMDA approach of \citeasnoun{ouazad2022mortgage}.
In the case of Collier county, the share misclassified is the same
in both the HMDA approach and the LLPW overestimate the share of conforming
loans. Overall, using HMDA loan amounts provides a lower share of
misclassifications.

We then turn to the regression results. Figure~\ref{fig:Rounding-and-Loan}
presents the distribution of the difference between the estimator
$\hat{\beta}^{H}$ obtained using the HDMA loan amounts as in \citeasnoun{ouazad2022mortgage}
and the estimator $\hat{\beta}^{*}$ on true data with no misclassification.
This is the blue line. The distribution of the difference $\hat{\beta}^{LL}-\hat{\beta}^{*}$
is the orange line. As the graph makes clear in both case 1 (subfigure
(a), Clay County, FL) and in case 2 (subfigure (c), Collier County,
FL), the \citeasnoun{lacour2022adverse} is exhibits higher standard
deviation, while \possessivecite{ouazad2022mortgage} approach yields
only a minor increase in standard deviation compared to the estimator
using the true classification. Subfigures (b) and (d) present the
distribution of the absolute difference $\vert\hat{\beta}^{s}-\hat{\beta}^{*}\vert$
for $s=H$ and for $s=LL$. In both cases, the absolute bias is higher
in the \citeasnoun{lacour2022adverse} than with \possessivecite{ouazad2022mortgage}
approach.

Figure~\ref{fig:sample_size} shows that the estimator based on \possessivecite{lacour2022adverse}
performs worse regardless of sample size. Subfigure (a) displays the
absolute deviation $\vert\hat{\beta}^{s}-\hat{\beta}^{*}\vert$ from
the true value for sample sizes ranging from 50 to 2000. The absolute
deviation is substantially higher at all $N$s. Subfigure (b) shows
that the estimator is also less precise. The standard deviation of
\possessivecite{lacour2022adverse} estimator is higher at all $N$s. 

\subsubsection*{Conclusion}

HMDA loan amounts are rounded to the nearest integer. Conforming loan
limits are typically at an exact multiple of thousands of dollars.
With HMDA, the approach of \citeasnoun{ouazad2022mortgage} yields
the smallest share of misclassifications, a lower absolute deviation,
and is more precise. While using private data with exact loan amounts
(e.g. Corelogic data) would address this issue, this is not what \citeasnoun{lacour2022adverse}
suggest, and their approach yields biased and imprecise estimators.

\subsection{Point \#2: The Role of High Cost Counties' Specific Conforming Loan
Limits\label{subsec:Point-=0000232:-The}}

\subsubsection*{The Geographic Distribution of High-Cost Counties}

Conforming loan limits vary by county and by year, as noted by \citeasnoun{ouazad2022mortgage}.
In any given year, roughly 100 to 200 counties (out of 3,143) are
deemed ``high-cost counties.'' Figure~\ref{fig:high_cost} presents
a map of conforming loan limit \$ by county, and delineates the boundaries
of states of the Atlantic coast and the Gulf of Mexico. These are
the states most exposed to hurricanes and are thus the focus of any
analysis of the impact of hurricanes on mortgage securitization. Figure~\ref{fig:high_cost}
suggests that most coastal counties are not high cost counties. Many
high cost counties are in the San Francisco Bay Area, the Los Angeles
area, and the Seattle area. A notable exception is the New York City
area, which both experienced Hurricane Sandy in 2012 and is a high-cost
area. Yet, there is no difference between \possessivecite{ouazad2022mortgage}
and \possessivecite{lacour2022adverse} conforming loan limits in
the New York metropolitan area. This is a first intuitive indication
that the quantitative importance of \possessivecite{lacour2022adverse}
point may be small. The next section investigates this point.

\section{Testing the Climate Securitization Hypothesis:\protect \\
Securitization Dynamics in the Aftermath of Natural Disasters\label{ssec:specification_main}}

We reestimate \possessivecite{ouazad2022mortgage} baseline specification
using the limits provided by \citeasnoun{lacour2022adverse}. The
results are presented on Figure~\ref{fig:results_replication_ouazad_kahn}
for all three outcome variables, approval, origination, and securitization
conditional on approval. They are similar to the original findings
in \possessivecite{ouazad2022mortgage} Figure 8. Tables \ref{tab:regression_approval}
to \ref{tab:reg_securization} provide further statistical tests for
each bandwidth. The null hypothesis here is that the lending standards
in the conforming segment are not differentially changing after a
billion-dollar disaster relative to the overall market. A \emph{rejection
}of the null hypothesis with a positive and statistically significant
estimate for approval or origination rates suggests that lending standards
are becoming more lenient in the conforming segment as compared to
the jumbo segment. This is what Figure \ref{fig:results_replication_ouazad_kahn}
suggests: approval probabilities increase gradually all the way to
$+6$ percentage points in year 3 (subfigure (a)), the origination
probabilities increase all the way to $+5$ percentage points in year
3 (subfigure (c)), and the probability of securitization conditional
on origination increases by approximately 12 percentage points in
year 4 (subfigure (e)). The bars indicate the double-clustered standard
errors at the ZIP and year levels. 

The timing is also similar to \citeasnoun{ouazad2022mortgage}: in
the first 3 years following a natural disaster, approval, origination
rates increase significantly and then taper off. In the third and
fourth year, securitization rates $P(\text{Securitized}\vert\text{Originated})$
increase in turn, suggesting that lenders are changing their securitization
practices in the conforming segment conditional on origination. For
each figure, the right-hand column is for the impact on the mortgage
market regardless of the conforming segment. There there is a systematic
decline in approval, origination, and securitization rates: approval
rates decline by up to 7 percentage points (subfigure (b)), origination
rates by 7 percentage points as well (subfigure (d)), and securitization
rates by up to 10 percentage points (subfigure (f)), indicating that
the origination and securitization activity in the jumbo segment tapers
off significantly after a natural disaster. This is likely due to
both the difficulty of securitizing jumbo loans in the private label
market and the reluctance of lenders to originate and hold.

Hence what likely explains the difference between \citeasnoun{lacour2022adverse}
and \citeasnoun{ouazad2022mortgage} is the significant errors of
\citeasnoun{lacour2022adverse} documented in Section~\ref{subsec:Significant-Errors-in}
and the incorrect rounding of conforming loan limits.

\subsubsection*{Inspecting the Mechanism: Treatment Effects at Each Distance of the
Conforming Loan Limit\label{subsec:What-Drives-Results}}

What drives the results of both \citeasnoun{ouazad2022mortgage} and
this paper? Is the effect driven by treatment effects far away from
the conforming loan limit, or by sharp discontinuities exactly at
the limit? The approach described in this section provides appealing
and intuitive graphical approaches to an understanding of the main
treatment effects, presented Figures \ref{fig:inspecting_the_mechanism}\textendash \ref{fig:inspecting_the_mechanism-2}.
Open source code is available at the link on the cover page.

To understand the method, consider a distance $\delta\in[-0.10,+0.10]$
to the conforming loan limit. We would like to estimate the impact
$\text{TE}(\delta)$ of a billion dollar event on approval probabilities
exactly at $\delta$. We also would like to control for year, ZIP,
and disaster confounders, as in the main regression. This is performed
by estimating the coefficients $\xi_{t}$ for $t=-4,-3,-2,0,1,2,3,4$
that minimize:
\begin{align}
\min & \sum_{it}\left(\text{Approval}_{it}-\sum_{t=-4}^{+4}\xi_{t}(\delta)\cdot\text{Treated}_{j(i)}\times\textrm{Time}_{t=y-y_{0}(d)}\right.\nonumber \\
 & \qquad\qquad\qquad\qquad-\textrm{Year}_{y(t,d)}-\text{Disaster}_{d}-\text{ZIP}_{j(i)}\Biggr)^{2}K\left(\frac{\Delta\text{Loan Amount}_{it}-\delta}{h}\right)\label{eq:approach_lpr}
\end{align}
while weighting observations by their distance $\Delta\text{Loan Amount}_{it}-\delta$.
Hence when $\delta=-0.10$, the coefficients $\xi_{t}(\delta)$ measure
the treatment effect for loan amounts 10\% below the conforming loan
limit. When $\delta=+0.10$, the coefficients $\xi_{t}(\delta)$ measure
the treatment effect for loan amounts 10\% above the conforming loan
limit.\emph{ }$K$ is a Gaussian kernel and we use the bandwidth $h=1\%$.
Robustness to bandwidth choice is presented on Tables~\ref{tab:regression_approval}\textendash \ref{tab:reg_securization}.
When considering $\delta<0$, the treatment effects are estimated
using observations in the conforming segment. When $\delta>0$ the
treatment effects are estimated using observations in the jumbo segment.

An advantage of this approach is it flexibility and its visual representation.
A drawback may be the difficulty of obtaining standard errors; this
is addressed in the next subsection where discontinuities at the conforming
loan limit are formally tested using double clustering at the Year
and ZIP levels.

Figure \ref{fig:inspecting_the_mechanism-1} presents the estimation
results for the approval rate, using 40 distances, and a bandwidth
of~1\%. Subfigure (a) is for the treatment effects at time $t=+1$,
(b) for time $t=+2$, (c) for time $t=+3$, (d) for time $t=+4$.
The graph suggest that most of the treatment effects occur at the
conforming loan limit. In time $t=+1$, the approval rate declines
for jumbo loans in the two bins at the immediate right of the limit,
while the treatment effects in the jumbo segment for loan amounts
from 2.5 to 10\% above the limit are similar to treatment effects
in the conforming segment. In time $t=+2$, the situation changes
drastically: the treatment effects drop sharply for almost all loan
amounts in the jumbo segment, and the discontinuity (drop) in approval
rates is visible at the conforming loan limit. A similar scenario
appears in time $t=+3$ (subfigure (c)). In time $t=+4$, we should
not expect a discontinuity in approval \emph{rates}, as documented
earlier and consistent with the baseline findings of both this paper
(Figure \ref{fig:results_replication_ouazad_kahn}) and \possessivecite{ouazad2022mortgage}
Figure 8.

Figure \ref{fig:inspecting_the_mechanism} presents the estimation
results when the outcome variable is the origination rate of mortgage
applications. Mortgages are originated when they are approved by the
lender and accepted by the borrower. The picture is qualitatively
similar to those for the approval rate (Figure \ref{fig:inspecting_the_mechanism-1}),
suggesting that it is not borrowers' behavior that is driving results,
but rather lenders. 

Figure \ref{fig:inspecting_the_mechanism} presents the estimation
results when the outcome variable is the securitization rate of mortgage
\emph{originations}. We should expect these results to be different
from those of the origination and approval rates, given that the securitization
probabilities conditional on origination picks up in years +3 and
+4. In practice, we observe that there are discontinuities in the
impact of billion dollar disasters on securitization probabilities
in each time period $t=+1$ to $t=+4$. The discontinuity in the treatment
effect become large: over +10 percentage points in time $t=+3$, over
+16 percentage points in time $t=+4$. The next subsection provides
a formal significance test of these discontinuities controlling for
the year, ZIP, disaster confounders, and the interaction of year,
ZIP, disaster f.e.s with the discontinuity as in \possessivecite{ouazad2022mortgage}.

\subsubsection*{Regression Discontinuity Bandwidth and Estimation Results}

The previous section~\ref{subsec:What-Drives-Results} suggested
that a significant part of the mechanism underlying the securitization
hypothesis is at the limit or in a window around the limit. This suggests
the use of the regression discontinuity framework, for which an extensive
literature provides methodological guidance, including \citeasnoun{imbens2008regression}
and \citeasnoun{cattaneo2022regression}. The approach here performs
such regression discontinuity while keeping the same of controls and
fixed effects that are present in the baseline specification \possessivecite{ouazad2022mortgage}:
year, ZIP, disaster, year$\times$Below the conforming loan limit.

The regression discontinuity estimator of the impact of a billion
dollar event is estimated as in \citeasnoun{cattaneo2022regression}.
We set $\delta=0$ (at the limit) in specification \ref{eq:approach_lpr}
and start by estimating the treatment effects exactly on the left
side of the conforming loan limit:
\begin{align}
\min & \sum_{it}\left(\text{Approval}_{it}-\sum_{t=-4}^{+4}\xi_{t}(\delta)\cdot\text{Treated}_{j(i)}\times\textrm{Time}_{t=y-y_{0}(d)}\right.\nonumber \\
 & \qquad\qquad\qquad\qquad-\textrm{Year}_{y(t,d)}-\text{Disaster}_{d}-\text{ZIP}_{j(i)}\Biggr)^{2}K\left(\frac{\Delta\text{Loan Amount}_{it}}{h}\right),\label{eq:approach_lpr-1}
\end{align}
for $\Delta\text{Loan Amount}_{it}<0$ (conforming segment). This
yields estimates of treatment effects $\xi_{t}(\delta=0^{-})$. We
do the same estimation, but setting $\delta=0$ and considering only
mortgages of the jumbo segment. This yields estimates of treatment
effects $\xi_{t}(\delta=0^{+})$. As in \citeasnoun{cattaneo2022regression},
the impact of a billion-dollar disaster on the discontinuity in approval
rates at the conforming loan limit is:
\begin{equation}
\tau_{RD}=\xi_{t}(\delta=0^{-})-\xi_{t}(\delta=0^{+})\label{eq:RD_te}
\end{equation}
The innovation here is that we estimate the impact of billion dollar
disasters on the regression discontinuity controlling for the $\textrm{Year}$
fixed effects, the $\textrm{Disaster}$ specific fixed effects, the
$\textrm{ZIP}$ code fixed effects. Since the fixed effects are different
on each side of the conforming loan limit, this also controls for
year $\times$ below limit, $ZIP$ $\times$below limit and disaster
$\times$ below limit confounders. 

This RD design of \citeasnoun{cattaneo2022regression} is straightforward
to implement. It corresponds to the regression:
\begin{align}
Outcome_{it} & =\alpha\cdot\text{Below\,Conforming\,Limit}_{ijy(t,d)}+\gamma\text{Below\,Conforming\,Limit}_{ijy(t,d)}\times\text{Treated}_{j(i)}\nonumber \\
 & \quad\quad+\sum_{t=-T}^{+T}\xi_{t}\cdot\text{Treated}_{j(i)}\times\text{Time}{}_{t=y-y_{0}(d)}\nonumber \\
 & \quad\quad+\sum_{t=-T}^{+T}\tau_{t}\cdot\text{Below\,Conforming\,Limit}_{ijy(t,d)}\times\text{Treated}_{j(i)}\times\text{Time}{}_{t}\nonumber \\
 & \quad\quad+\sum_{y=1995}^{2016}\zeta_{y}\cdot\text{Below\,Conforming\,Limit}_{ijy(t,d)}\times\text{Year}{}_{y(t)}\nonumber \\
 & \quad\quad+\sum_{d}\chi_{d}\cdot\text{Below\,Conforming\,Limit}_{ijy(t,d)}\times\text{Disaster}_{d}\nonumber \\
 & \quad\quad+\sum_{j}z_{j}\cdot\text{Below\,Conforming\,Limit}_{ijy(t,d)}\times\text{ZIP}_{j(i)}\nonumber \\
 & \quad\quad+\text{Year}{}_{y(t,d)}+\text{Disaster}{}_{d}+\text{ZIP}{}_{j(i)}+\varepsilon_{it},\label{eq:main_specification}
\end{align}
weighted by the distance of each loan to the conforming loan limit
$K\Biggl(\frac{\Delta\textrm{Loan Amount}_{it}}{h}\Biggr)$. The parameters
of interest are the treatment effects $\tau_{t}$ for $t=1,2,3,4$.
They measure the discontinuity in treatment effects at the conforming
loan limit.

We use a Gaussian kernel for $K(\cdot)$, popular in a number of seminal
papers such as \citeasnoun{dinardo1996labor}, \citeasnoun{connor2012efficient},
\citeasnoun{duclos2004polarization}, \citeasnoun{barone2008garch}.
This fixed effect panel regression is similar to the main specification
of \citeasnoun{ouazad2022mortgage}, where observations are weighted
by the values $K_{i}=K\left(\frac{\log L_{i}-\log\overline{L}^{*}}{h}\right)$
and the bandwidth $h$ varies between 1\% and 20\%. The standard errors
are double-clustered by ZIP and year.

Tables \ref{tab:regression_approval}\textendash \ref{tab:reg_securization}
present the results. For each table, the upper panel is for bandwidths
$h$ of 1\% to 4\%, in increments of 1 ppt. The lower panel is for
bandwidths $h$ of 5, 10, 15, 20\%. Table \ref{tab:regression_approval}
is for the approval rate, Table \ref{tab:reg_originated} is for the
origination rate, Table \ref{tab:reg_securization} for the securitization
rate in the universe of originated mortgages. 

Each table reports both the impacts for the mortgage market ($\text{Treated}_{j(i)}\times\text{Time}{}_{t=y-y_{0}(d)}$
indicator variables) and for the conforming market ($\text{Below\,Conforming\,Limit}_{ijy(t,d)}\times\text{Treated}_{j(i)}\times\text{Time}{}_{t}$
indicator variables). 

Results suggest that the bandwidth of 1\% to 3\% maximizes the mean
squared error for the approval and origination outcome variables respectively.
Results suggest that billion dollar disasters increase the discontinuity
in approval probabilities by up to 6.48\% ({*}{*}), increase the discontinuity
in origination probabilities by 6.08\% ({*}{*}), and the discontinuity
in securitization probabilities by up to 17.7\% ({*}{*}{*}). The timing
of the effects also matches the timings of the main regression: an
increase in approval and origination rates, followed by an increase
in securitization rates. The code is available at \emph{\href{https://www.ouazad.com/paper/code_2023_05.zip}{https://www.ouazad.com/paper/code\_2023\_05.zip}}.

\section{Policy Trade-Offs: Market Pricing of Climate Risk and Bluelining\label{sec:Policy-Trade-Offs:-Market}}

\subsection{Flexible Market-Based Guarantee Fees}

When lenders securitize their mortgages by selling them to the Government
Sponsored Enterprises Fannie Mae and Freddie Mac, they transfer the
risk of delinquency and default in exchange for the payment of a guarantee
fee. These guarantee fees are presented in a Loan Level Performance
Adjustment matrix, as a number of basis points of the balance of the
loan. Guarantee fee reports of the Federal Housing Finance Agency
(FHFA) suggest that g-fees are independent of climate risk. 

Under the \emph{Climate Securitization Hypothesis}, lenders may take
advantage of securitization whenever either their estimates of losses
for a specific mortgage are higher than g-fees, when lenders are risk
averse \textendash{} and thus their risk neutral probability of losses
is higher than the g-fees \textendash , or when private label securitization
offers a greater income than government securitization. This may happen
as (1) g-fees do not depend on climate risk including forecasts of
drought, wildfires, hurricane storm surges, riverine flooding and
(2) lenders may develop expertise, human capital, and data science
to assess climate risk at a granular level. 

Guarantee fees are not set by a competitive process of auctions or
the meeting of bid and ask on a market as in \citeasnoun{duffie2005over}.
Rather, guarantee fees are set by the Federal Housing Finance Agency
as the outcome of an administrative and political process. \citeasnoun{deritis2014general}
outlines three rules for setting g-fees at an appropriate level: (1)
ensuring an appropriate level of capital, and \citeasnoun{layton2023}
suggests that the average g-fee has been between 0.45 and 0.49\% since
2014, lower than the level that allow Fannie Mae and Freddie Mac to
reach its capital standard. (2) g-fees with acceptable redistributive
properties, such as lower g-fees for lower income households; this
paper suggests that g-fees may be adjusted to be a function of both
climate risk and other parameters such as FICO, LTV, DTI. The pass-through
of such g-fees into mortgage interest rates would incentivize adaptation
efforts~\cite{kahn2021adapting} in the form of greater screening~
and greater self-protection~\cite{ehrlich1972market}. (3) g-fees
should be set taking into account the elasticity of the supply of
loans to the agencies; yet, recent increases in the market share of
GSE loans combined with the decline in both unsecuritized first liens
and private-label MBSs suggest that the supply of loans has low elasticity
w.r.t. to the g-fees at their current level.

\begin{figure}[H]
\caption{Loan Level Price Adjustment matrix on May 5th, 2023}
\label{fig:llpa}

\emph{This table presents the guarantee fees (g-fees) charged by Fannie
Mae for the securitization of purchase money loans. Additional g-fees
may be charged for condos, ARMs, investment properties, second homes,
high balance loans, and high DTI ratios above 40\%. }

\begin{center}

\includegraphics[scale=0.75]{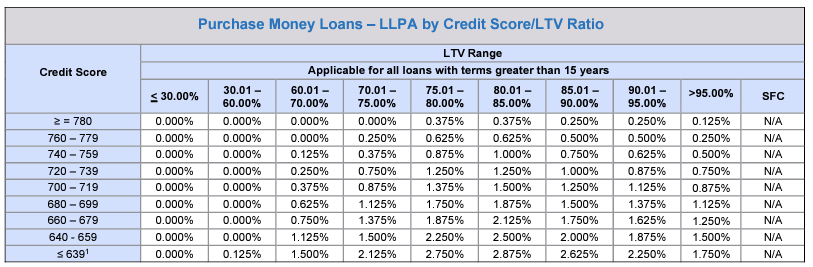}

\end{center}

\emph{Source: https://singlefamily.fanniemae.com/media/9391/display
.}
\end{figure}

Aside from guarantee fees, the impact of climate risk on the agencies
may be mitigated by (1)~flood insurance policies, (2)~private mortgage
insurance, (3)~credit risk transfers such as Connecticut Avenue Securities.
We examine these in turn.

The 10-K filing of Fannie Mae for the fiscal year ended December 31,
2022 suggests that flood insurance policies covering homes under the
National Flood Insurance Program may have a limited impact on credit
losses.
\begin{quote}
\emph{Only a small portion of loans in our guaranty book of business
as of December 31, 2022 was located in a Special Flood Hazard Area,
for which we require flood insurance: 3.3\% of loans in our single-family
guaranty book of business and 6.8\% of loans in our multifamily guaranty
book of business. We believe that only a small portion of borrowers
in most places outside of a Special Flood Hazard Area obtain flood
insurance. The risk of significant flooding in places outside of a
Special Flood Hazard Area (that is, in places where we do not require
flood insurance) is expected to increase in the coming years as a
result of climate change.}
\end{quote}
\citeasnoun{cohen2021storm} suggests that natural disasters such
as hurricane Sandy provide new news about the location of flood risk
beyond the boundaries of the Special Flood Hazard Areas.

Private Mortgage Insurance covers mortgages with loan-to-value (LTV)
ratios above 80\%. The minimum balance covered for fixed-rate mortgages
with a 30-year maturity is 6\% for mortgages with an LTV between 80
and 85\%, 12\% for mortgages between 85 and 90\%, and between 16 and
35\% for mortgages with LTVs above 90\%. \citeasnoun{deritis2014general}
suggests that on average PMI covers 27.5\% of the balance. \citeasnoun{bhutta2022moral}
studies the PMI market and suggests that there is little variation
in premia across places or across private mortgage insurers, and thus
competition in this market is primarily in volume rather than pricing.
Hence PMI pricing incentives are not correlated with climate risk.
Finally, 2022 10-K filings suggest that ``\emph{Although our primary
mortgage insurer counterparties currently approved to write new business
must meet risk-based asset requirements, there is still a risk that
these counterparties may fail to fulfill their obligations to pay
our claims under insurance policies.}'' Hence the impact of climate
risk on the agencies' net income and capital depends on the magnitude
of the impact of climate risk on delinquencies and defaults.

The third and most recent approach to hedging credit losses is the
Credit Risk Transfer program. \citeasnoun{gete2022climate} collect
CRT yield data and study the response of such yields to differential
exposure to hurricanes Irma and Harvey. These CRT yields have the
potential of providing market-implied g-fees,\footnote{\citeasnoun{calabria2023shelter} suggests major flaws in the CRT
program, as they provide insufficient coverage for credit losses.} and the authors suggest that they would be 10\% higher than the current
g-fees in counties most exposed to hurricanes and 35\% lower in inland
counties. Translated into the current LLPA matrix of Figure~\ref{fig:llpa},
this suggests that this would be between a 3.75 and 28.75 basis point
increase in counties exposed to hurricanes; and a decline between
13 basis points and 100 basis points in inland counties. This paper
thus suggests the possibility of adding a third dimension to the LLPA
matrix, which would provide both redistribution away from coastal
areas and incentives for households to locate in safer areas. 

\subsection{A Challenging Trade-Off Between Redlining and the Price Signal of
Higher Flood Risk}

While the efficiency of the mortgage and securitization markets would
suggest flexible market-based g-fees, one concern is that such pricing
would entail a new form of redlining, the \emph{bluelining} of hurricane-exposed
areas.\footnote{An extensive literature discusses the historical practice of mortgage
redlining \cite{aaronson2021effects,fishback2020race,fishback2022new}.
\citeasnoun{aaronson2021effects} is an empirical analysis of the
discontinuities implied by the Home Owners Loan Corportation's redlining
maps. \citeasnoun{fishback2020race} and \citeasnoun{fishback2022new}
examines the evidence on redlining of \citeasnoun{aaronson2021effects}. } \citeasnoun{aaronson2021effects} finds substantial welfare consequences
of redlining beyond interest rates and the approval rates of mortgage
applications: lower homeownership rates, house values, and higher
segregation. Thus a key question raised by \emph{bluelining} is the
trade-off between the signal of increased flood risk, which incentivizes
moving out of harm's way vs. the lasting impacts on affordability
and homeownership. Such policy trade-off is at the frontier of policy
and research discussions.

A key empirical question is whether those areas most at risk of flooding
are also areas with lower income households and minority households.
Figure \ref{fig:demographics_census} presents the descriptive (correlational)
relationship between Census demographics and First Street Foundation's
flood risk factor at the Zip level. In these figures we consider four
simple metrics: the share white, the median household income, the
share on cash public assistance or food stamps/SNAP, and the rent
as share of household income. These descriptive plots are not causal,
and they also do not control for state or MSA fixed effects, thus
the sorting within states and within metropolitan areas may differ
from the US-wide sorting depicted in the chart.

The Figure portrays a nuanced set of facts. First, areas at risk of
flooding according to First Street have lower median household income.
Second, these areas tend to have higher shares of whites. Third, these
households spend more on rent as a fraction of household income, and
they are also more likely to be on cash public assistance, food stamps
or SNAP. This suggests that a geographic pricing of g-fees may have
distributional consequences that differ markedly from the historical
legacy of FHA or HOLC redlining: households at risk of flooding are
more likely to be lower income white households.

\section{Conclusion: Adaptation to Climate Risk and the Securitization Market\emph{\label{sec:Conclusion:-Adaptation-and}}}

The Climate Securitization Hypothesis posits that natural disasters
increase the probability that lenders originate loans sold to government
sponsored securitizers. An emerging natural disaster literature explores
how shocks affects many aspects of the real economy~\cite{boustan2020effect}.
In the real economy, margins of adjustment include migration~\cite{deryugina2018economic,smith2022adjusting},
home price declines~\cite{ortega2018rising,cohen2021storm}, and
public finance dynamics~\cite{healy2009myopic,jerch2023local}. In
the financial sector, as investment and consumption are typically
leveraged~\cite{geanakoplos2010leverage} , margins of adjustment
in response to climate risk exposure include the LTV ratios~\cite{sastry2021bears,bakkensen2023leveraging},
interest rates and yields~\cite{goldsmith2022sea,nguyen2022climate},
approval rates, amortization structures of loans. In the financial
sector, risk is hedged using risk transfers~\cite{gete2022climate}
and derivative products~\cite{ouazad2022investors}. In financial
markets, natural disasters may also affect market microstructure and
liquidity \cite{rehse2019effects}.

A consistent finding in this empirical literature is that natural
disasters induce shifts in investment, consumption, and leverage as
both households and investors reoptimize in the aftermath of these
events. Such natural disasters increase uncertainty~\cite{baker2020using}.
Natural disasters may also introduce ambiguity about the distribution
of future probability distributions, making assets exposed to such
ambiguous risks less valuable. \footnote{The finance literature suggests that ambiguity is priced~\cite{brenner2018asset}.}

In the aftermath of a natural disaster, physical assets in the affected
area can become less valuable and cash flows may decline. In the mortgage
market, evidence suggests a higher probability of loan delinquency
and default risk in the aftermath of such shocks~\cite{ratcliffe2020bad,issler2021housing,holtermans2022climate,biswas2023california,ho2023we,ouazaddiscussion}.
Hikes in insurance premia and losses of earning opportunities~\cite{indaco2021hurricanes}
may cause cash flow shocks that are one of the triggers (one of the
double hurdles) of mortgage delinquencies~\cite{ganong2023borrowers}.
The increased probability of delinquency and foreclosure may be higher
if the spatial correlation of the shock is such that many neighboring
home owners are simultaneously considering defaulting on their loans,
causing foreclosure externalities~\cite{gerardi2015foreclosure}.

Under the current rules, bank and non-bank lenders in exposed areas
have increased incentives to ``originate and distribute'' loans.
This is the Climate Securitization Hypothesis. The securitization
option is particularly valuable when losses are either correlated
in the lender's portfolio or if such losses are correlated with the
lender's net income. During a time of more intense natural disasters,
this issue takes on a greater importance. If lenders are aware that
they can securitize such loans, this weakens their screening incentives
and households' incentive to invest in self-protection.

\citeasnoun{lacour2022adverse} acknowledge this microeconomic logic
but claim that the empirical approach does not provide evidence supporting
the CSH. In this paper, we have examined their claims and have focused
on both the statistical validity and the economic content of their
statistical criticisms. In our re-examinination of the evidence, we
have made three main points: first, the construction errors of the
event study in \citeasnoun{lacour2022adverse} are very serious and
render their results uninterpretable. Hurricane years are miscoded.
Recent literature provides guidance on the appropriate construction
of difference-in-difference event studies~\cite{callaway2021difference}.
Second, lenders have incentives to originate conforming and jumbo
loans even at small distances of the loan amounts to the conforming
loan limit, suggesting that not all mortgages in the vicinity of the
limit should be classified as conforming. Third, careful and granular
analysis of the discontinuities in approval, origination and securitization
probabilities suggests that lenders are significantly more likely
to originate conforming loans than jumbo loans in the aftermath of
a natural disaster. This identification strategy by regression discontinuity
design (RDD) at the conforming loan limit identifies the \emph{option
value of securitization}, i.e. the impact of the securitization option
on the net income of mortgage originations for lenders.

Our re-examination of the evidence provides confidence in the recent
sample evidence supporting the CSH. As future disasters take place,
we conjecture that nimble financial markets will play a key role in
the climate change adaptation process. Financial engineering helps
in the diversification of risk and in designing new assets that complete
markets. 

First, financial markets will improve the design of pools that either
diversify climate risk or provide selected exposure to climate risk
factors. Climate risk will change the design of Agency and Private
Label Mortgage-Backed Securities. They may also affect the design
of To Be Announced (TBA) markets \cite{vickery2013tba} for agency
Mortgage-Backed Securities as those currently do not feature climate
risk among the observable set of characteristics of traded pools.

Second, financial markets can complete the market with securities
that provide cash flows contingent on the realization of risk. This
includes catastrophe bonds, insurance linked securities, and weather
derivatives. 

\bibliographystyle{agsm}
\bibliography{ouazad_kahn_2023}

\pagebreak{}

\clearpage{}

\begin{figure}

\caption{Rounding and Loan Amounts: Why LaCour-Little et al.'s (2022) Approach
Systematically Biases the Counts of Conforming Loans in Two Cases
\label{fig:Rounding-and-Loan}}
\label{fig:rounding}

{\footnotesize{}These figures are an illustration of the rounding
of loan amounts to the nearest integer. The true loan amount $L^{*}$
is on the horizontal axis. $L$, the observed loan amount is on the
vertical axis. $\overline{L}^{*}$ the true conforming loan limit,
is the green line. The conforming loan limit rounded up as in \citeasnoun{lacour2022adverse}
is the red line. The \citeasnoun{lacour2022adverse} approach yields
a systematic overestimation of the number of conforming loans. In
contrast, the approach of \citeasnoun{ouazad2022mortgage} yields
either an overestimation (case 1) or an underestimation (case 2) of
the number of conforming loans.}{\footnotesize\par}

\begin{center}

\subfloat[Case \#1: Clay County, Florida, Limit \$424,100]{

\centering{}\includegraphics[scale=0.32]{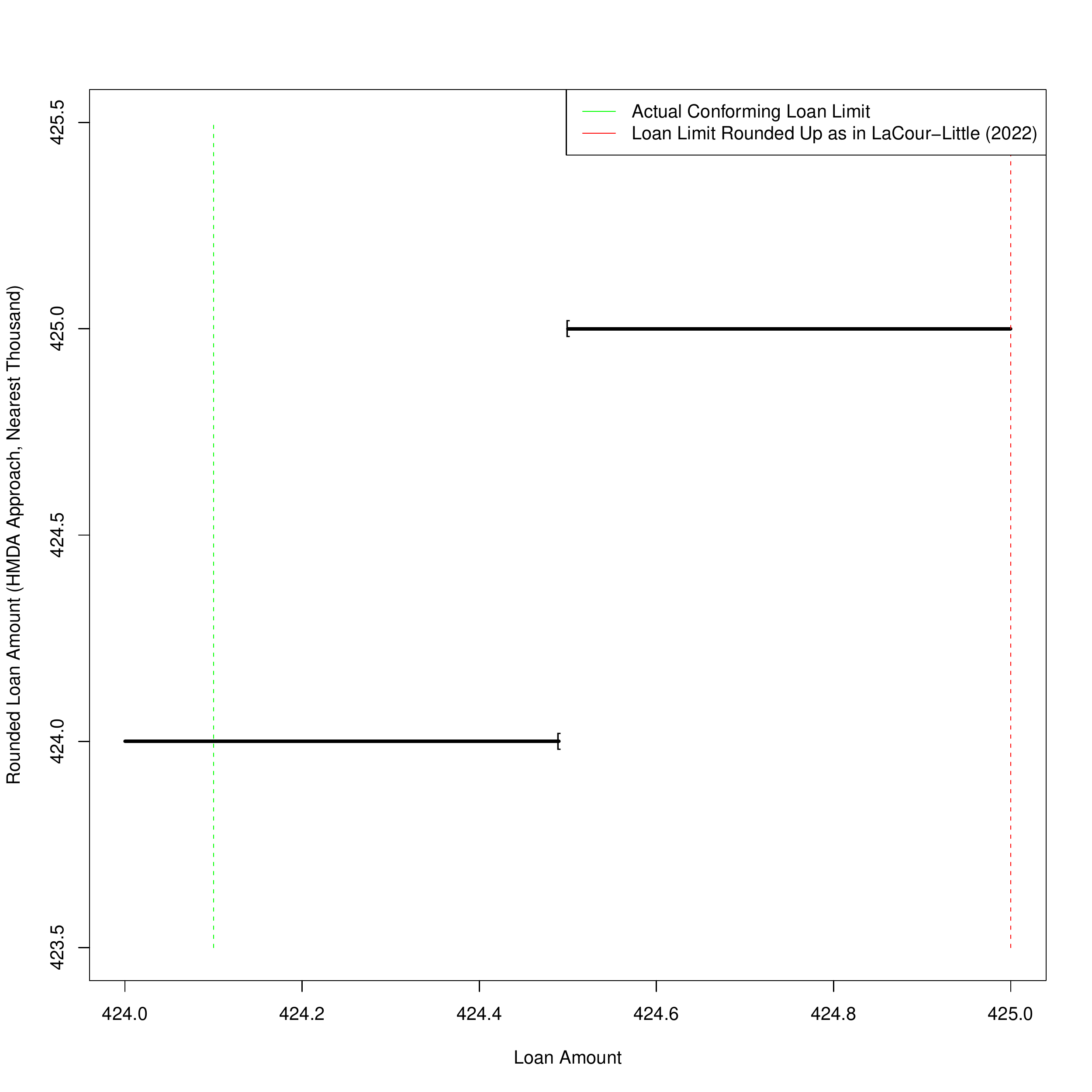}}

\subfloat[Case \#2: Collier County, Florida, Limit \$450,800]{
\centering{}\includegraphics[scale=0.32]{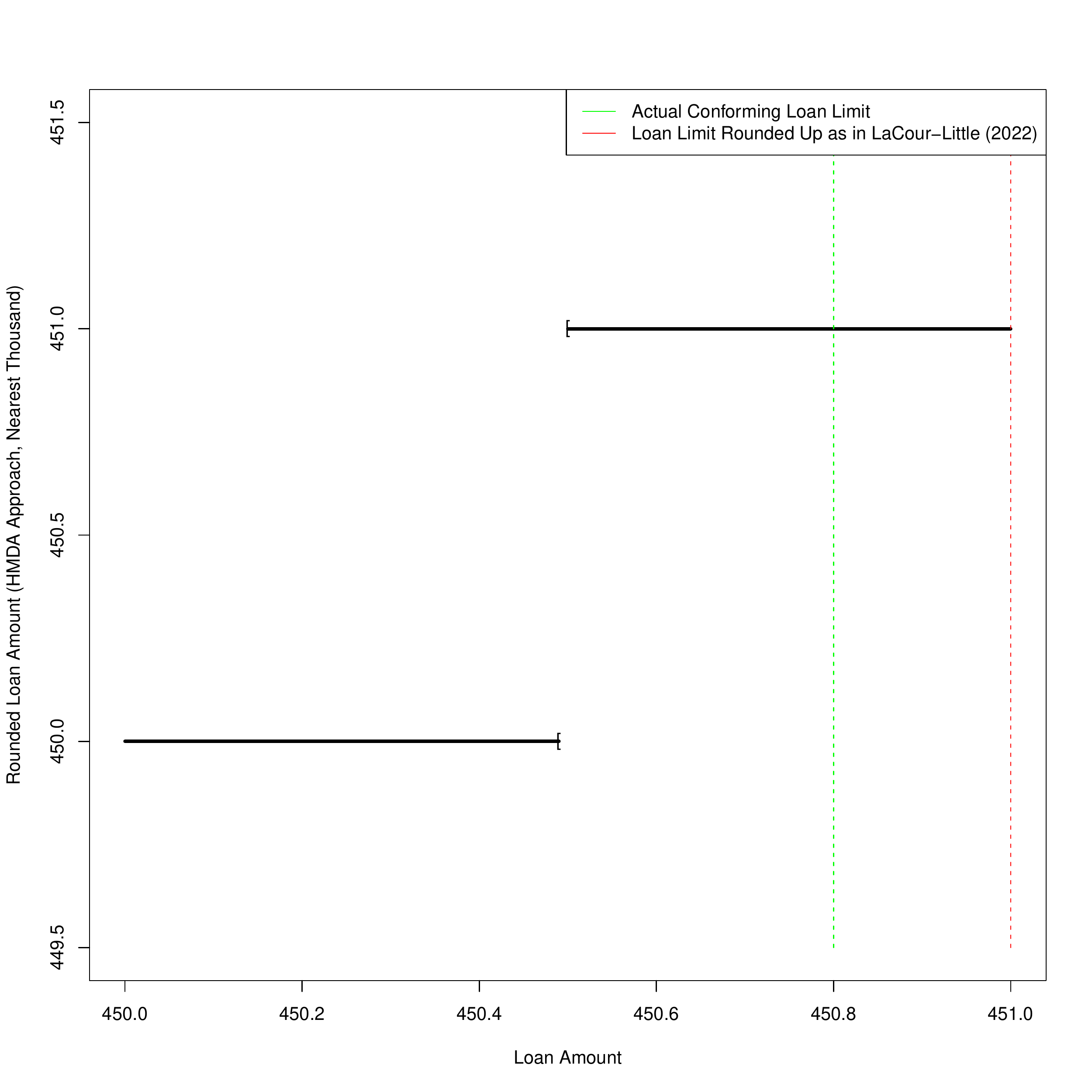}}

\end{center}
\end{figure}

\pagebreak{}

\clearpage{}

\begin{figure}
\caption{Bias and Imprecision of the LaCour-Little et al. (2022) rounding approach
with Cross-Sectional Bunching Estimators \textendash{} Bias and Standard
Deviation vs Number of Observations\label{fig:sample_size}}

{\footnotesize{}The top panel present the evolution of the absolute
deviation $\vert\hat{\beta}^{s}-\hat{\beta}^{*}\vert$ with respect
to the number of observations in the window around the conforming
loan limit, in the Monte Carlo Simulation. The orange line is using
the \citeasnoun{lacour2022adverse} approach. The blue line uses \possessivecite{ouazad2022mortgage}
approach. The bottom panel presents the standard deviation $SD(\hat{\beta}^{s})$
of the estimators. Same color code. \possessivecite{lacour2022adverse}
yields higher absolute deviations and higher variances for all sample
sizes.}{\footnotesize\par}

\begin{center}

\subfloat[Absolute Deviation of the Cross-Sectional Bunching Estimators from
the True Value]{

\includegraphics[scale=0.33]{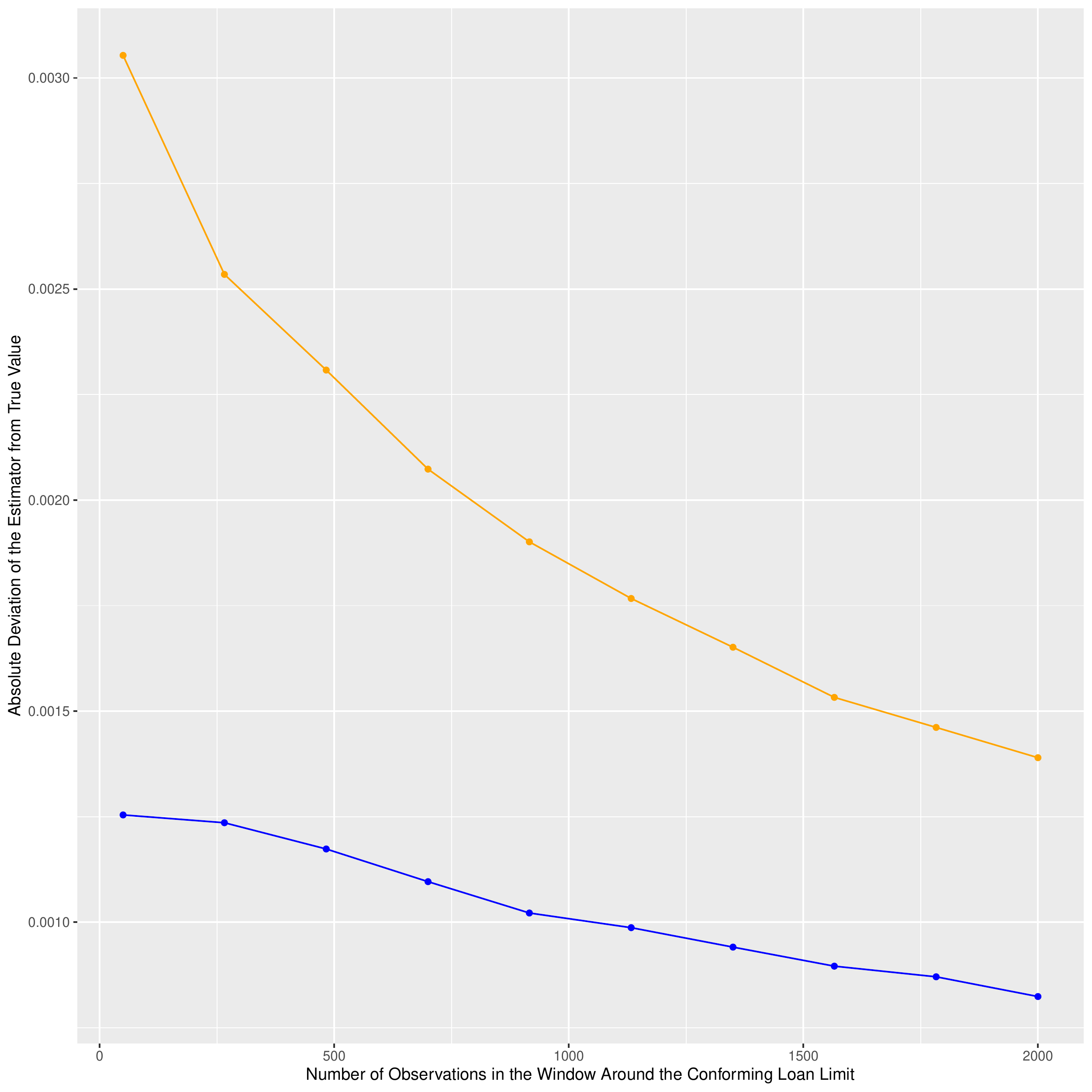}}

\subfloat[Standard Deviation of the Cross-Sectional Bunching Estimators]{

\includegraphics[scale=0.33]{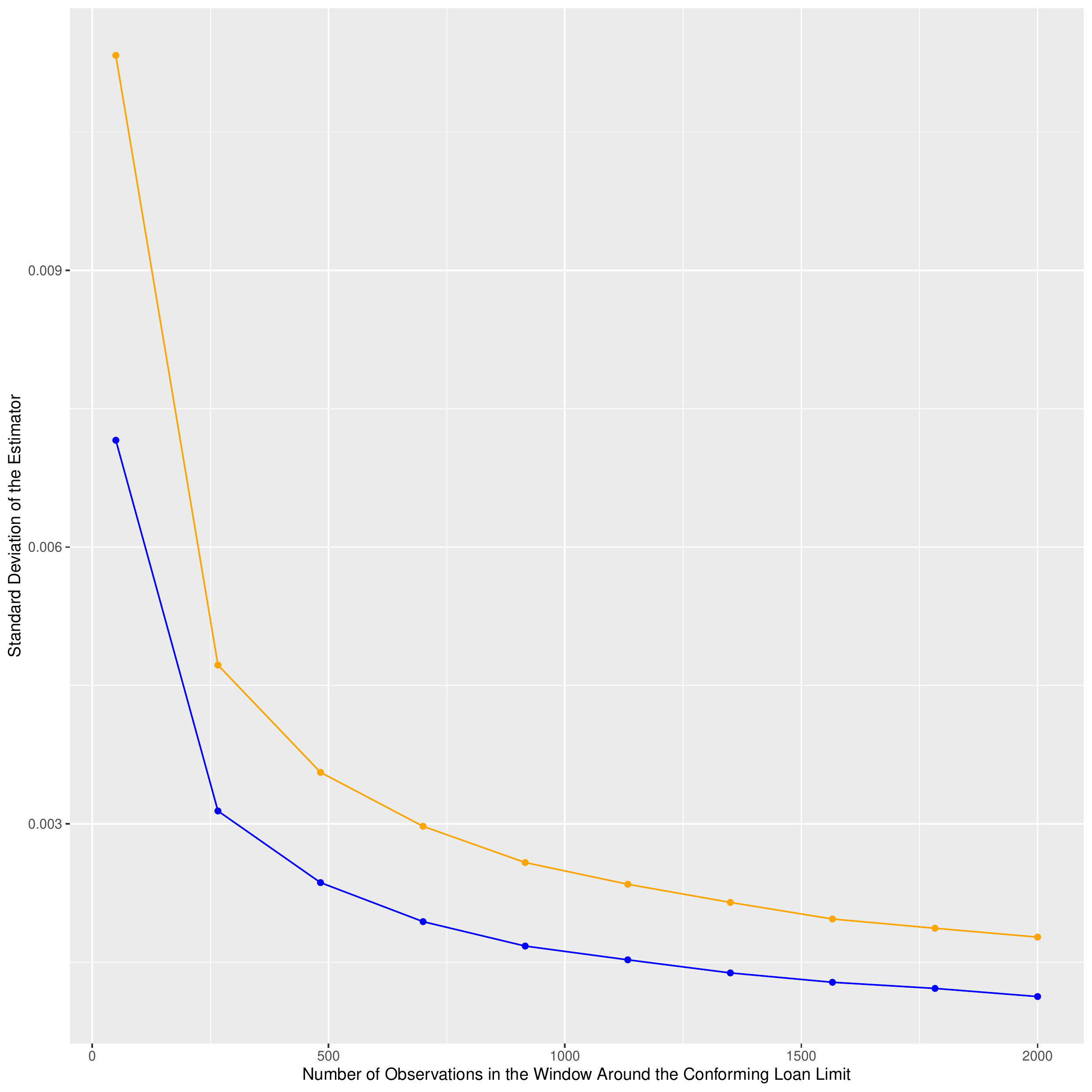}}

{\footnotesize{}The parameters of the Monte Carlo simulations are
$N\in[50,2000]$, $\alpha=0.2$, $\beta=0.1$.}{\footnotesize\par}

\end{center}
\end{figure}
\pagebreak{}

\clearpage{}
\begin{figure}
\caption{Rounding and Loan Amounts: A Monte-Carlo Analysis of Variance and
Bias}
\label{fig:rounding-monte-carlo}

\emph{\footnotesize{}These figures present the distribution of the
difference between the cross-sectional bunching estimator $\hat{\beta}^{s}$
with misclassification and the estimator $\hat{\beta}^{*}$ on the
true data. We consider two possible misclassifications: $s=H$ when
using the classification of conforming loans using the observed HMDA
loan amount and the actual conforming loan limit, as in}{\footnotesize{}
\citeasnoun{ouazad2022mortgage}}\emph{\footnotesize{}; $s=LL$ when
the using the LaCour-Little classification based on rounding the conforming
loan limit to the highest integer. The LaCour-Little approach leads
to higher variance and higher absolute bias.}{\footnotesize\par}

\begin{center}

\subfloat[Case \#1, Clay County, FL, Distribution of $\hat{\beta}^{s}-\hat{\beta}^{*}$]{
\centering{}\includegraphics[scale=0.32]{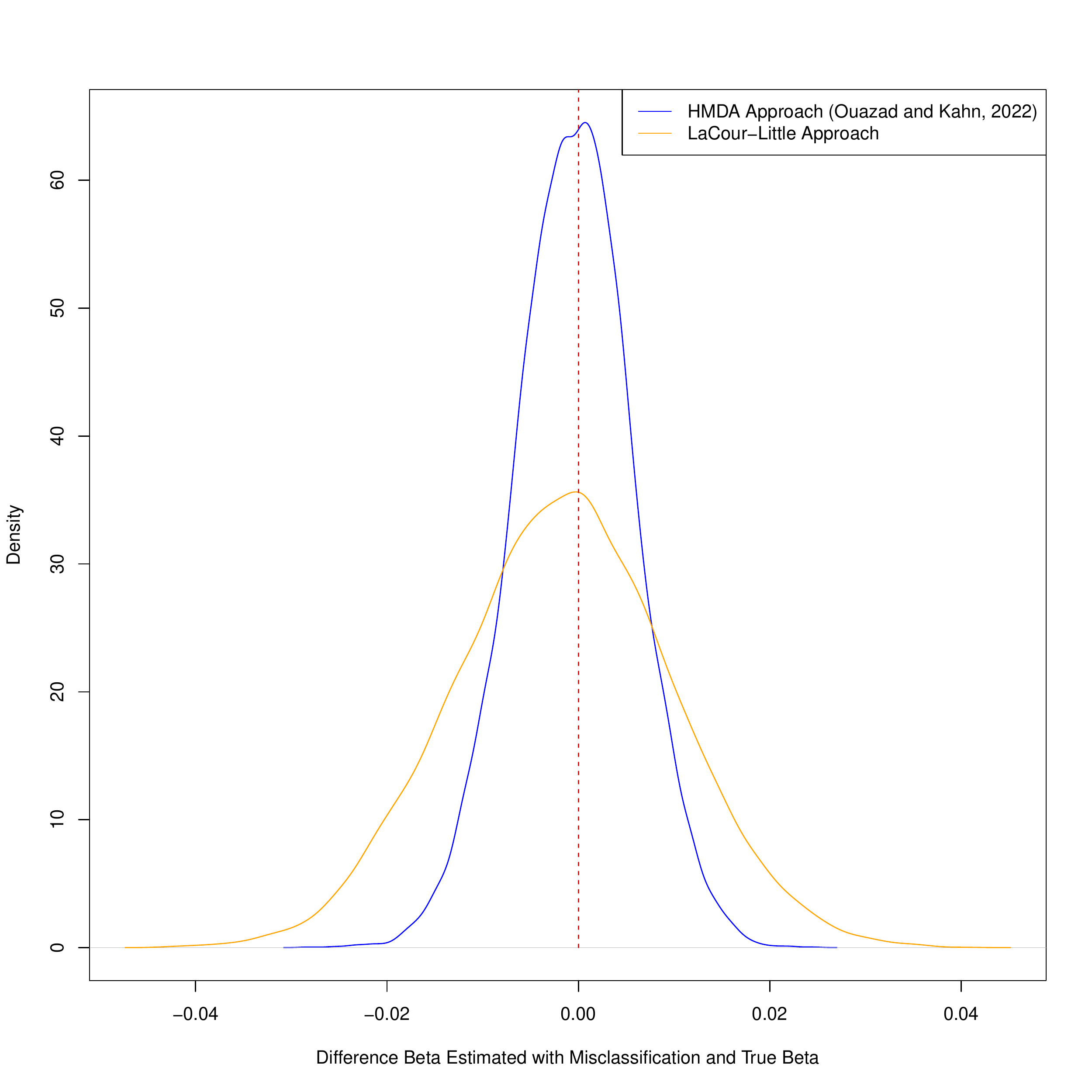}}\subfloat[Case \#1, Clay County, FL, Distribution of $|\hat{\beta}^{s}-\hat{\beta}^{*}|$]{
\centering{}\includegraphics[scale=0.32]{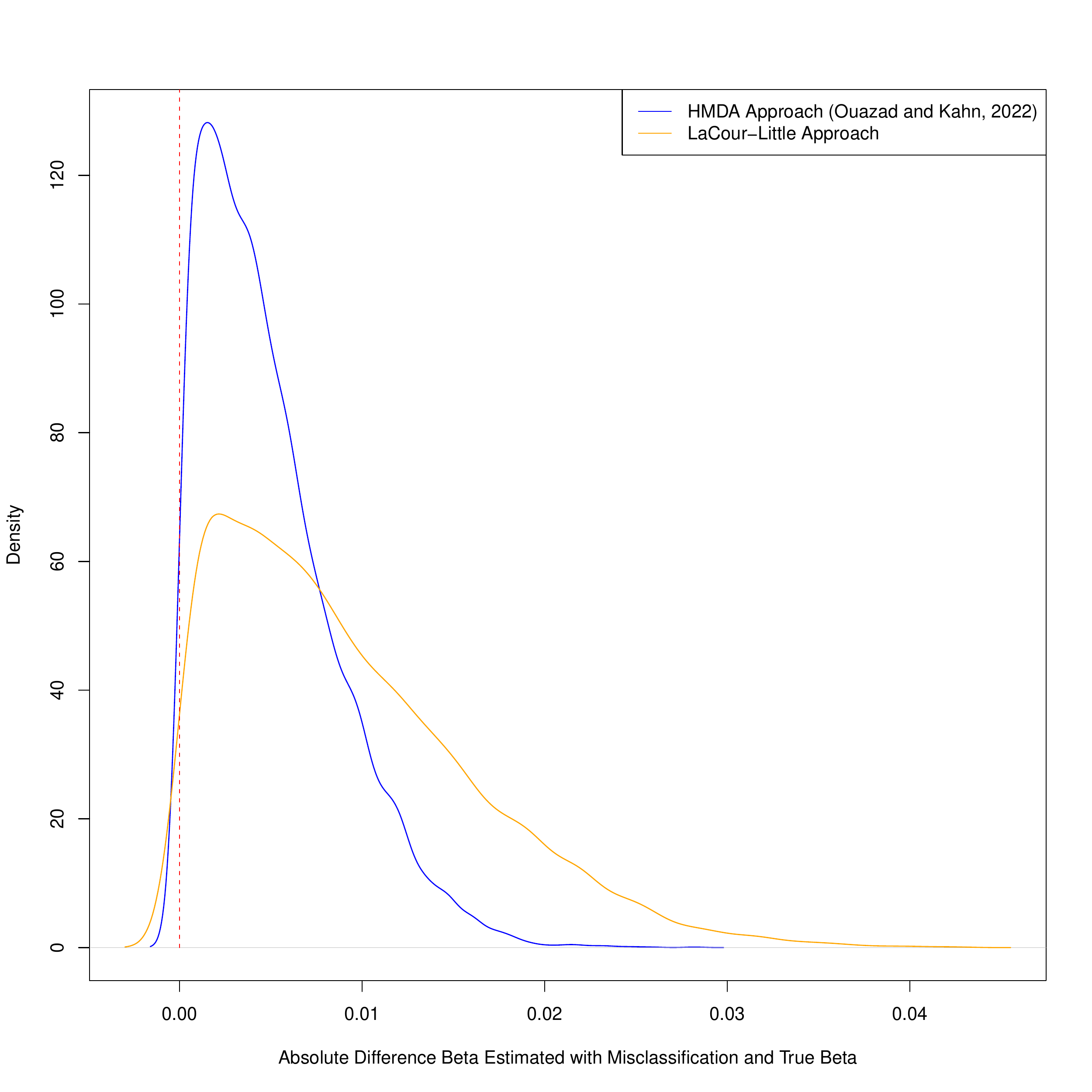}}

\subfloat[Case \#2, Collier County, FL, Distribution of $\hat{\beta}^{s}-\hat{\beta}^{*}$]{
\centering{}\includegraphics[scale=0.32]{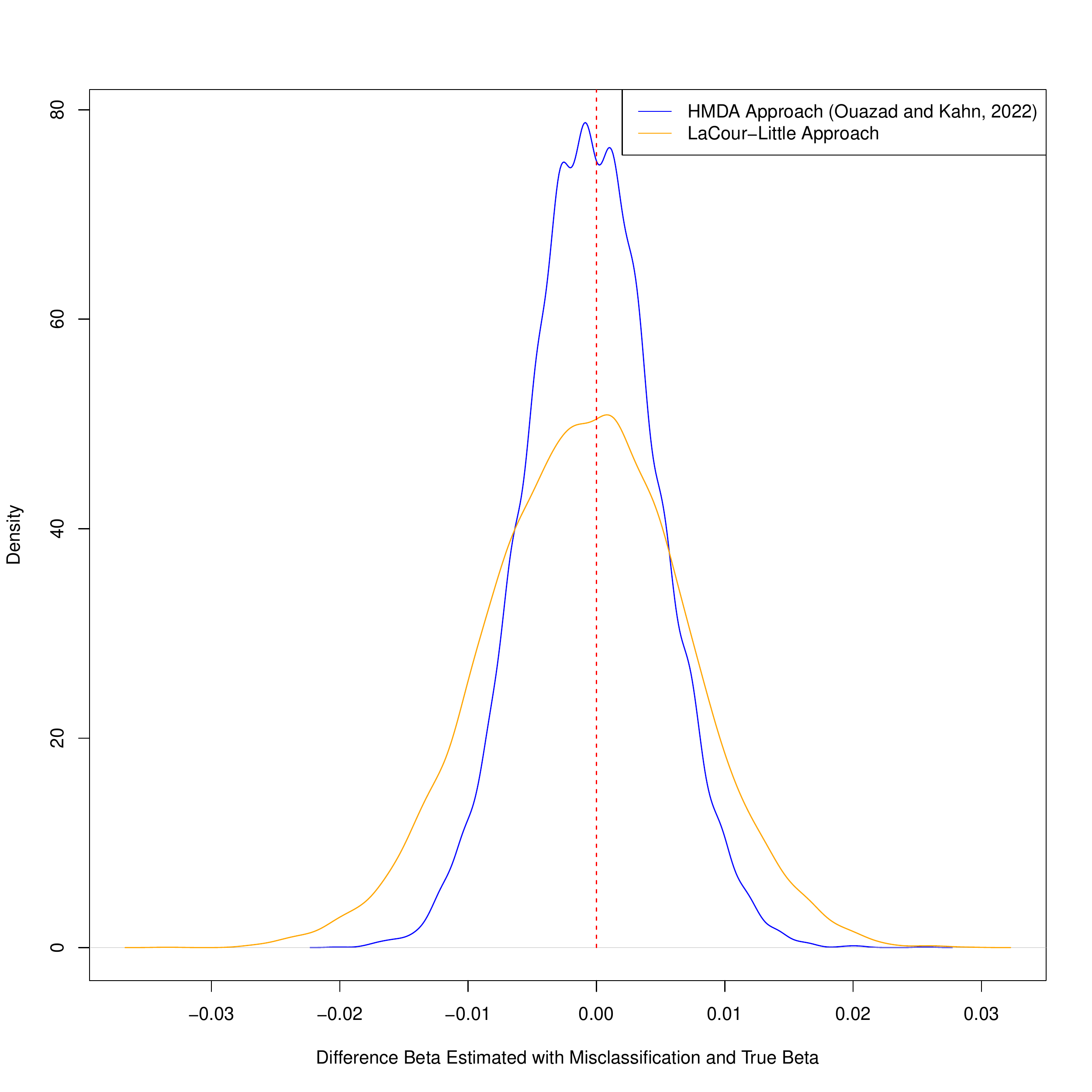}}\subfloat[Case \#2, Collier County, FL, Distribution of $|\hat{\beta}^{s}-\hat{\beta}^{*}|$]{
\centering{}\includegraphics[scale=0.32]{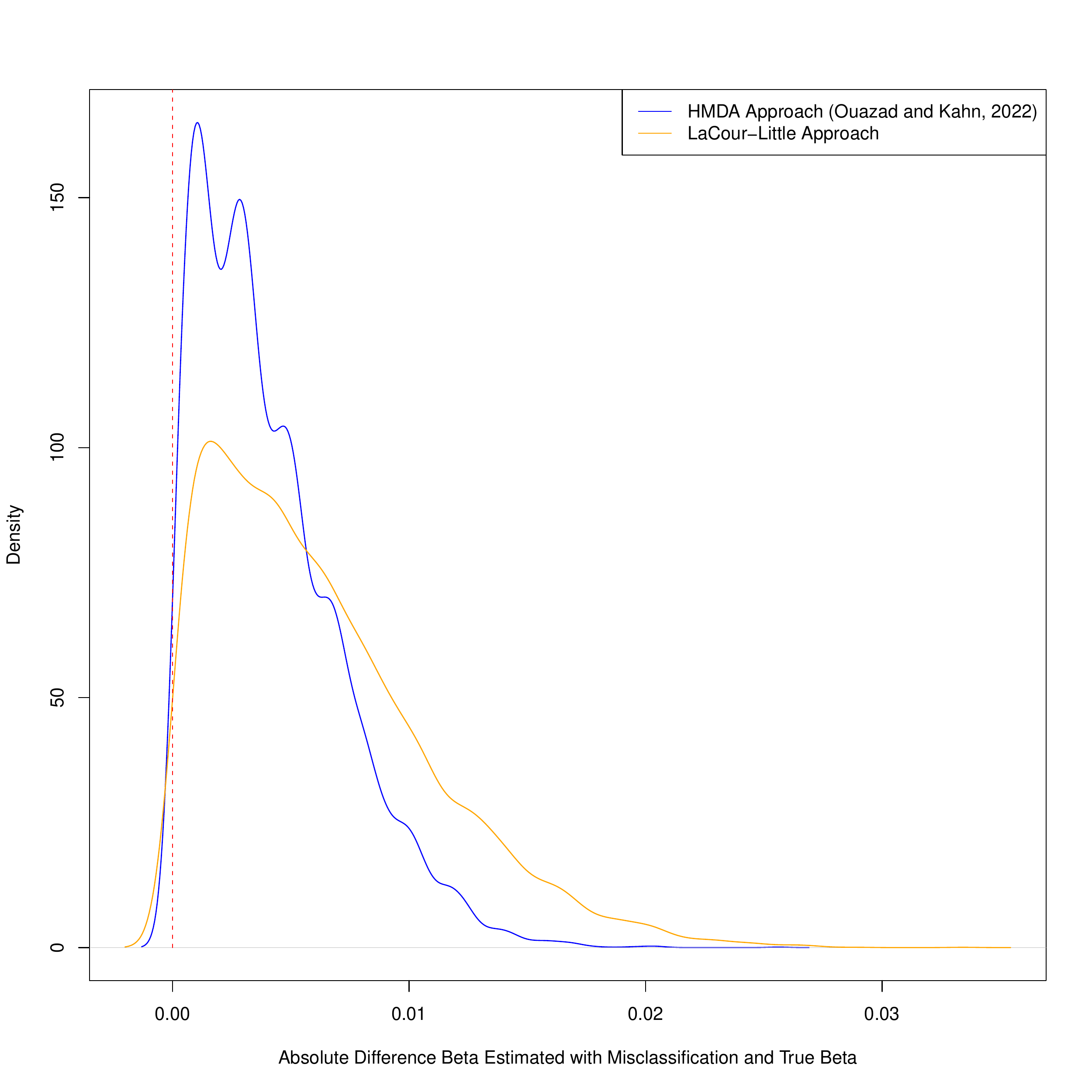}}

\bigskip{}

{\footnotesize{}The parameters of the Monte Carlo simulations are
$N=1,000$, $\alpha=0.2$, $\beta=0.1$.}{\footnotesize\par}

\end{center}
\end{figure}

\pagebreak{}

\clearpage{}

\begin{figure}
\caption{Geographic Distribution of Conforming Loan Limits and High Cost Counties}
\label{fig:high_cost}

\emph{This map presents the county-level conforming loan limit in
2016 (insights are similar when mapping other years post HERA). The
white dotted line are the boundaries of the Atlantic states and states
of the Gulf of Mexico, where most hurricanes occur and is the focus
of} \citeasnoun{ouazad2022mortgage}\emph{. Most high-cost counties
are outside the area of the Atlantic states and the Gulf of Mexico,
with the exception of the New York-Newark-Jersey City, NY-NJ-PA metro
area. Yet there both \citeasnoun{lacour2022adverse} and \citeasnoun{ouazad2022mortgage}
find significant impacts of hurricane exposure on origination probabilities
in the conforming segment.}

\begin{center}

\includegraphics[scale=0.7]{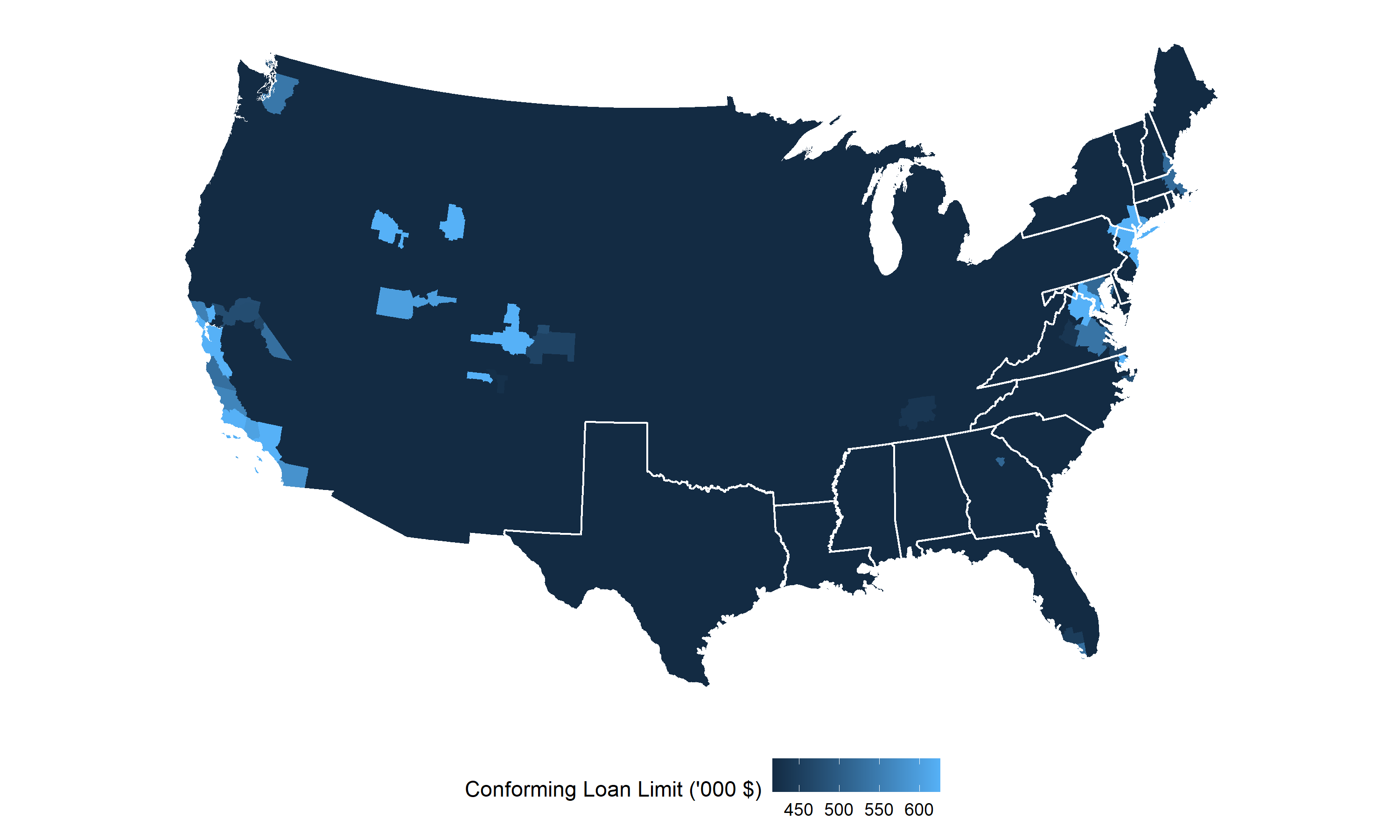}

\end{center}

\emph{Sources: County shapefile (2014) from the US Census Bureau,
conforming loan limits as in }\emph{\footnotesize{}\citeasnoun{lacour2022adverse}. }{\footnotesize\par}
\end{figure}

\pagebreak{}

\clearpage{}

\newgeometry{left=0.3in,right=0.3in,bottom=0.3in,top=0.3in}

\begin{figure}
\caption{Treatment Effects \textendash{} Replication of Ouazad and Kahn (2022)
with Limits Provided by LaCour-Little (2022) }
\label{fig:results_replication_ouazad_kahn}

\emph{These four graphics present the impact of a billion-dollar event
on the approval, origination rates and securitization conditional
on origination for mortgage applications (Section \ref{ssec:specification_main}).
The left column present the impact on conforming loans only (the coefficients
of the Below Limit} $\times$ \emph{Treated} $\times$ \emph{Time
$k$). The right column presents the impact for the entire mortgage
market in the window around the conforming loan limit. Confidence
intervals at 95\% obtained by double-clustering standard errors at
ZIP and year levels. These results are similar to }\possessivecite{ouazad2022mortgage}\emph{
Figure 8 and the source code is available at \href{https://github.com/aouazad/Mortgage-Securitization-Natural-Disasters-Reply.git}{this link}.} 

\begin{center}

\subfloat[Approval Probabilities, Conforming Market]{\centering{}\includegraphics[scale=0.35]{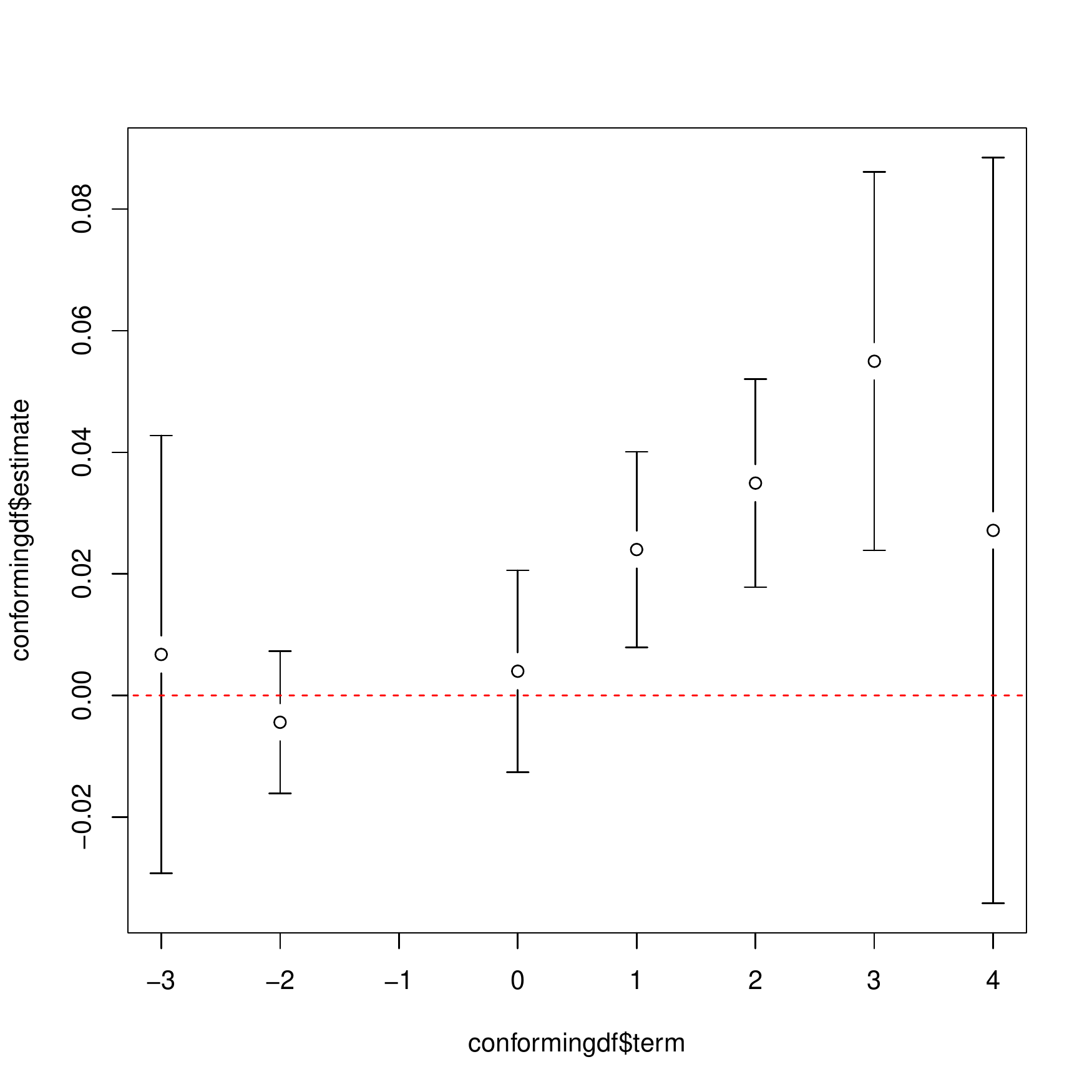}}~~~~~~~~~~~~\subfloat[Approval Probabilities, Overall Market]{\centering{}\includegraphics[scale=0.35]{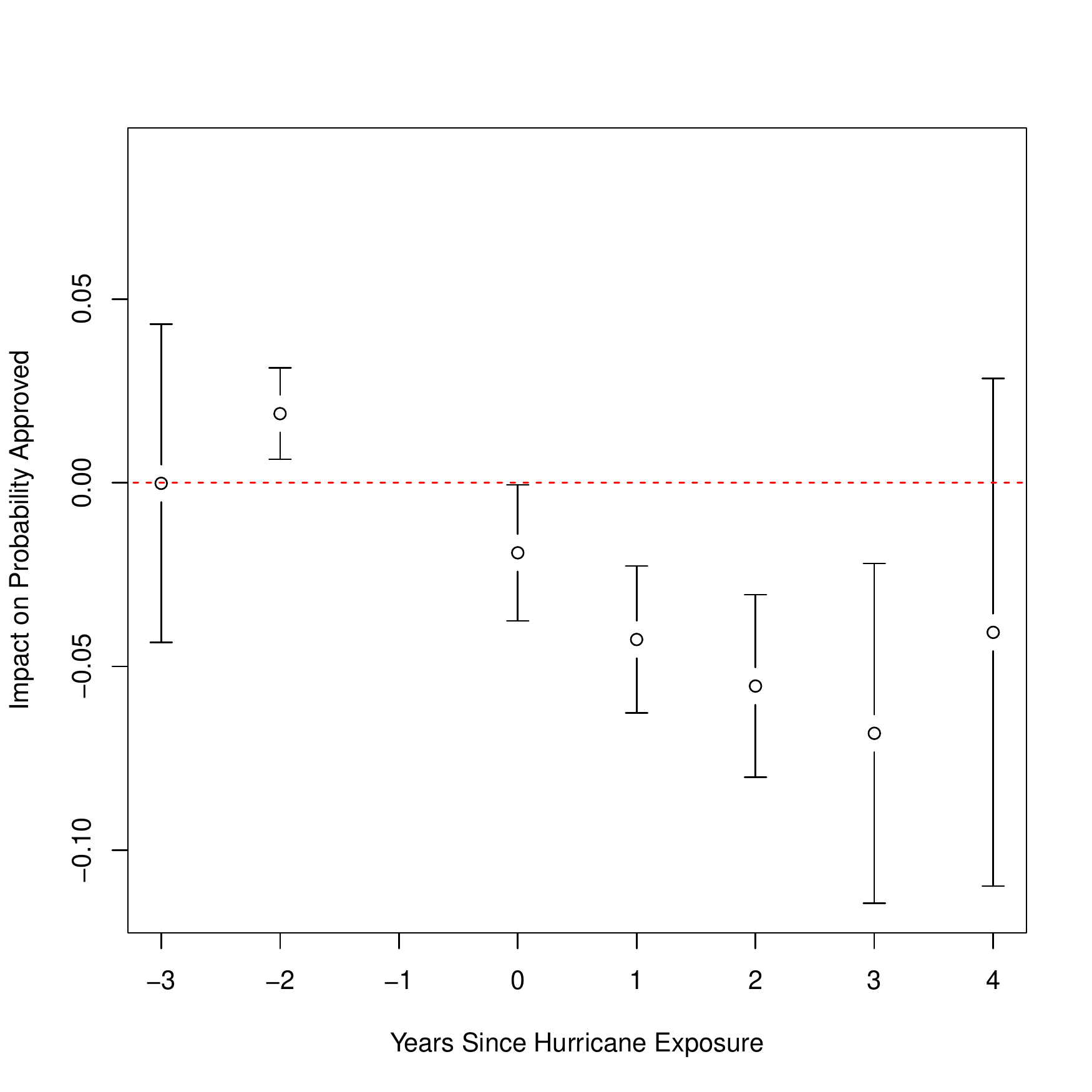} }

\bigskip{}

\subfloat[Origination Probabilities, Conforming Market]{\begin{centering}
\par\end{centering}
\centering{}\includegraphics[scale=0.35]{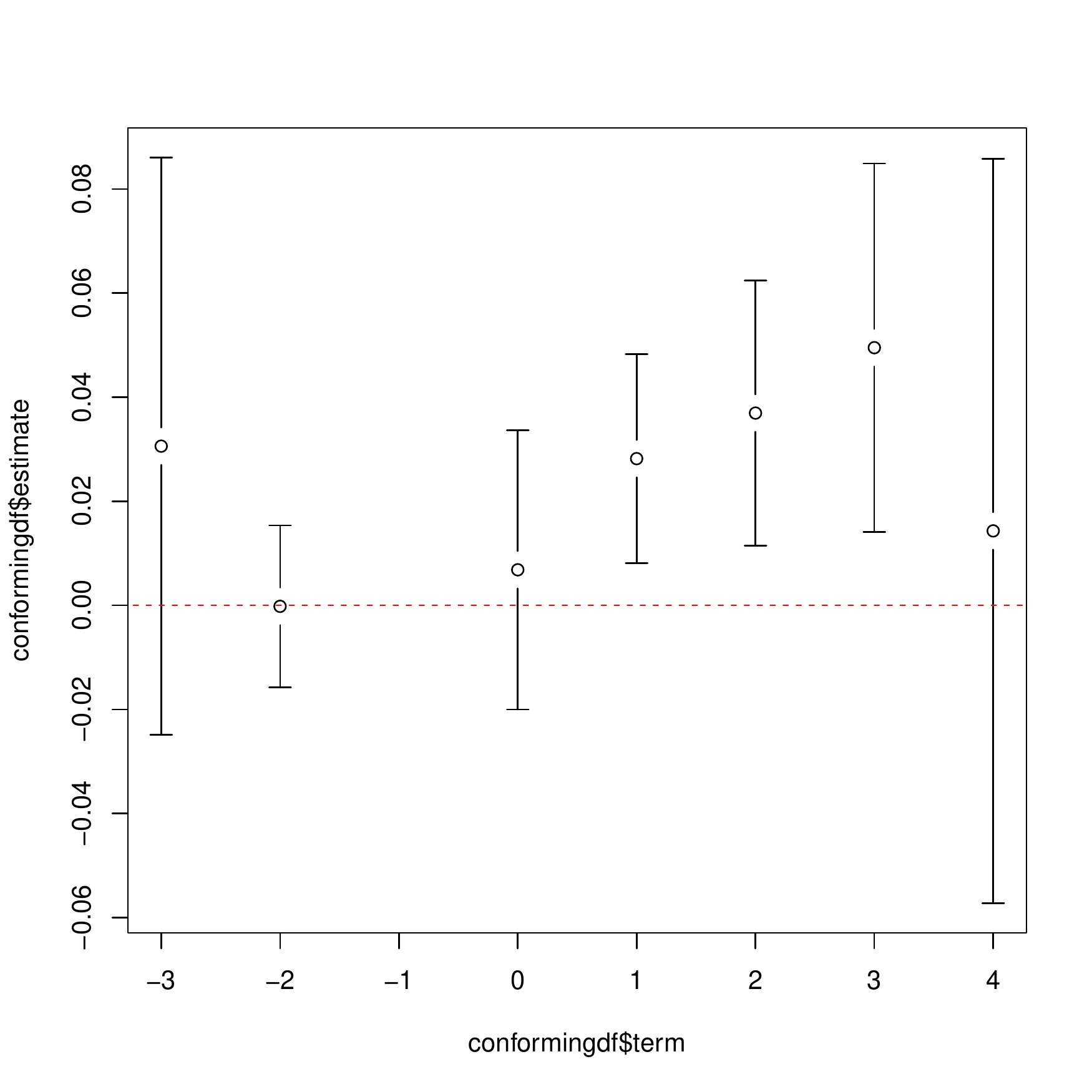}}~~~~~~~~~~~~\subfloat[Origination Probabilities, Overall Market]{\centering{}\includegraphics[scale=0.35]{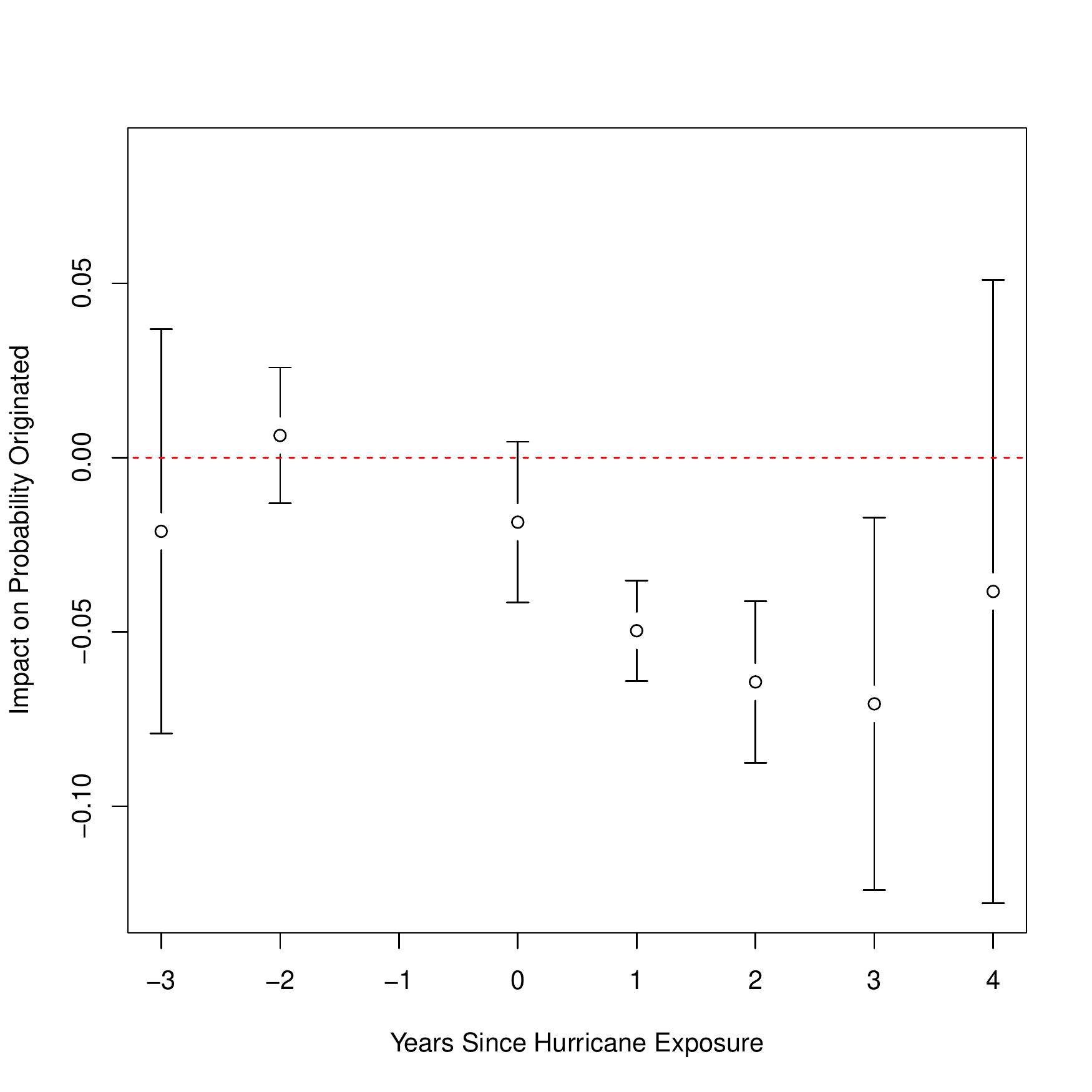}}\bigskip{}

\subfloat[P(Securitization$\vert$Origination), Conforming Market]{\centering{}\includegraphics[scale=0.35]{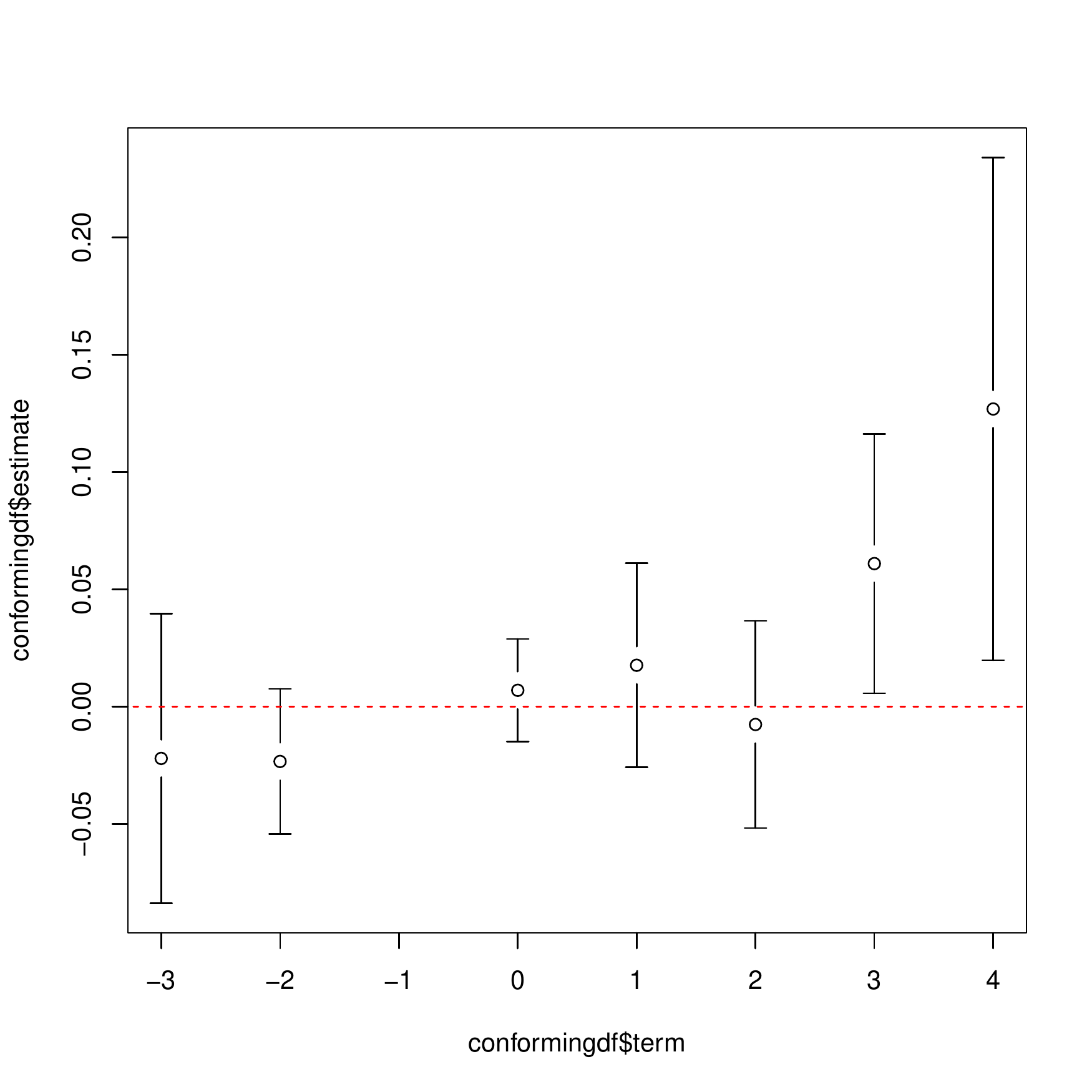}}~~~~~~~~~~~~\subfloat[P(Securitization$\vert$Origination), Overall Market]{\centering{}\includegraphics[scale=0.35]{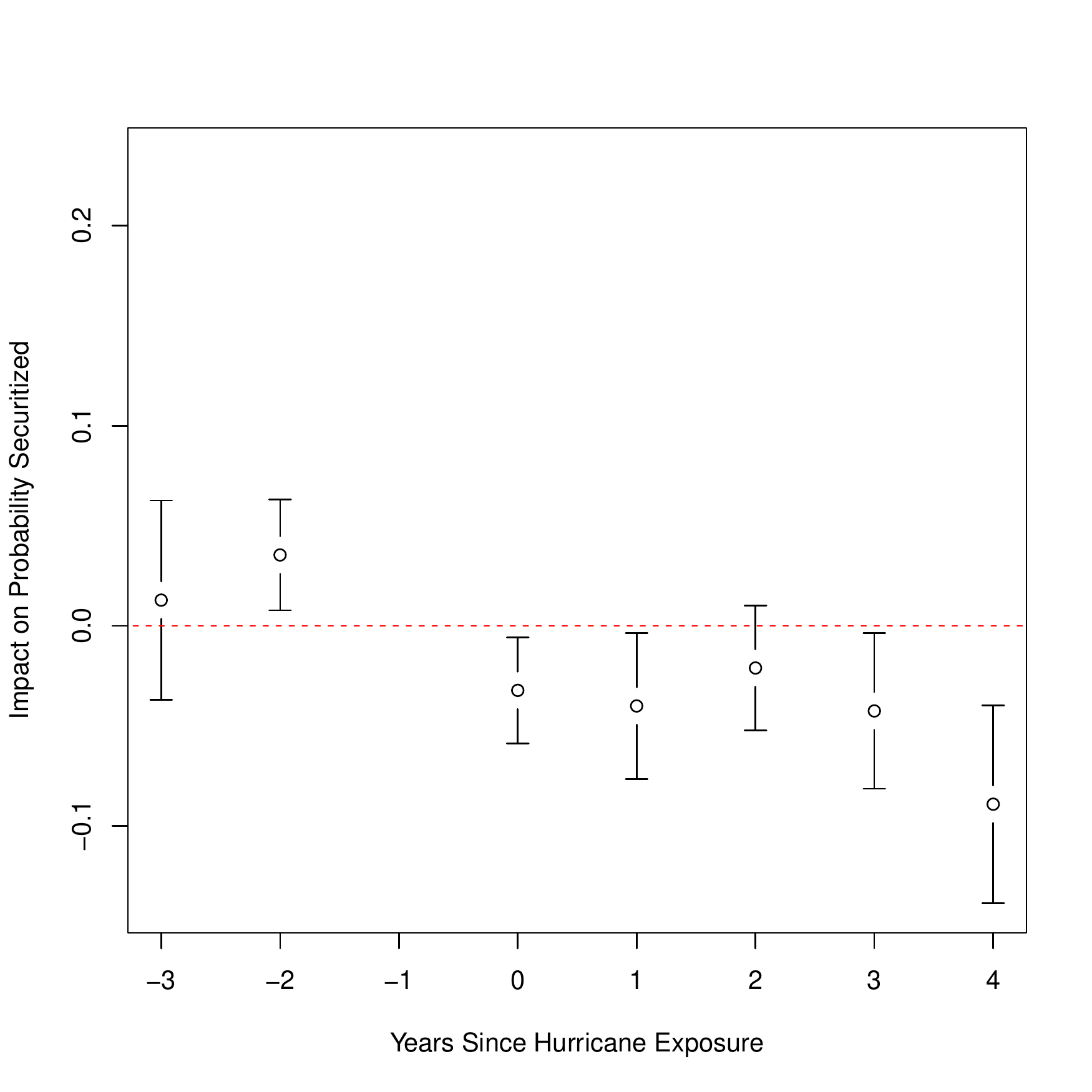}}

\end{center}\bigskip{}

\emph{Open source code available at \href{https://github.com/aouazad/Mortgage-Securitization-Natural-Disasters-Reply.git}{this link}.}
\end{figure}

\clearpage\pagebreak{}

\begin{figure}
\caption{Inspecting the Mechanism: Treatment Effect by Distance to the Conforming
Loan Limit}
\label{fig:inspecting_the_mechanism}

\emph{The graphs below inspect the mechanism driving the results presented
in Figure \ref{fig:results_replication_ouazad_kahn} using local polynomial
regressions. Each point in Subfigure (a), (b), (c), (d) is a separate
estimation of the impact of a billion dollar disaster on origination
rates. For each point, we weigh observations according to the distance
to the conforming loan limit.
\[
\text{Originated}_{it}=\sum_{t=-4}^{+4}\xi_{t}\cdot\text{Treated}_{j(i)}\times\textrm{Time}_{t=y-y_{0}(d)}+\textrm{Year}_{y(t,d)}+\text{Disaster}_{d}+\text{ZIP}_{j(i)}+\varepsilon_{it}
\]
}

\emph{The weights are the kernel $K((\Delta\log\text{Loan\,Amount}-\text{Distance\,to\,Conforming\,Loan\,Limit})/h)$,
where each of the 40 regressions uses a different }Distance to the
Conforming Loan Limit\emph{ $\in[-0.10,0.10]$. These regressions
display flexible results free of assumptions on rounding or on the
choice of the window. Figures suggest that (a)~the discontinuity
at the limit is key to the results, (b)~the discontinuity is sharp
in time $t=1,2,3$, and smoother in $t=4$. Tests of statistical significance
are performed on Table \ref{tab:reg_originated}.}

\bigskip{}

\begin{center}

\subfloat[Originated, time $t+1$]{

\includegraphics[scale=0.45]{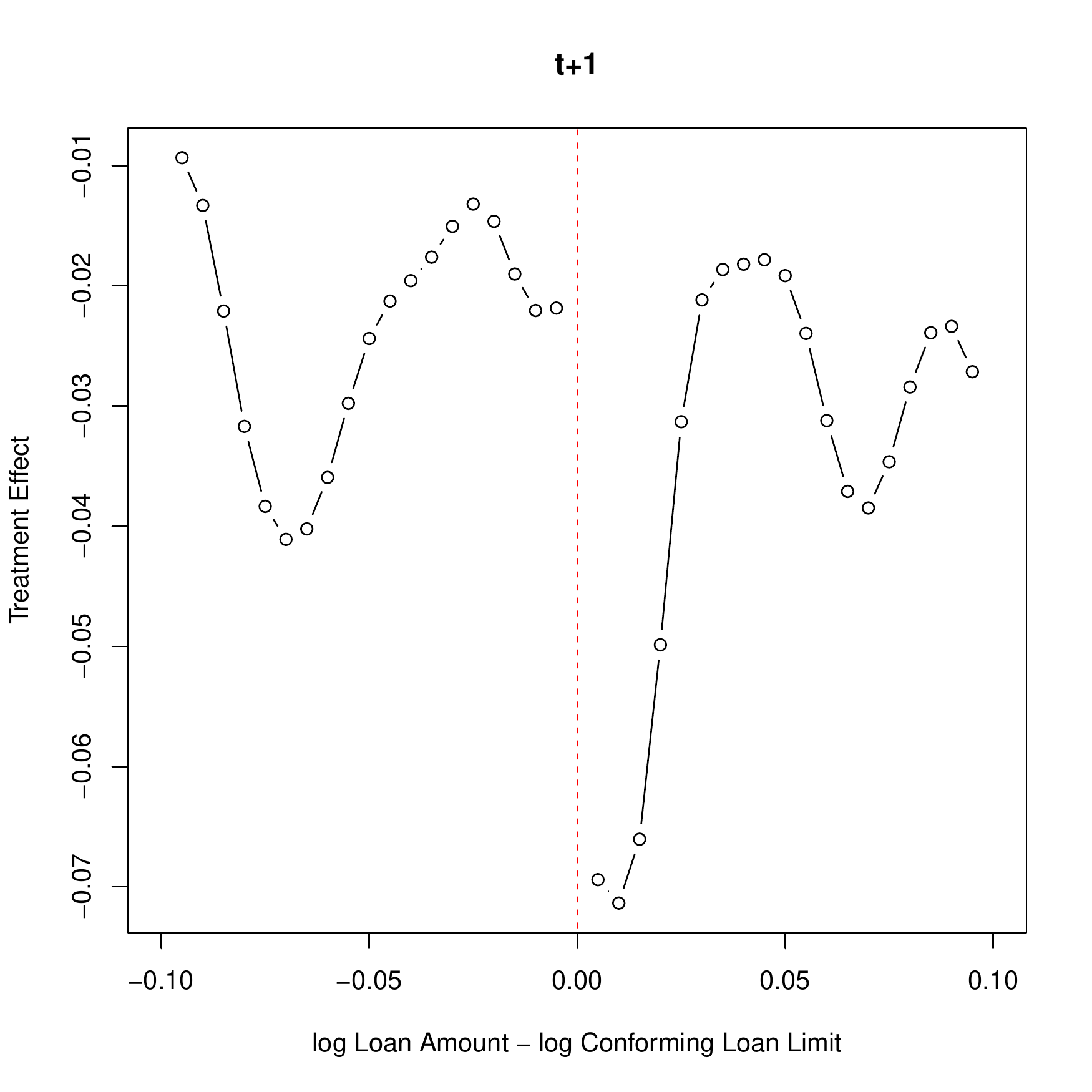}}\subfloat[Originated, time $t+2$]{

\includegraphics[scale=0.45]{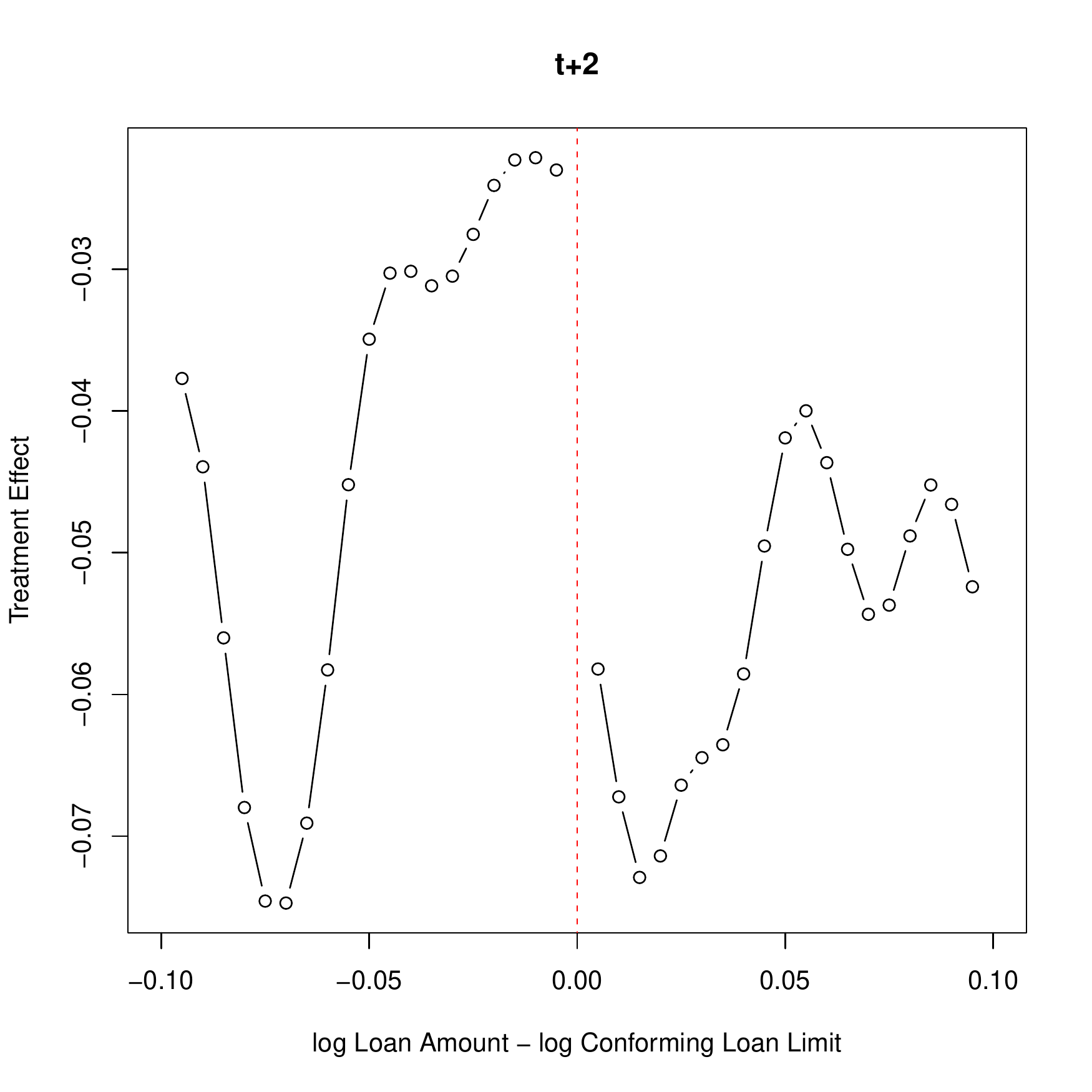}}

\subfloat[Originated, time $t+3$]{

\includegraphics[scale=0.45]{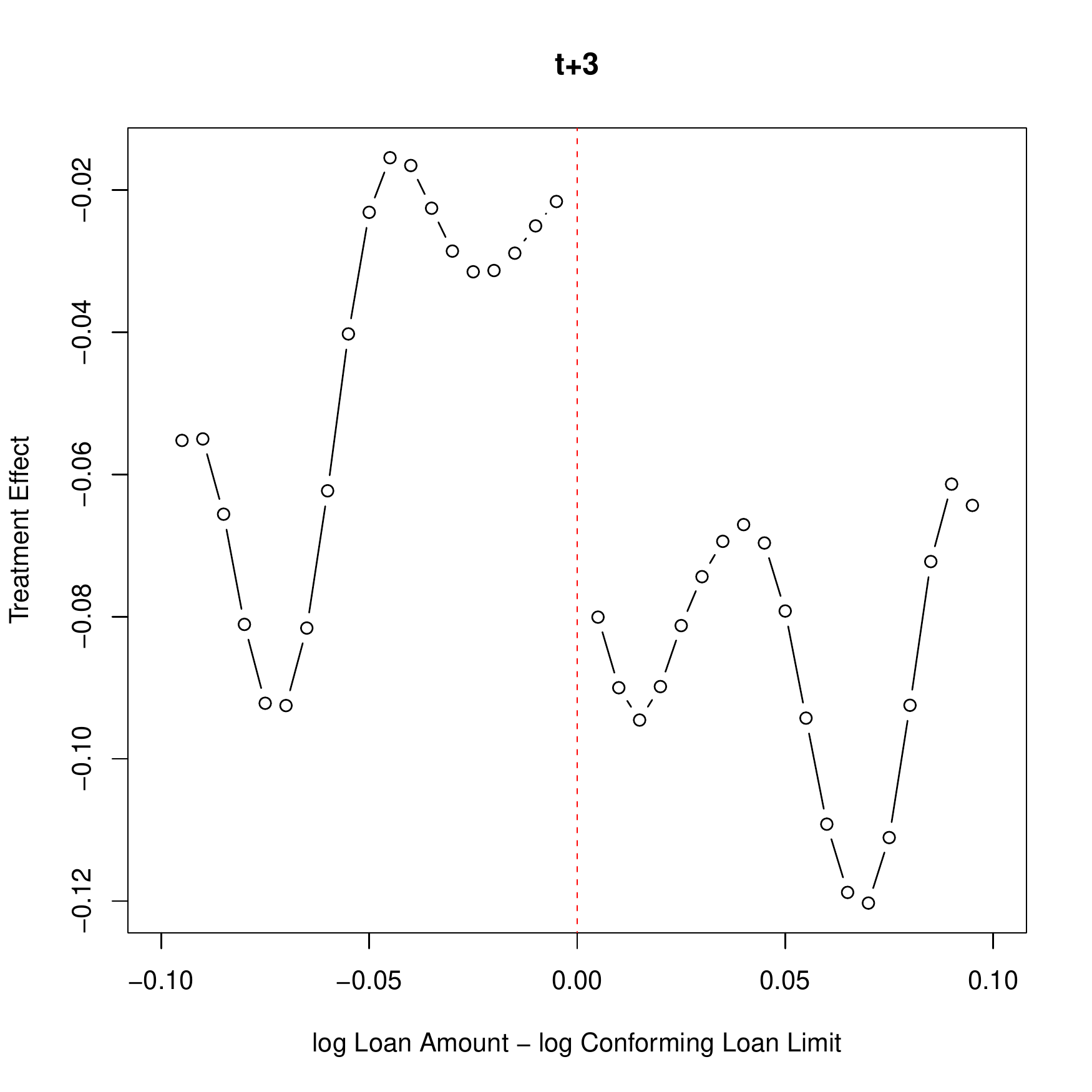}}\subfloat[Originated, time $t+4$]{

\includegraphics[scale=0.45]{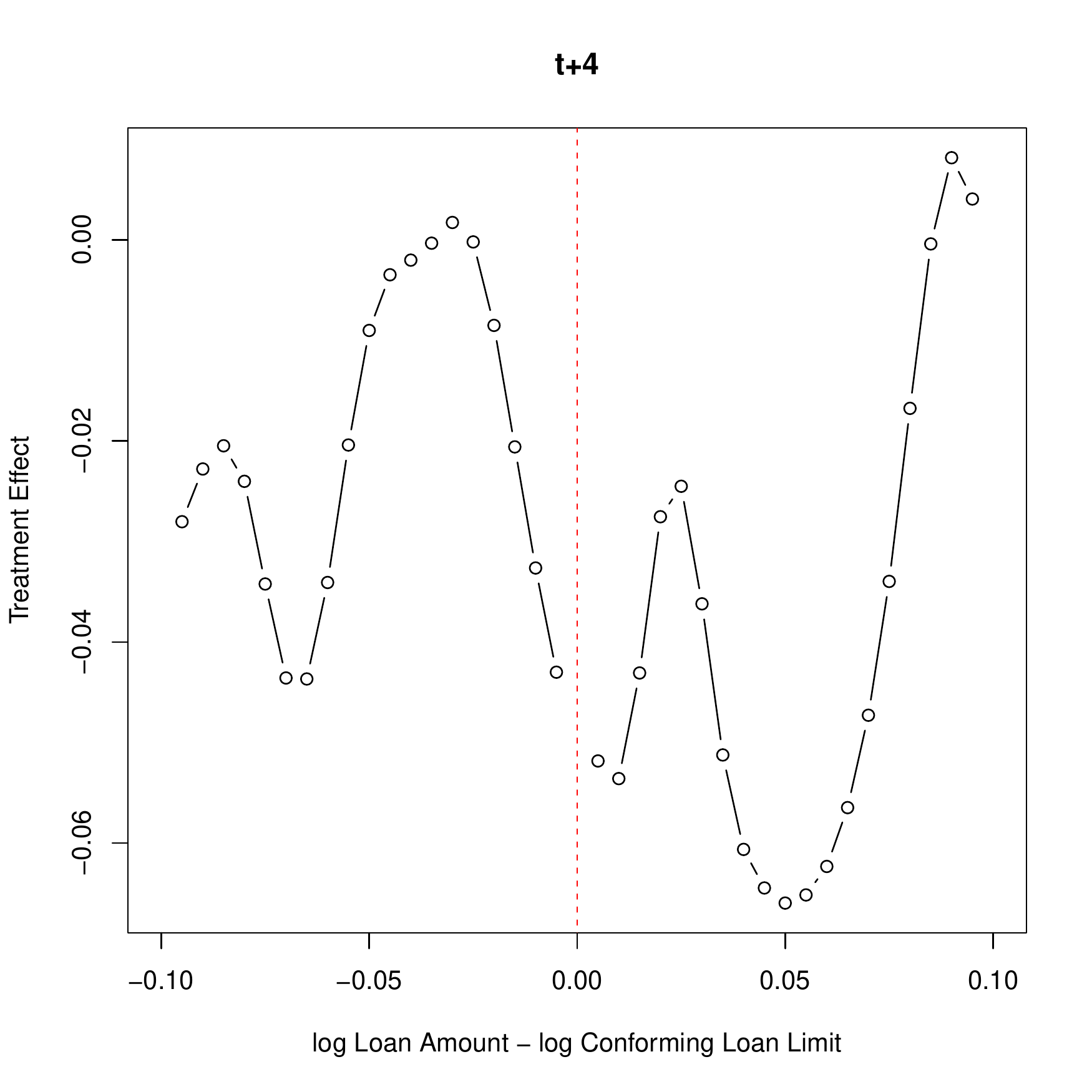}

}

\end{center}

\bigskip{}

\emph{Open source code available at \href{https://github.com/aouazad/Mortgage-Securitization-Natural-Disasters-Reply.git}{this link}.}
\end{figure}

\clearpage\pagebreak{}

\begin{figure}
\caption{Inspecting the Mechanism: Treatment Effect by Distance to the Conforming
Loan Limit}
\label{fig:inspecting_the_mechanism-1}

\emph{The graphs below inspect the mechanism driving the results presented
in Figure \ref{fig:results_replication_ouazad_kahn} using local polynomial
regressions. Each point in Subfigure (a), (b), (c), (d) is a separate
estimation of the impact of a billion dollar disaster on approval
rates. For each point, we weigh observations according to the distance
to the conforming loan limit.
\[
\text{Approval}_{it}=\sum_{t=-4}^{+4}\xi_{t}\cdot\text{Treated}_{j(i)}\times\textrm{Time}_{t=y-y_{0}(d)}+\textrm{Year}_{y(t,d)}+\text{Disaster}_{d}+\text{ZIP}_{j(i)}+\varepsilon_{it}
\]
}

\emph{The weights are the kernel $K((\Delta\log\text{Loan\,Amount}-\text{Distance\,to\,Conforming\,Loan\,Limit})/h)$,
where each of the 40 regressions uses a different }Distance to the
Conforming Loan Limit\emph{ $\in[-0.10,0.10]$ and the bandwidth $h=1\%$.
These regressions display flexible results free of assumptions on
rounding or on the choice of the window. Figures suggest that (a)~the
discontinuity at the limit is key to the results, (b)~the discontinuity
is sharp in time $t=1,2,3$, and smoother in $t=4$. Tests of statistical
significance are performed on Table \ref{tab:regression_approval}.}

\bigskip{}

\begin{center}

\subfloat[Approved, time $t+1$]{

\includegraphics[scale=0.45]{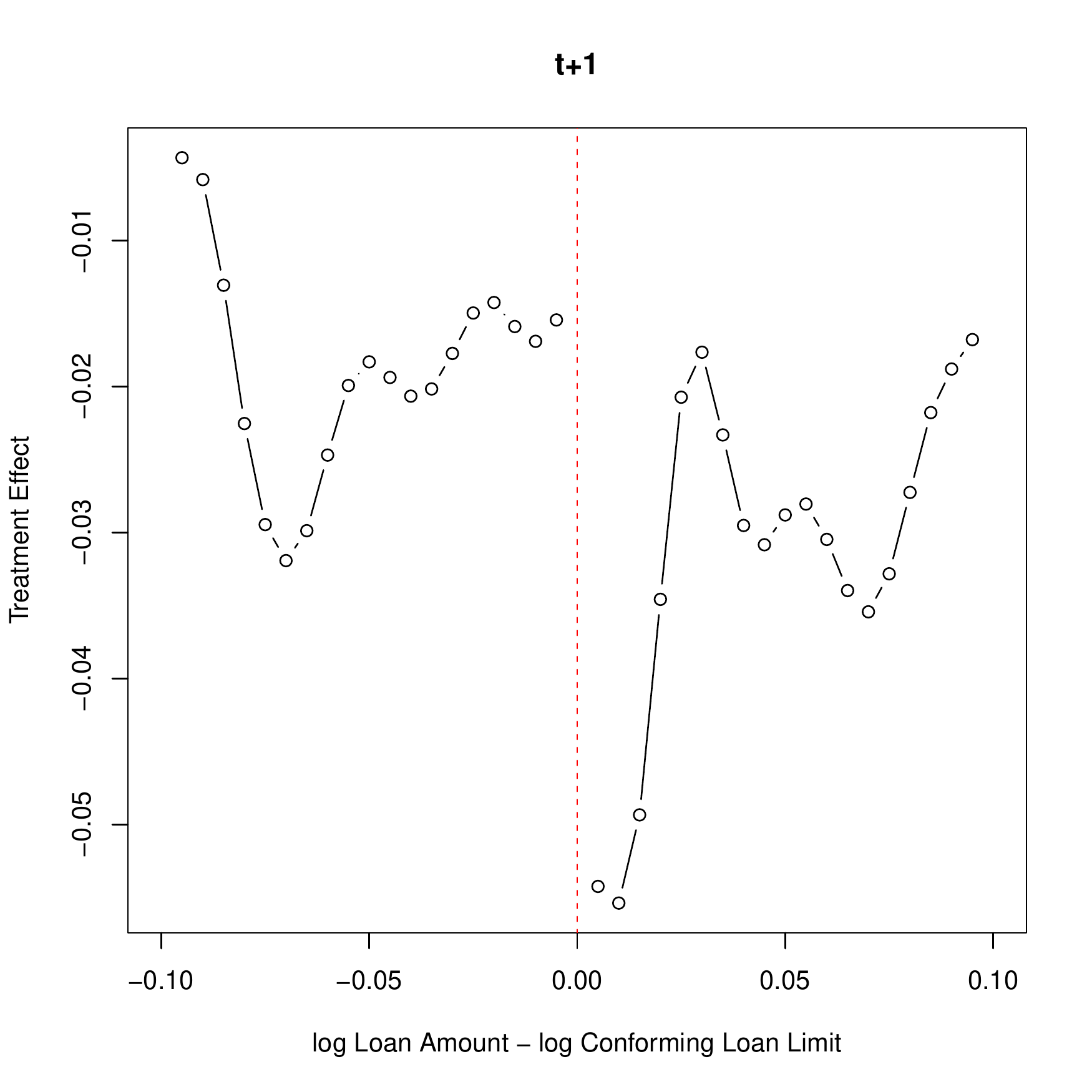}}\subfloat[Approved, time $t+2$]{

\includegraphics[scale=0.45]{figures/RD_pictures_originated_time2}}

\subfloat[Approved, time $t+3$]{

\includegraphics[scale=0.45]{figures/RD_pictures_originated_time3}}\subfloat[Approved, time $t+4$]{

\includegraphics[scale=0.45]{figures/RD_pictures_originated_time4}}

\end{center}

\bigskip{}

\emph{Open source code available at \href{https://github.com/aouazad/Mortgage-Securitization-Natural-Disasters-Reply.git}{this link}.}
\end{figure}
\clearpage\pagebreak{}
\begin{figure}
\caption{What Drives Results? Inspecting the Mechanism: Treatment Effect by
Distance to the Conforming Loan Limit}
\label{fig:inspecting_the_mechanism-2}

\emph{The graphs below inspect the mechanism driving the results presented
in Figure \ref{fig:results_replication_ouazad_kahn} using local polynomial
regressions. Each point in Subfigure (a), (b), (c), (d) is a separate
estimation of the impact of a billion dollar disaster on GSE securitization
rates for the sample of originated mortgages. For each point, we weigh
observations according to the distance to the conforming loan limit.
\[
\text{Securitized Conditional on Origination}_{it}=\sum_{t=-4}^{+4}\xi_{t}\cdot\text{Treated}_{j(i)}\times\textrm{Time}_{t=y-y_{0}(d)}+\textrm{Year}_{y(t,d)}+\text{Disaster}_{d}+\text{ZIP}_{j(i)}+\varepsilon_{it}
\]
}

\emph{The weights are the kernel $K((\log\text{Loan\,Amount}-\text{Distance\,to\,Conforming\,Loan\,Limit})/h)$,
where each of the 40 regressions uses a different }Distance to the
Conforming Loan Limit\emph{ $\in[-0.10,0.10]$. These regressions
display flexible results free of assumptions on rounding or on the
choice of the window. Figures suggest that (a)~the discontinuity
at the limit is key to the results, (b)~the discontinuity is sharp
and growing in every time period $t=1\dots4$. Tests of statistical
significance are performed on Table \ref{tab:reg_securization}.}

\bigskip{}

\begin{center}

\subfloat[Securitized Conditional on Originated, time $t+1$]{

\includegraphics[scale=0.45]{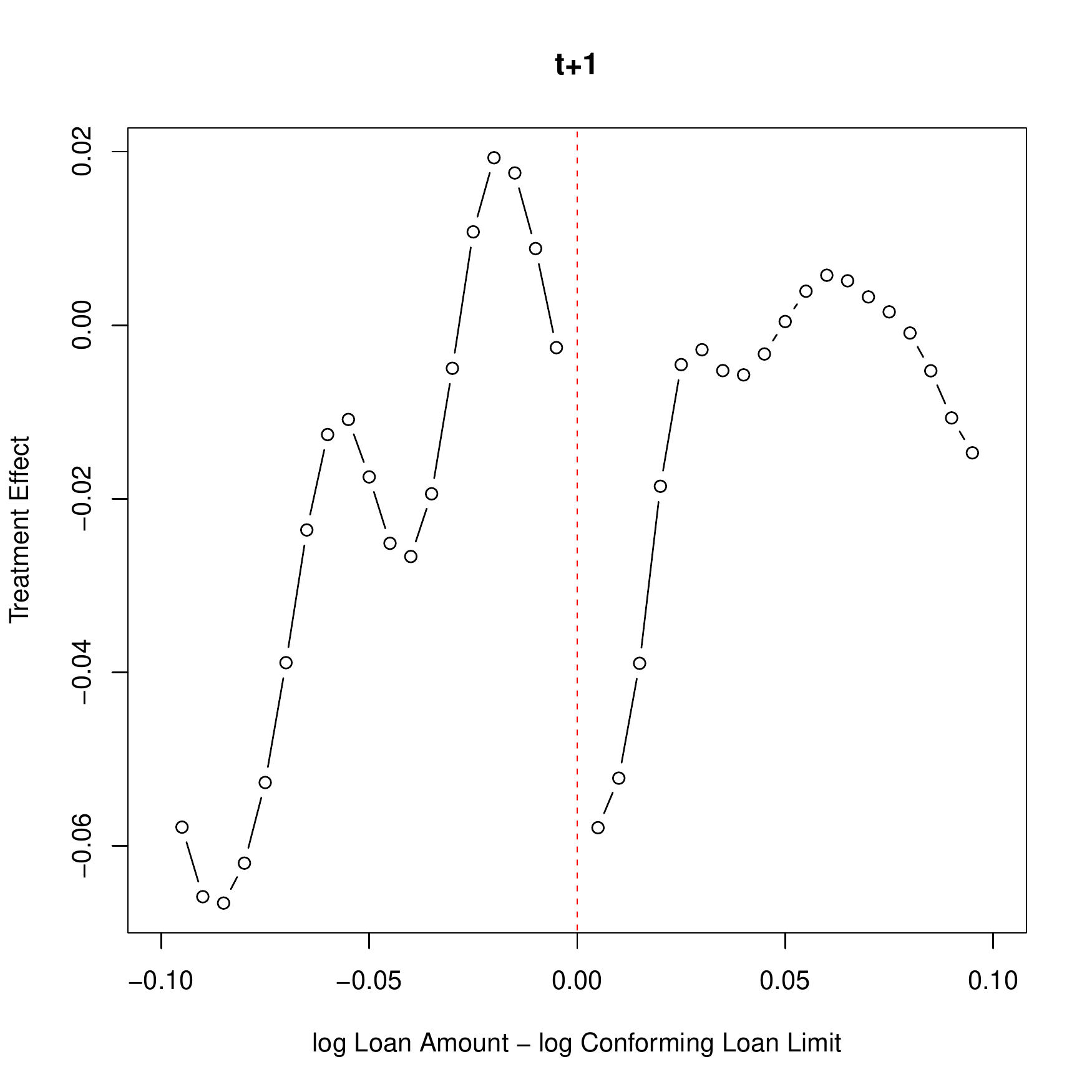}}\subfloat[Securitized Conditional on Originated, time $t+2$]{

\includegraphics[scale=0.45]{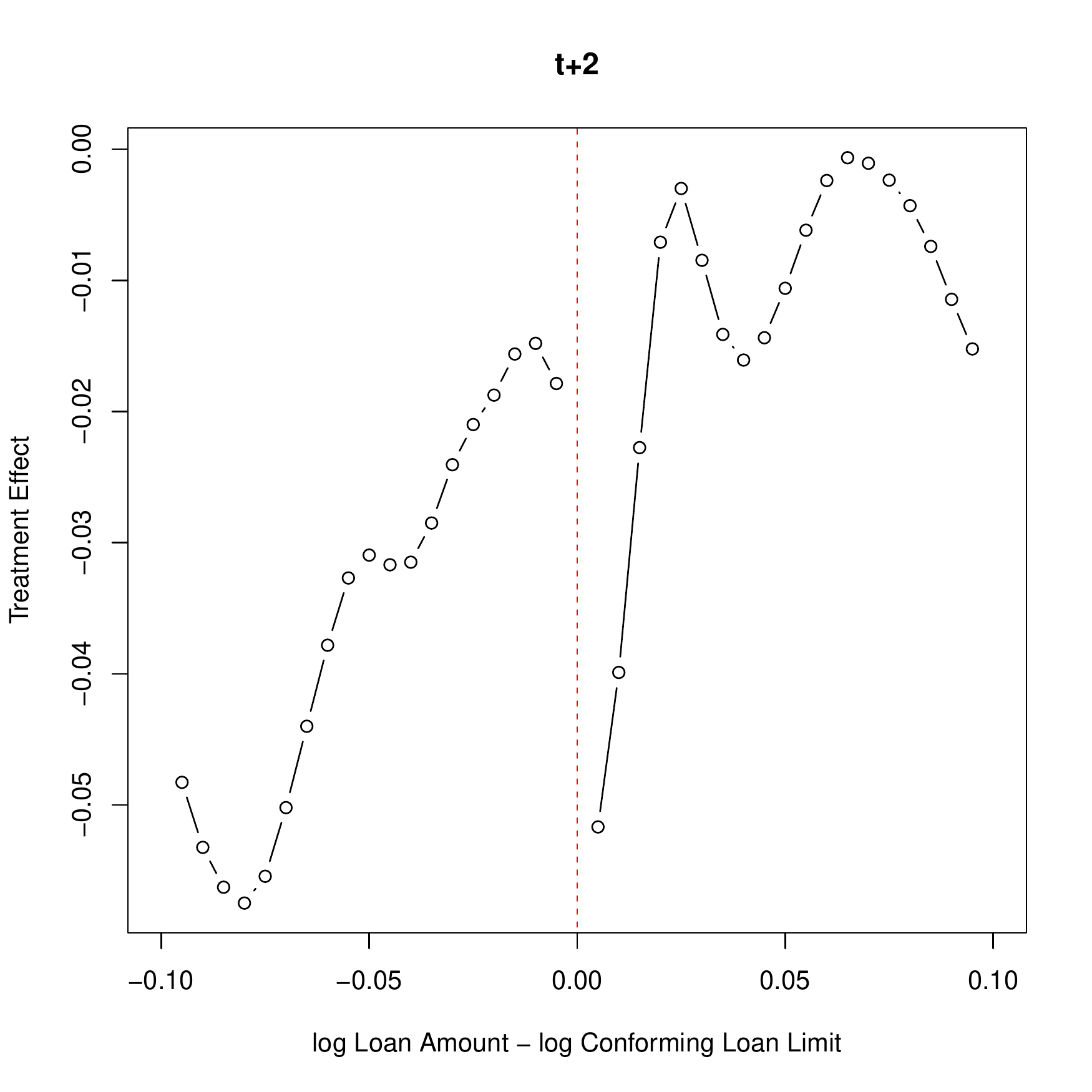}}

\subfloat[Securitized Conditional on Originated, time $t+3$]{

\includegraphics[scale=0.45]{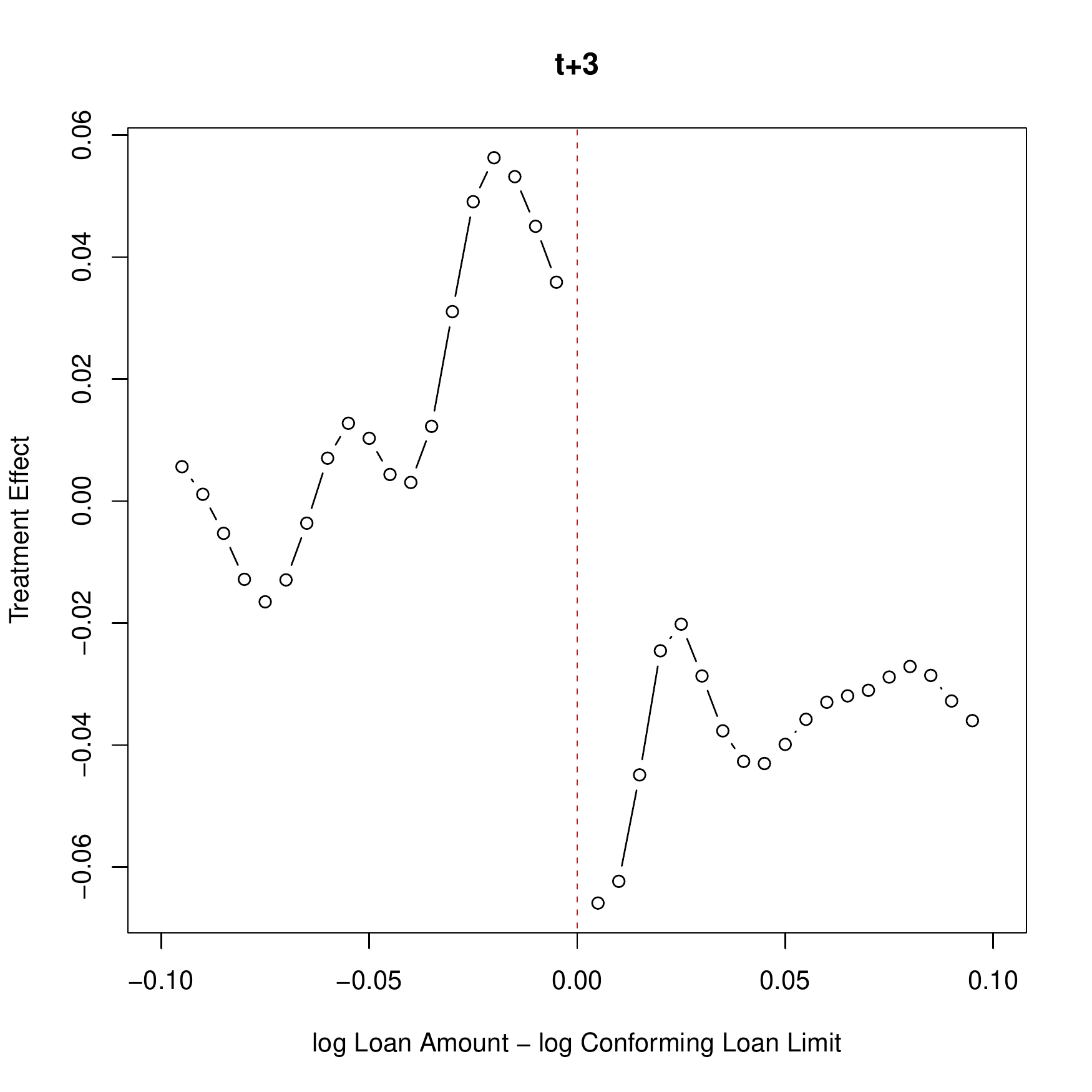}}\subfloat[Securitized Conditional on Originated, time $t+4$]{

\includegraphics[scale=0.45]{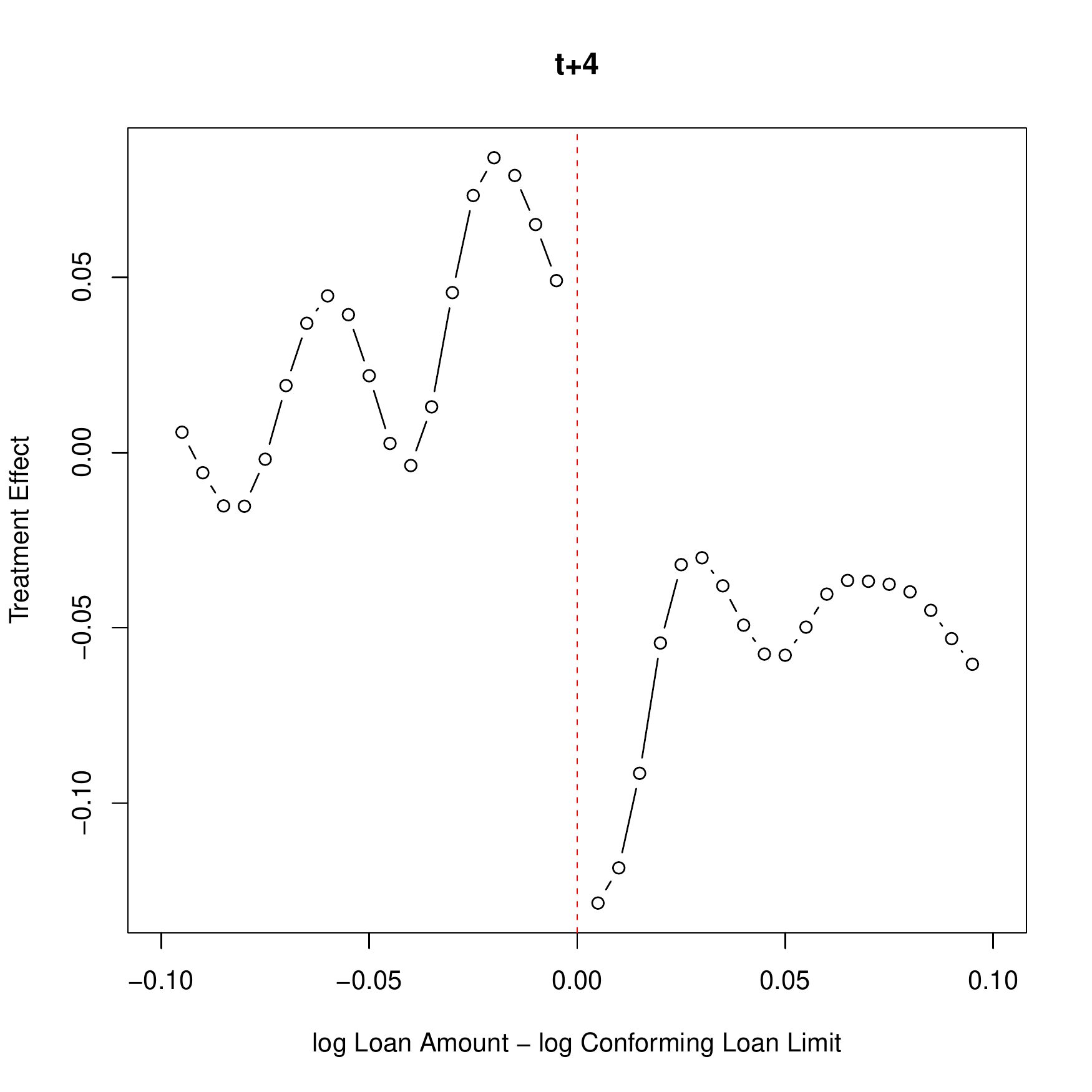}}

\end{center}

\bigskip{}

\emph{Open source code available at \href{https://github.com/aouazad/Mortgage-Securitization-Natural-Disasters-Reply.git}{this link}.}
\end{figure}

\clearpage\pagebreak{}

\begin{figure}
\caption{Would a Geographic Pricing of G-Fees Imply Redlining? First Street
Risk Factor and Zip-Level Demographics}
\label{fig:demographics_census}

\emph{These four figures display census demographics at the ZIP level
against the average ZIP-level First Street Flood Risk Score. Demographics
from the 2020 Census accessed through the National Historical Geographic
Information System of the University of Minnesota. Each code indicates
the corresponding Census table. Each plot is binned using 40 quantiles
of the risk score.}

\begin{center}

\subfloat[Household Income (AMR8E001)]{

\includegraphics[scale=0.45]{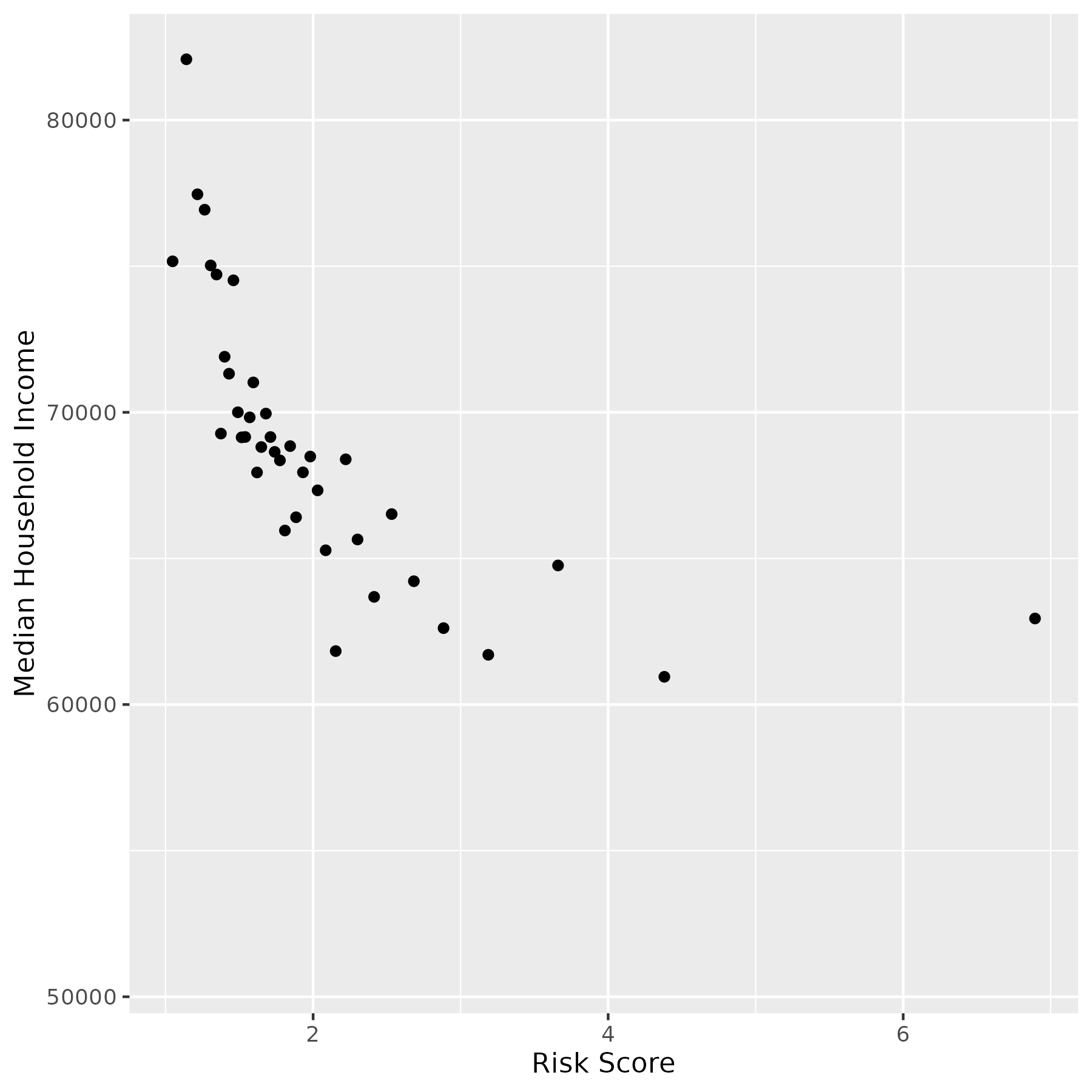}}~~~~~\subfloat[Share White (B02001)]{

\includegraphics[scale=0.45]{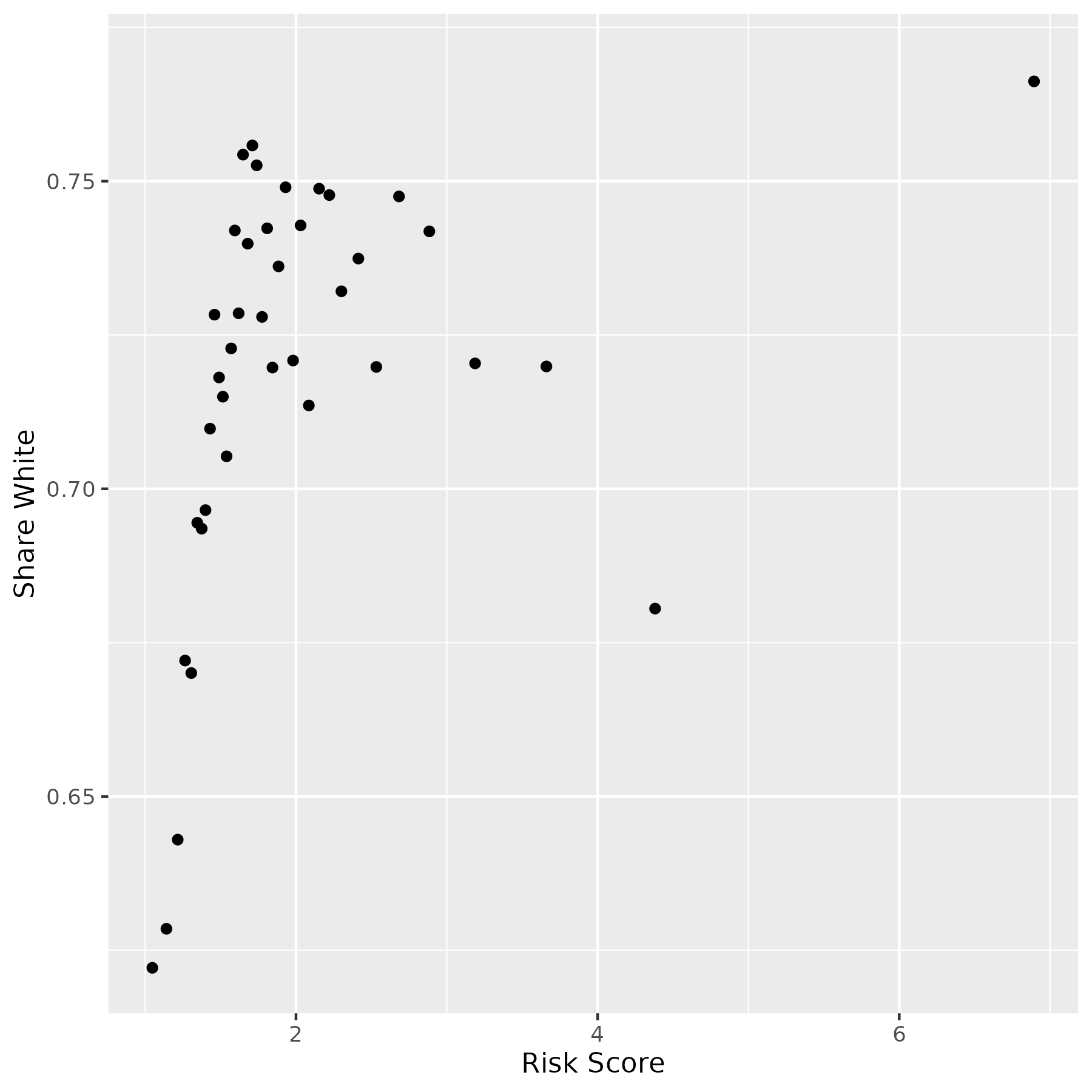}}

\subfloat[Share on Cash Public Assistance or Food Stamps/SNAP (B19058)]{

\includegraphics[scale=0.45]{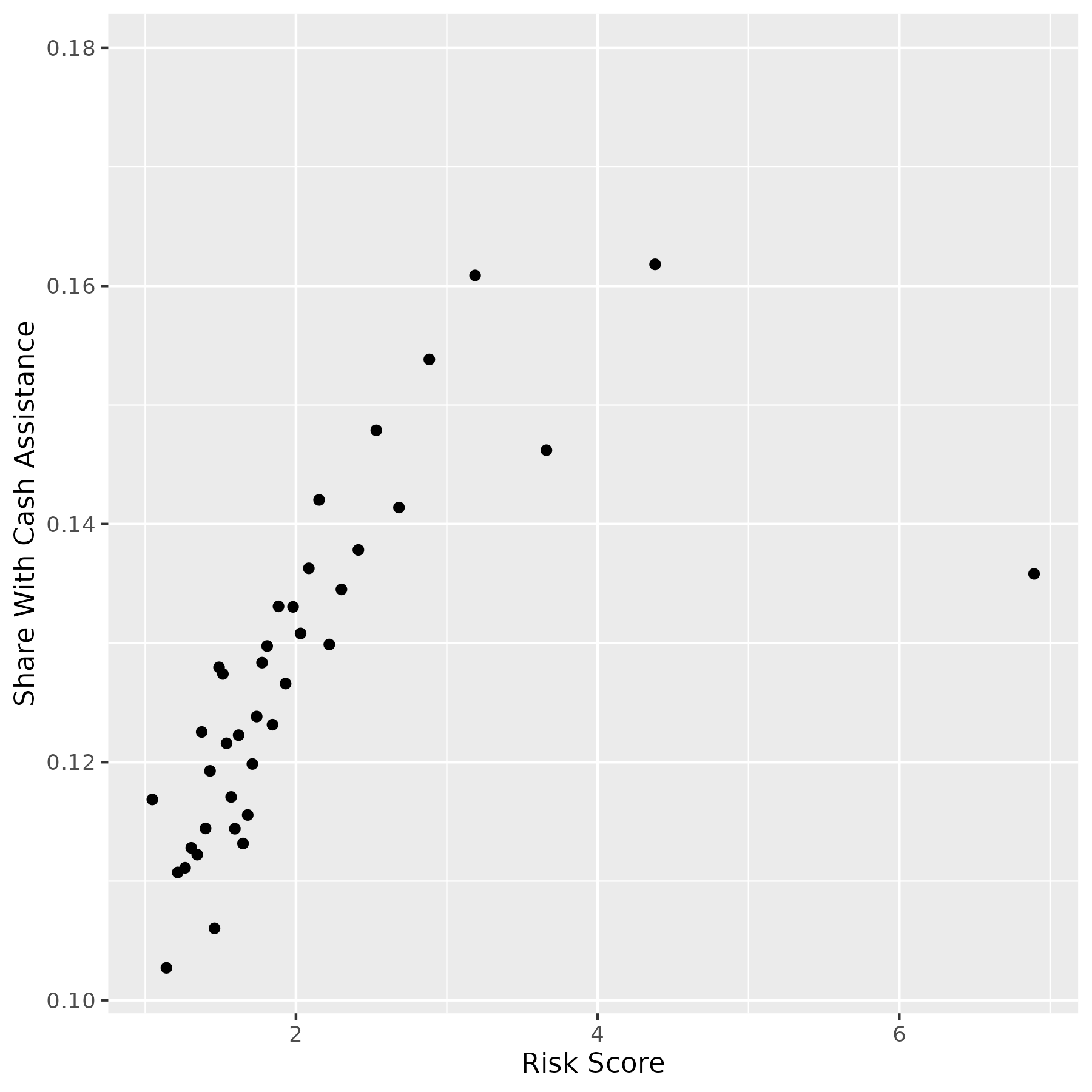}}~~~~~\subfloat[Rent as Share of Household Income (B25071)]{

\includegraphics[scale=0.45]{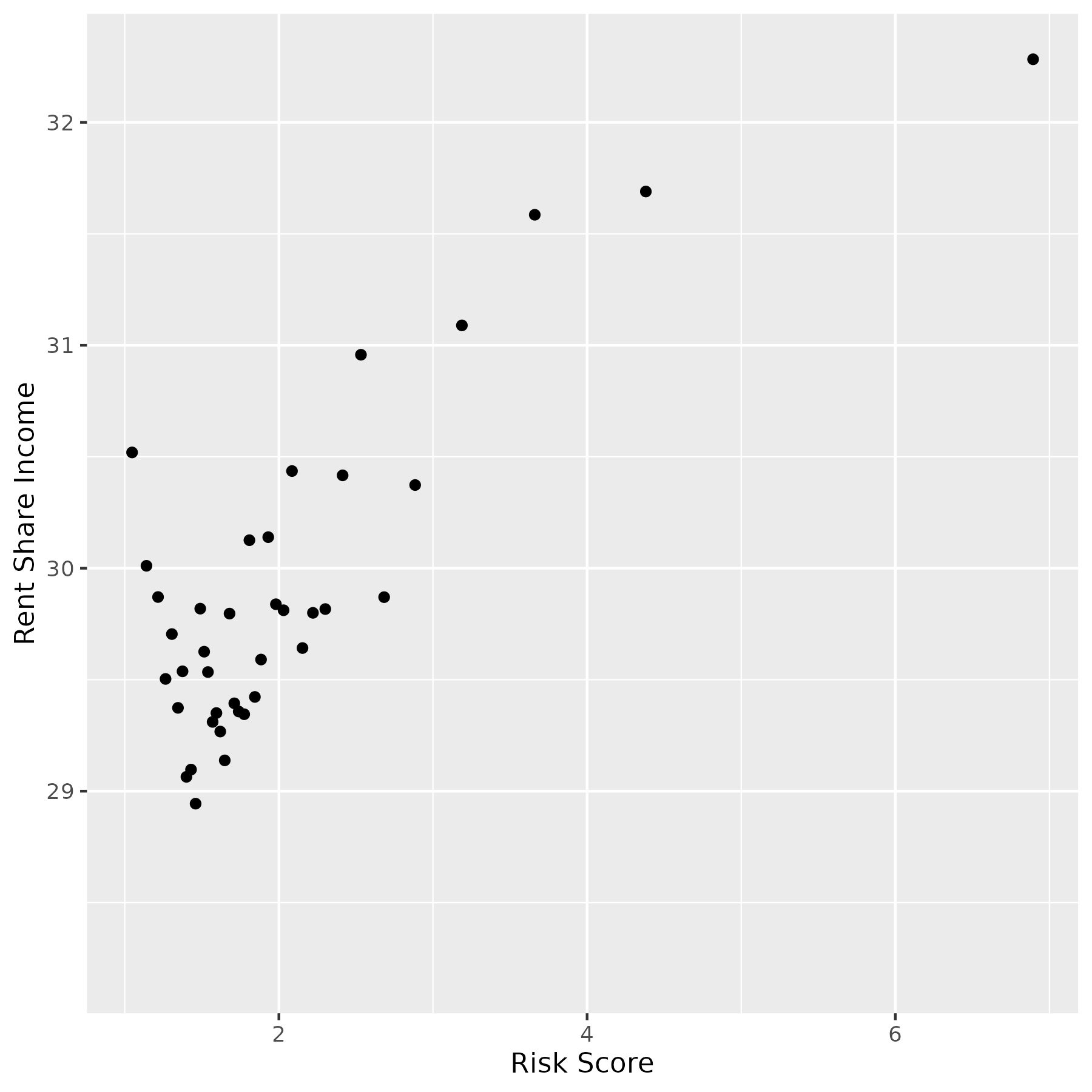}}

\end{center}
\end{figure}

\clearpage\pagebreak{}

\begin{table}

\caption{Regression Discontinuity Estimate of the Impact of Billion-Dollar
Disasters on Approval Probabilities}
\label{tab:regression_approval}

\begin{center}
\scriptsize
\input{tables/regression_result_approved_1_to_8_no_placebo.tex}
\end{center}
\end{table}

\clearpage\pagebreak{}
\begin{table}
\caption{Regression Discontinuity Estimate of the Impact of Billion-Dollar
Disasters on Origination Probabilities}
\label{tab:reg_originated}

\begin{center}
\scriptsize
\input{tables/regression_result_originated_1_to_8_no_placebo.tex}
\end{center}
\end{table}

\clearpage\pagebreak{}
\begin{table}
\caption{Regression Discontinuity Estimate of the Impact of Billion-Dollar
Disasters on Securitization Probabilities, Conditional on Origination}

\label{tab:reg_securization}

\begin{center}
\scriptsize
\input{tables/regression_result_securitized_1_to_8_no_placebo.tex}
\end{center}
\end{table}

\end{document}

%% file: tables/RD_miscoding_1.tex
\begin{tabular}{lcccc}
    \toprule
    Dependent Variable: & \multicolumn{4}{c}{1 if Incorrect Treatment Year}\\
    Model:                            & (1)            & (2)            & (3)            & (4)\\  
    \midrule
    \emph{Variables}\\
    (Intercept)                       & 0.0741$^{***}$ & 0.0507$^{*}$   & 0.0391         & 0.0298\\   
                                      & (0.0255)       & (0.0277)       & (0.0231)       & (0.0272)\\  
                                      \\ 
    Below Conforming Limit            & 0.0541$^{***}$ & 0.0847$^{***}$ & 0.0999$^{***}$ & 0.1121$^{***}$\\   
                                      & (0.0184)       & (0.0283)       & (0.0268)       & (0.0322)\\   
\\
    $\Delta \log(\textrm{Loan Amount})$    &                & 7.980$^{**}$   & 9.818$^{***}$  & 11.29$^{***}$\\   
                                      &                & (3.373)        & (3.229)        & (3.866)\\
                                      \\   
    $\Delta \log(\textrm{Loan Amount})^2$    &                &                & 4.686$^{**}$   & 5.353$^{**}$\\   
                                      &                &                & (1.796)        & (1.956)\\
                                      \\   
    $\Delta \log(\textrm{Loan Amount})^3$    &                &                &                & -3.504\\   
                                      &                &                &                & (2.195)\\   
                                      \\
    \midrule
    \emph{Fit statistics}\\
    Observations                      & 173,870        & 173,870        & 173,870        & 173,870\\  
    R$^2$                             & 0.00518        & 0.00710        & 0.00812        & 0.00870\\  
    Adjusted R$^2$                    & 0.00517        & 0.00709        & 0.00810        & 0.00867\\  
    \midrule \midrule
    \multicolumn{5}{l}{\emph{Clustered (ZCTA5 \& as\_of\_year) standard-errors in parentheses}}\\
    \multicolumn{5}{l}{\emph{Signif. Codes: ***: 0.01, **: 0.05, *: 0.1}}\\
 \end{tabular}

%% file: tables/regression_result_approved_1_to_8_no_placebo.tex
\begin{tabular*}{0.8\textwidth}{@{\extracolsep{\fill}}lcccc}
   \toprule
   Dependent Variable: & \multicolumn{4}{c}{Approved = 0,1}\\
   Bandwidth:                                          & $\pm1\%$            & $\pm2\%$            & $\pm3\%$            & $\pm4\%$  \\  
   \midrule
    Treated $\times$ Time 0                          & -0.0195$^{**}$  & -0.0213$^{**}$  & -0.0209$^{*}$   & -0.0198$^{*}$\\   
                                                     & (0.0084)        & (0.0098)        & (0.0102)        & (0.0101)\\   
    Treated $\times$ Time +1                          & -0.0507$^{***}$ & -0.0486$^{***}$ & -0.0455$^{***}$ & -0.0432$^{***}$\\   
                                                     & (0.0111)        & (0.0105)        & (0.0109)        & (0.0111)\\   
    Treated $\times$ Time +2                          & -0.0598$^{**}$  & -0.0596$^{***}$ & -0.0575$^{***}$ & -0.0548$^{***}$\\   
                                                     & (0.0220)        & (0.0156)        & (0.0141)        & (0.0136)\\   
    Treated $\times$ Time +3                          & -0.0755$^{**}$  & -0.0719$^{**}$  & -0.0717$^{***}$ & -0.0731$^{***}$\\   
                                                     & (0.0300)        & (0.0245)        & (0.0232)        & (0.0225)\\   
    Treated $\times$ Time +4                          & -0.0565         & -0.0478         & -0.0457         & -0.0451\\   
                                                     & (0.0461)        & (0.0383)        & (0.0337)        & (0.0303)\\   
   Treated $\times$ Time 0 $\times$ Below Limit      & -0.0037         & 0.0011          & 0.0027          & 0.0028\\   
                                                     & (0.0141)        & (0.0115)        & (0.0101)        & (0.0096)\\   
   Treated $\times$ Time 1 $\times$ Below Limit      & 0.0383$^{**}$   & 0.0348$^{***}$  & 0.0300$^{***}$  & 0.0268$^{***}$\\   
                                                     & (0.0149)        & (0.0080)        & (0.0061)        & (0.0067)\\   
   Treated $\times$ Time 2 $\times$ Below Limit      & 0.0526$^{*}$    & 0.0501$^{***}$  & 0.0425$^{***}$  & 0.0360$^{***}$\\   
                                                     & (0.0275)        & (0.0162)        & (0.0117)        & (0.0100)\\   
   Treated $\times$ Time 3 $\times$ Below Limit      & 0.0648$^{**}$   & 0.0609$^{***}$  & 0.0583$^{***}$  & 0.0570$^{***}$\\   
                                                     & (0.0270)        & (0.0188)        & (0.0160)        & (0.0146)\\   
   Treated $\times$ Time 4 $\times$ Below Limit      & 0.0208          & 0.0264          & 0.0284          & 0.0291\\   
                                                     & (0.0430)        & (0.0351)        & (0.0297)        & (0.0259)\\   
   \midrule
   Year $\times$ Below Limit                         & Yes             & Yes             & Yes             & Yes\\  
   Zip $\times$ Below Limit                          & Yes             & Yes             & Yes             & Yes\\  
   Disaster $\times$ Below Limit                     & Yes             & Yes             & Yes             & Yes\\  
   \midrule
   Observations                                        & 2,572,574       & 2,572,574       & 2,572,574       & 2,572,574\\  
   R$^2$                                               & 0.07088         & 0.07076         & 0.06983         & 0.06887\\  
   \midrule
   Dependent Variable: & \multicolumn{4}{c}{Approved = 0,1}\\
   Bandwidth:                                         & $\pm5\%$             & $\pm10\%$           & $\pm15\%$            & $\pm20\%$   \\  
   \midrule 
    Treated $\times$ Time 0                            & -0.0182$^{*}$   & -0.0127         & -0.0115         & -0.0111\\   
                                                      & (0.0097)        & (0.0075)        & (0.0067)        & (0.0064)\\   
    Treated $\times$ Time +1                            & -0.0408$^{***}$ & -0.0316$^{***}$ & -0.0287$^{***}$ & -0.0277$^{***}$\\   
                                                       & (0.0107)        & (0.0086)        & (0.0080)        & (0.0078)\\   
    Treated $\times$ Time +2                            & -0.0524$^{***}$ & -0.0450$^{***}$ & -0.0428$^{***}$ & -0.0420$^{***}$\\   
                                                       & (0.0131)        & (0.0127)        & (0.0133)        & (0.0135)\\   
    Treated $\times$ Time +3                            & -0.0737$^{***}$ & -0.0740$^{***}$ & -0.0743$^{***}$ & -0.0745$^{***}$\\   
                                                       & (0.0218)        & (0.0215)        & (0.0218)        & (0.0220)\\   
    Treated $\times$ Time +4                            & -0.0442         & -0.0396         & -0.0384         & -0.0381\\   
                                                       & (0.0281)        & (0.0252)        & (0.0249)        & (0.0249)\\   
   Treated $\times$ Time 0 $\times$ Below Limit      & 0.0022          & -0.0019         & -0.0036         & -0.0042\\   
                                                       & (0.0089)        & (0.0062)        & (0.0054)        & (0.0051)\\   
   Treated $\times$ Time 1 $\times$ Below Limit      & 0.0242$^{***}$  & 0.0140$^{**}$   & 0.0103$^{*}$    & 0.0088\\   
                                                       & (0.0066)        & (0.0056)        & (0.0055)        & (0.0056)\\   
   Treated $\times$ Time 2 $\times$ Below Limit      & 0.0310$^{***}$  & 0.0176$^{***}$  & 0.0134$^{***}$  & 0.0118$^{**}$\\   
                                                       & (0.0087)        & (0.0047)        & (0.0045)        & (0.0048)\\   
   Treated $\times$ Time 3 $\times$ Below Limit      & 0.0553$^{***}$  & 0.0465$^{***}$  & 0.0428$^{***}$  & 0.0413$^{***}$\\   
                                                       & (0.0136)        & (0.0105)        & (0.0094)        & (0.0090)\\   
   Treated $\times$ Time 4 $\times$ Below Limit      & 0.0289          & 0.0256          & 0.0241          & 0.0236\\   
                                                       & (0.0234)        & (0.0194)        & (0.0186)        & (0.0183)\\   
   \midrule
   Year $\times$ Below Limit                           & Yes             & Yes             & Yes             & Yes\\  
   Zip $\times$ Below Limit                            & Yes             & Yes             & Yes             & Yes\\  
   Disaster $\times$ Below Limit                       & Yes             & Yes             & Yes             & Yes\\  
   \midrule
   Observations                                        & 2,572,574       & 2,572,574       & 2,572,574       & 2,572,574\\  
   R$^2$                                               & 0.06790         & 0.06482         & 0.06413         & 0.06398\\  
   \bottomrule
   \multicolumn{5}{l}{\emph{Clustered (ZCTA5CE10 \& year) standard-errors in parentheses}}\\
   \multicolumn{5}{l}{\emph{Signif. Codes: ***: 0.01, **: 0.05, *: 0.1}}\\
\end{tabular*}

%% file: tables/regression_result_originated_1_to_8_no_placebo.tex
\begin{tabular*}{0.8\textwidth}{@{\extracolsep{\fill}}lcccc}
   \toprule
   Dependent Variable: & \multicolumn{4}{c}{Originated = 0,1}\\
   Bandwidth:                                          & $\pm1\%$            & $\pm2\%$            & $\pm3\%$            & $\pm4\%$  \\  
   \midrule
    Treated $\times$ Time 0                         & -0.0153         & -0.0204         & -0.0211         & -0.0202\\   
                                                    & (0.0146)        & (0.0136)        & (0.0128)        & (0.0124)\\   
    Treated $\times$ Time +1                        & -0.0655$^{***}$ & -0.0627$^{***}$ & -0.0568$^{***}$ & -0.0527$^{***}$\\   
                                                    & (0.0155)        & (0.0114)        & (0.0094)        & (0.0091)\\   
    Treated $\times$ Time +2                        & -0.0491         & -0.0649$^{***}$ & -0.0665$^{***}$ & -0.0653$^{***}$\\   
                                                    & (0.0336)        & (0.0176)        & (0.0135)        & (0.0131)\\   
    Treated $\times$ Time +3                        & -0.0700$^{**}$  & -0.0848$^{**}$  & -0.0869$^{**}$  & -0.0891$^{***}$\\   
                                                    & (0.0328)        & (0.0325)        & (0.0311)        & (0.0301)\\   
    Treated $\times$ Time +4                        & -0.0431         & -0.0455         & -0.0477         & -0.0488\\   
                                                    & (0.0532)        & (0.0491)        & (0.0436)        & (0.0385)\\
   Treated $\times$ Time 0 $\times$ Below Limit     & -0.0135         & -0.0021         & 0.0027          & 0.0039\\   
                                                    & (0.0258)        & (0.0215)        & (0.0177)        & (0.0155)\\   
   Treated $\times$ Time 1 $\times$ Below Limit     & 0.0457          & 0.0427$^{**}$   & 0.0356$^{**}$   & 0.0304$^{**}$\\   
                                                    & (0.0261)        & (0.0177)        & (0.0125)        & (0.0106)\\   
   Treated $\times$ Time 2 $\times$ Below Limit     & 0.0246          & 0.0408          & 0.0393$^{**}$   & 0.0349$^{**}$\\   
                                                    & (0.0434)        & (0.0244)        & (0.0171)        & (0.0143)\\   
   Treated $\times$ Time 3 $\times$ Below Limit     & 0.0501          & 0.0607$^{**}$   & 0.0608$^{**}$   & 0.0604$^{***}$\\   
                                                    & (0.0320)        & (0.0262)        & (0.0218)        & (0.0194)\\   
   Treated $\times$ Time 4 $\times$ Below Limit     & -0.0083         & 0.0060          & 0.0140          & 0.0177\\   
                                                    & (0.0508)        & (0.0432)        & (0.0355)        & (0.0290)\\   
   \midrule
   Year $\times$ Below Limit                                & Yes             & Yes             & Yes             & Yes\\  
   Zip $\times$ Below Limit                                 & Yes             & Yes             & Yes             & Yes\\  
   Disaster $\times$ Below Limit                            & Yes             & Yes             & Yes             & Yes\\  
   \midrule
   Observations                                        & 2,572,574       & 2,572,574       & 2,572,574       & 2,572,574\\  
   R$^2$                                               & 0.06821         & 0.06919         & 0.06900         & 0.06849\\  
   \midrule
   Dependent Variable: & \multicolumn{4}{c}{Originated = 0,1}\\
   Bandwidth:                                         & $\pm5\%$             & $\pm10\%$           & $\pm15\%$            & $\pm20\%$   \\  
   \midrule
   \emph{Variables}\\
    Treated $\times$ Time 0                         & -0.0186         & -0.0114         & -0.0089         & -0.0079\\   
                                                    & (0.0120)        & (0.0087)        & (0.0072)        & (0.0066)\\   
    Treated $\times$ Time +1                        & -0.0495$^{***}$ & -0.0382$^{***}$ & -0.0341$^{***}$ & -0.0325$^{***}$\\   
                                                    & (0.0088)        & (0.0073)        & (0.0070)        & (0.0070)\\   
    Treated $\times$ Time +2                        & -0.0641$^{***}$ & -0.0611$^{***}$ & -0.0601$^{***}$ & -0.0598$^{***}$\\   
                                                    & (0.0131)        & (0.0138)        & (0.0146)        & (0.0149)\\   
    Treated $\times$ Time +3                        & -0.0901$^{***}$ & -0.0903$^{***}$ & -0.0907$^{***}$ & -0.0909$^{***}$\\   
                                                    & (0.0292)        & (0.0282)        & (0.0283)        & (0.0284)\\   
    Treated $\times$ Time +4                        & -0.0480         & -0.0396         & -0.0362         & -0.0350\\   
                                                    & (0.0348)        & (0.0281)        & (0.0268)        & (0.0264)\\ 
   Treated $\times$ Time 0 $\times$ Below Limit     & 0.0036          & -0.0022         & -0.0052         & -0.0065\\   
                                                    & (0.0140)        & (0.0091)        & (0.0077)        & (0.0073)\\   
   Treated $\times$ Time 1 $\times$ Below Limit     & 0.0267$^{**}$   & 0.0143          & 0.0095          & 0.0075\\   
                                                    & (0.0100)        & (0.0085)        & (0.0086)        & (0.0088)\\   
   Treated $\times$ Time 2 $\times$ Below Limit     & 0.0312$^{**}$   & 0.0230$^{**}$   & 0.0207$^{**}$   & 0.0199$^{**}$\\   
                                                    & (0.0132)        & (0.0097)        & (0.0090)        & (0.0091)\\   
   Treated $\times$ Time 3 $\times$ Below Limit     & 0.0586$^{***}$  & 0.0483$^{***}$  & 0.0446$^{**}$   & 0.0432$^{**}$\\   
                                                    & (0.0177)        & (0.0151)        & (0.0152)        & (0.0154)\\   
   Treated $\times$ Time 4 $\times$ Below Limit     & 0.0184          & 0.0115          & 0.0069          & 0.0048\\   
                                                    & (0.0244)        & (0.0156)        & (0.0140)        & (0.0136)\\   
   \midrule
   Year $\times$ Below Limit                        & Yes             & Yes             & Yes             & Yes\\  
   Zip $\times$ Below Limit                         & Yes             & Yes             & Yes             & Yes\\  
   Disaster $\times$ Below Limit                    & Yes             & Yes             & Yes             & Yes\\  
   \midrule
   Observations                                        & 2,572,574       & 2,572,574       & 2,572,574       & 2,572,574\\  
   R$^2$                                               & 0.06785         & 0.06584         & 0.06548         & 0.06545\\  
   \bottomrule
   \multicolumn{5}{l}{\emph{Clustered (ZCTA5CE10 \& year) standard-errors in parentheses}}\\
   \multicolumn{5}{l}{\emph{Signif. Codes: ***: 0.01, **: 0.05, *: 0.1}}\\
\end{tabular*}

%% file: tables/regression_result_securitized_1_to_8_no_placebo.tex
\begin{tabular*}{0.8\textwidth}{@{\extracolsep{\fill}}lcccc}
   \toprule
   Dependent Variable: & \multicolumn{4}{c}{Securitized = 0,1}\\
   Bandwidth:                                          & $\pm1\%$            & $\pm2\%$            & $\pm3\%$            & $\pm4\%$  \\  
   \midrule
    Treated $\times$ Time 0                          & -0.0183         & -0.0253         & -0.0290$^{*}$   & -0.0296$^{*}$\\   
                                                     & (0.0150)        & (0.0149)        & (0.0146)        & (0.0141)\\   
    Treated $\times$ Time +1                         & -0.0604$^{***}$ & -0.0594$^{***}$ & -0.0560$^{**}$  & -0.0506$^{**}$\\   
                                                     & (0.0176)        & (0.0191)        & (0.0198)        & (0.0197)\\   
    Treated $\times$ Time +2                         & -0.0572$^{***}$ & -0.0483$^{*}$   & -0.0436$^{*}$   & -0.0384\\   
                                                     & (0.0180)        & (0.0228)        & (0.0232)        & (0.0227)\\   
    Treated $\times$ Time +3                         & -0.0594$^{*}$   & -0.0671$^{**}$  & -0.0661$^{**}$  & -0.0627$^{**}$\\   
                                                     & (0.0307)        & (0.0287)        & (0.0281)        & (0.0277)\\   
    Treated $\times$ Time +4                         & -0.1310$^{***}$ & -0.1272$^{***}$ & -0.1199$^{***}$ & -0.1116$^{***}$\\   
                                                     & (0.0438)        & (0.0316)        & (0.0285)        & (0.0269)\\   
   Treated $\times$ Time 0 $\times$ Below Limit      & 0.0083          & 0.0193          & 0.0208          & 0.0193\\   
                                                     & (0.0292)        & (0.0232)        & (0.0183)        & (0.0155)\\   
   Treated $\times$ Time 1 $\times$ Below Limit      & 0.0466          & 0.0567$^{*}$    & 0.0496          & 0.0398\\   
                                                     & (0.0314)        & (0.0309)        & (0.0299)        & (0.0288)\\   
   Treated $\times$ Time 2 $\times$ Below Limit      & 0.0342          & 0.0290          & 0.0202          & 0.0118\\   
                                                     & (0.0234)        & (0.0259)        & (0.0268)        & (0.0269)\\   
   Treated $\times$ Time 3 $\times$ Below Limit      & 0.0872$^{**}$   & 0.1042$^{***}$  & 0.1002$^{***}$  & 0.0933$^{***}$\\   
                                                     & (0.0372)        & (0.0262)        & (0.0249)        & (0.0246)\\   
   Treated $\times$ Time 4 $\times$ Below Limit      & 0.1663$^{***}$  & 0.1773$^{***}$  & 0.1688$^{***}$  & 0.1583$^{***}$\\   
                                                     & (0.0440)        & (0.0413)        & (0.0418)        & (0.0432)\\   
   \midrule
   Year $\times$ Below Limit                             & Yes             & Yes             & Yes             & Yes\\  
   Zip $\times$ Below Limit                              & Yes             & Yes             & Yes             & Yes\\  
   Disaster $\times$ Below Limit                         & Yes             & Yes             & Yes             & Yes\\  
   \midrule
   Observations                                        & 2,049,035       & 2,049,035       & 2,049,035       & 2,049,035\\  
   R$^2$                                               & 0.06641         & 0.07239         & 0.07727         & 0.08178\\  
   \midrule 
   Dependent Variable: & \multicolumn{4}{c}{Securitized = 0,1}\\
   Bandwidth:                                         & $\pm5\%$             & $\pm10\%$           & $\pm15\%$            & $\pm20\%$   \\  
   \midrule
    Treated $\times$ Time 0                         & -0.0292$^{**}$  & -0.0271$^{*}$         & -0.0263               & -0.0259\\   
                                                    & (0.0136)        & (0.0147)              & (0.0158)              & (0.0163)\\   
    Treated $\times$ Time +1                        & -0.0456$^{**}$  & -0.0340$^{*}$         & -0.0309               & -0.0297\\   
                                                    & (0.0192)        & (0.0187)              & (0.0192)              & (0.0195)\\   
    Treated $\times$ Time +2                        & -0.0342         & -0.0270               & -0.0254               & -0.0249\\   
                                                    & (0.0222)        & (0.0221)              & (0.0225)              & (0.0226)\\   
    Treated $\times$ Time +3                        & -0.0594$^{**}$  & -0.0520$^{*}$         & -0.0501$^{*}$         & -0.0494$^{*}$\\   
                                                    & (0.0273)        & (0.0273)              & (0.0277)              & (0.0279)\\   
    Treated $\times$ Time +4                        & -0.1045$^{***}$ & -0.0882$^{***}$       & -0.0830$^{***}$       & -0.0809$^{**}$\\   
                                                    & (0.0262)        & (0.0267)              & (0.0274)              & (0.0277)\\   
   Treated $\times$ Time 0 $\times$ Below Limit     & 0.0174          & 0.0125                & 0.0117                & 0.0115\\   
                                                    & (0.0136)        & (0.0117)              & (0.0121)              & (0.0124)\\   
   Treated $\times$ Time 1 $\times$ Below Limit     & 0.0313          & 0.0157                & 0.0139                & 0.0136\\   
                                                    & (0.0273)        & (0.0226)              & (0.0216)              & (0.0214)\\   
   Treated $\times$ Time 2 $\times$ Below Limit     & 0.0057          & -0.0024               & -0.0027               & -0.0027\\   
                                                    & (0.0267)        & (0.0243)              & (0.0229)              & (0.0224)\\   
   Treated $\times$ Time 3 $\times$ Below Limit     & 0.0867$^{***}$  & 0.0688$^{**}$         & 0.0638$^{**}$         & 0.0621$^{**}$\\   
                                                    & (0.0246)        & (0.0263)              & (0.0276)              & (0.0282)\\   
   Treated $\times$ Time 4 $\times$ Below Limit     & 0.1487$^{***}$  & 0.1226$^{**}$         & 0.1146$^{**}$         & 0.1115$^{*}$\\   
                                                    & (0.0448)        & (0.0502)              & (0.0523)              & (0.0532)\\   
   \midrule
   Year $\times$ Below Limit                            & Yes             & Yes                   & Yes                   & Yes\\  
   Zip $\times$ Below Limit                             & Yes             & Yes                   & Yes                   & Yes\\  
   Disaster $\times$ Below Limit                        & Yes             & Yes                   & Yes                   & Yes\\  
   \midrule
   Observations                                        & 2,049,035       & 2,049,035             & 2,049,035             & 2,049,035\\  
   R$^2$                                               & 0.08595         & 0.09977               & 0.10505               & 0.10724\\  
   \midrule \midrule
   \multicolumn{5}{l}{\emph{Clustered (ZCTA5CE10 \& year) standard-errors in parentheses}}\\
   \multicolumn{5}{l}{\emph{Signif. Codes: ***: 0.01, **: 0.05, *: 0.1}}\\
\end{tabular*}